\documentclass{article}
\usepackage{amssymb}
\usepackage{amsmath}
\usepackage{amsthm}
\usepackage{booktabs}
\usepackage{longtable}
\usepackage{ifpdf}
\usepackage[colorlinks=true, linkcolor=black, citecolor=black, urlcolor=blue]{hyperref}

\numberwithin{equation}{section}

\def\QTR#1#2{{\csname#1\endcsname#2}}{\relax}

\newtheorem*{thm}{Theorem}
\newtheorem{remark}{Remark}

\begin{document}

\title{Null Geodesic Congruences, Asymptotically-Flat Spacetimes and
  Their  Physical Interpretation}

\def\headertitle{Null Geodesic Congruences}

\author{Timothy M.\ Adamo \\ Mathematical Institute, \\University of Oxford, U.K. \\ \small\href{mailto:adamo@maths.ox.ac.uk}{\texttt{adamo@maths.ox.ac.uk}}
\and
Ezra T.\ Newman \\ Department of Physics and Astronomy, \\University of Pittsburgh, U.S.A. \\ \small\href{mailto:newman@pitt.edu}{\texttt{newman@pitt.edu}}
\and
Carlos Kozameh \\ Facultad de Matem\'atica, Astronom\'ia y F\'isica, \\ Universidad Nacional de C\'ordoba, Argentina \\ \small \href{mailto:kozameh@famaf.unc.edu.ar}{\texttt{kozameh@famaf.unc.edu.ar}}
}

\date{}
\maketitle

\begin{abstract}
A priori, there is nothing very special about shear-free or
asymptotically shear-free null geodesic congruences. Surprisingly,
however, they turn out to possess a large number of fascinating
geometric properties and to be closely related, in the context of
general relativity, to a variety of physically significant effects. It
is the purpose of this paper to try to fully develop these issues.

This work starts with a detailed exposition of the theory of shear-free and
asymptotically shear-free null geodesic congruences, i.e., congruences with
shear that vanishes at future conformal null infinity. A major portion of
the exposition lies in the analysis of the space of regular shear-free and
asymptotically shear-free null geodesic congruences. This analysis leads to
the space of complex analytic curves in an auxiliary four-complex
dimensional space, $\mathcal{H}$-space. They in turn play a dominant
role in the applications.

The applications center around the problem of extracting interior physical
properties of an asymptotically-flat spacetime directly from the asymptotic
gravitational (and Maxwell) field itself, in analogy with the determination
of total charge by an integral over the Maxwell field at infinity or the
identification of the interior mass (and its loss) by (Bondi's) integrals of
the Weyl tensor, also at infinity.

More specifically, we will see that the asymptotically shear-free congruences
lead us to an asymptotic definition of the center-of-mass and its equations
of motion. This includes a kinematic meaning, in terms of the center-of-mass
motion, for the Bondi three-momentum. In addition, we obtain insights into
intrinsic spin and, in general, angular momentum, including an
angular-momentum--conservation law with well-defined flux terms. When
a Maxwell field is present, the asymptotically shear-free congruences
allow us to determine/define at infinity a center-of-charge world line
and intrinsic magnetic dipole moment.
\end{abstract}


\newpage

\subsection*{\textit{Update: January 2012}}

These revisions were done exclusively by Adamo and Newman, therefore
the author order was slightly changed. There was nothing essentially wrong in the
earlier version, but we have included several new results (in the text
and in appendices), corrected an error of interpretation in~\ref{gauge-invariance}, and (the main reason for the revision) we found
much easier ways of doing some of the long calculations with very much
simpler arguments.
Roughly the changes are:  Aside from a few word changes,
\begin{enumerate}
\item We have extended the glossary at the beginning by including
        a Table~\ref{units} for the geometric quantities.

\item In Section~\ref{real-cuts-I} we have added in a fair amount
        to the discussion of real structures associated with the
        complex world lines.

\item The major revisions are in Section~\ref{results}.  Here we
        basically rewrote the entire section; i.e., the derivation of
        our major results, using much simpler arguments. This greatly
        shortened the derivation and associated argument.
        
\item In the Section~\ref{gauge-invariance}, we corrected an error in the discussion of representation theory.
        
\item We added items in Section~\ref{conclusion}
        and changed some of the wording in Section~\ref{background}.
        
\item 19 new references were added.
        
\item The Acknowledgements~\ref{acknowledgments} were changed.
        
\item Two new appendices (\ref{appendixE}, \ref{appendixF})
        were added presenting new material found since the previous
        version.
\end{enumerate}








\newpage

\tableofcontents

\newpage

\section{Introduction}

Though from the very earliest days of Lorentzian geometries, families of
null geodesics (null geodesic congruences (NGCs)) were obviously known to
exist, it nevertheless took many years for their significance to be
realized. It was from the seminal work of Bondi~\cite{Bondi}, with the
introduction of null surfaces and their associated null geodesics used for
the study of gravitational radiation, that the importance
of NGCs became recognized. To analyze the differential structure of such
congruences, Sachs~\cite{Sachs} introduced the fundamental `tools', known
as the optical parameters, namely, the divergence, the twist (or curl) and
the shear of the congruence. From the optical parameters one then could
classify congruences by the vanishing (or the asymptotic vanishing) of one
or more of these parameters. All the different classes exist in flat space
but, in general, only special classes exist in arbitrary spacetimes. For
example, in flat space, divergence-free congruences always exist, but for
non-flat vacuum spacetimes they exist only in the case of certain high
symmetries. On the other hand, twist-free congruences (null surface-forming
congruences) exist in all Lorentzian spacetimes. General vacuum spacetimes 
\textit{do not} allow shear-free congruences, though all
\textit{asymptotically-flat spacetimes} do allow
\textit{asymptotically} shear-free congruences, a natural
generalization of shear-free congruences, to exist.

Our primary topic of study will be the cases of shear-free and
asymptotically shear-free NGCs. In flat space the general shear-free
congruences have been extensively studied. However, only recently has the
special family of \textit{regular} congruences been investigated. In
general, as mentioned above, vacuum (or Einstein--Maxwell) metrics do
not possess shear-free congruences; the exceptions being the
algebraically-special metrics, all of which contain one or two such
congruences. On the other hand, all asymptotically-flat spacetimes
possess large numbers of \textit{regular asymptotically} shear-free
congruences.  By a `regular congruence' we mean a NGC that has all of its null geodesics coming from the interior of the space-time and intersecting with future null infinity; none of its geodesics lie on future null infinity.  This condition on the congruences play a fundamental role in the present work.

A priori there does not appear to be anything very special about
shear-free or asymptotically shear-free NGCs. However, over the years,
simply by observing a variety of topics, such as the classification of
Maxwell and gravitational fields (algebraically-special metrics),
twistor theory, $\mathcal{H}$-space theory and asymptotically-flat
spacetimes, there have been more and more reasons to consider them to
be of considerable importance. One of the earliest examples of this is
Robinson's~\cite{Robinson} demonstration that a necessary condition for a
curved spacetime to admit a null solution of Maxwell's equation is
that there be, in that space, a congruence of null, \textit{shear-free
  geodesics}. Recent results have shown that the
regular congruences -- both the shear-free and the asymptotically
shear-free congruences -- have certain very attractive and surprising
properties; each congruence is determined by a complex analytic curve
in the auxiliary complex space that is referred to as
$\mathcal{H}$-space. For asymptotically-flat spacetimes, some of these
curves contain a great deal of physical information about the
spacetime itself~\cite{PhysicalContent, UCF, RTmetrics}.

It is the main purpose of this work to give a relatively complete
discussion of these issues. However, to do so requires a digression.

A major research topic in general relativity (GR) for many years has been
the study of asymptotically flat spacetimes. Originally, the term
`asymptotically flat' was associated with gravitational fields, arising from
finite bounded sources, where infinity was approached along space-like
directions (e.g.,~\cite{Spacelike1,Spacelike2}). Then the very beautiful
work of Bondi~\cite{Bondi} showed that a richer and more meaningful idea
to be associated with `asymptotically flat' was to study gravitational
fields in which infinity was approached along null directions. This led to an
understanding of gravitational radiation via the Bondi energy-momentum loss
theorem, one of the profound results in GR. The Bondi energy-momentum loss
theorem, in turn, was the catalyst for the entire contemporary subject of
gravitational radiation and gravitational wave detectors. The fuzzy idea of
where and what is infinity was clarified and made more specific by the work
of Penrose~\cite{Scri1,Scri2} with the introduction of the conformal
compactification (via the rescaling of the metric) of spacetime, whereby
infinity was added as a boundary and brought into a finite spacetime
region. Penrose's infinity or spacetime boundary, referred to as Scri
or $\mathfrak{I}$, has many sub-regions: future null infinity,
$\mathfrak{I}^{+}$; past null infinity, $\mathfrak{I}^{-}$; future and
past timelike infinity, $\boldsymbol{I}^{+}$ and
$\boldsymbol{I}^{-}$; and spacelike infinity, $\boldsymbol{I}^{0}$~\cite{FrauendienerLR}.
In the present work, $\mathfrak{I}^{+}$ and its neighborhood will be
our primary arena for study.

A basic question for us is what information about the interior of the
spacetime can be obtained from a study of the asymptotic gravitational
field; that is, what can be learned from the remnant of the full field that
now `lives' or is determined on $\mathfrak{I}^{+}$? This quest is analogous
to obtaining the total interior electric charge or the electromagnetic
multipole moments directly from the asymptotic Maxwell field, i.e., the
Maxwell field at $\mathfrak{I}^{+}$, or the Bondi energy-momentum
four-vector from the gravitational field (Weyl tensor) at $\mathfrak{I}^{+}$. However, the ideas described and developed here are not really in the mainstream
of GR; they may lie outside the usual interest and knowledge of many researchers.
Nevertheless, they are strictly within GR: no new physics is introduced;
only the vacuum Einstein or Einstein--Maxwell equations are used. The ideas
come simply from observing (discovering) certain unusual and previously
overlooked features of solutions to the Einstein equations and their
asymptotic behavior.

These observations, as mentioned earlier, centered on the
realization of the remarkable properties and importance of the special
families of null geodesics: the regular shear-free and asymptotically
shear-free NGCs.

The most crucial and striking of these overlooked features (mentioned now
but fully developed later) are the following: in flat space every regular
shear-free NGC is determined by the \textit{arbitrary choice of a complex
analytic world line} in complex Minkowski space, $\mathbb{M}_{\mathbb{C}}$.
Furthermore and more surprising, for \textit{every asymptotically-flat
spacetime}, every regular asymptotically shear-free NGC is determined by
the given Bondi shear (given for the spacetime itself) and by the choice of
an arbitrary complex analytic world line in an auxiliary complex
four-dimensional space, $\mathcal{H}$-space, endowed with a complex
Ricci-flat metric. In other words, the space of regular shear-free
and asymptotically shear-free NGCs are both determined by arbitrary analytic
curves in $\mathbb{M}_{\mathbb{C}}$ and $\mathcal{H}$-space respectively~\cite{PhysicalContent, UCF, Footprints}.

Eventually, a \textit{unique} complex world line in this space is
singled out, with both the real and imaginary parts being given physical
meaning. The detailed explanation for the determination of this world line
is technical and reserved for a later discussion. However, a rough intuitive
idea can be given in the following manner.

The idea is a generalization of the trivial procedure in electrostatics
of first defining the electric dipole moment, relative to an origin, and
then shifting the origin so that the dipole moment vanishes and thus obtaining
the center of charge. Instead, we define, on $\mathfrak{I}^{+}$, with
specific Bondi coordinates and tetrad, the complex mass dipole moment (the
real mass dipole plus `$i$' times angular momentum) from certain
components of the asymptotic Weyl tensor. (The choice of the specific Bondi
system is the analogue of the choice of origin in the electrostatic case.)
Then, knowing how the asymptotic Weyl tensor transforms under a change of
tetrad and coordinates, one sees how the complex mass dipole moment changes
when the tetrad is rotated to one defined from the asymptotically shear-free
congruence. By setting the transformed complex mass dipole moment to zero,
the unique complex world line, identified as the complex center of mass, is
obtained. A similar process can be used in Einstein--Maxwell theory to obtain a complex center of charge.

This procedure, certainly unusual and perhaps
appearing ambiguous, does logically hold together. The real
justification for these identifications comes not from this logical
structure though, but rather from the observed equivalence of the derived results
from these identifications with well-known classical mechanical and
electrodynamical relations. These derived results involve both kinematical
and dynamical relations. Though they will be discussed at length later, we
mention that they range from a kinematic expression for the Bondi momentum
of the form, $P=Mv+\ldots$; a derivation of Newton's second law, $F=Ma$; and a
conservation law for angular momentum with a well-known angular momentum
flux, to the prediction of the Dirac value of the gyromagnetic ratio. We
note that, for the charged spinning particle metric~\cite{KerrNewman}, the
imaginary part of the world line is indeed the spin angular momentum, a
special case of our results.

A major early clue that shear-free NGCs were important in GR was the
discovery of the (vacuum or Einstein--Maxwell) algebraically
special metrics. These metrics are defined by the algebraic degeneracy
in their principle null vectors, which form (by the Goldberg--Sachs
theorem~\cite{GoldbergSachs}) a null congruence which is both
\textit{geodesic and shear-free}. For the asymptotically-flat
algebraically-special metrics, this shear-free congruence (a very
special congruence from the set of asymptotically shear-free
congruences) determines a unique world line in the associated
auxiliary complex $\mathcal{H}$-space. This shear-free
congruence (with its associated complex world line) is a special case
of the above argument of transforming to the complex center of
mass. Our general asymptotically-flat situation is, thus, a
generalization of the algebraically-special case. Much of the analysis leading 
to the transformation of the complex dipoles
in the case of the general asymptotically-flat spaces arose from
generalizing the case of the algebraically-special metrics.

To get a rough feeling (first in flat space) of how the curves in $\mathbb{M}_{\mathbb{C}}$ are connected with the shear-free congruences, we first point
out that the shear-free congruences are split into two classes: the twisting
congruences and the twist-free ones. The regular twist-free ones are simply
the null geodesics (the generators) of the light cones with apex on an
arbitrary timelike Minkowski space world line. Observing backwards along
these geodesics from afar, one `sees' the world line. The regular twisting
congruences are generated in the following manner: consider the
complexification of Minkowski space, $\mathbb{M}_{\mathbb{C}}$. Choose an
arbitrary complex (analytic) world line in $\mathbb{M}_{\mathbb{C}}$ and
construct its family of \textit{complex light cones}. The projection into
the real Minkowski space, $\mathbb{M}$, of the complex geodesics (the
generators of these complex cones), yields the real shear-free twisting NGCs~\cite{AdamoNewman4}. The twist contains or `remembers' the apex on the
complex world line, and looking backwards via these geodesics, one appears `to
see' the complex world line. In the case of asymptotically shear-free
congruences in curved spacetime, one cannot trace the geodesics back to a
complex world line. However, one can have the illusion (i.e., a virtual
image) that the congruence is coming from a complex world line. It is from
this property that we can refer to the asymptotically shear-free congruences
as lying on \textit{generalized} light cones. There is a duality between the
real twisting congruences and the complex congruences coming from the
complex world line: knowledge of one determines the other.

The analysis of the geometry of the asymptotically shear-free NGCs is
greatly facilitated by the introduction of Good-Cut Functions (GCFs). Each
GCF is a complex slicing of $\mathfrak{I}^{+}$ from which the
associated asymptotically shear-free NGC and world line can be easily
obtained. For the special world line and congruence that leads to the
complex center of mass, there is a unique GCF that is referred
to as the Universal-Cut Function (UCF).

Information about a variety of objects is contained in and can be easily
calculated from the UCF: the unique complex world line; the direction
of each geodesic of the congruence arriving at $\mathfrak{I}^{+}$; and the
Bondi asymptotic shear of the spacetime. The ideas behind the GCFs and UCF
are due to some very pretty mathematics arising from the study of the `good-cut equation' and
its complex four-dimensional solution space, $\mathcal{H}$-space~\cite{Hspace, HspaceR}. In flat space almost every asymptotically vanishing
Maxwell field determines its own Universal Cut Function, where the
associated world line determines both the center of charge and the magnetic
dipole moment. In general, for Einstein--Maxwell fields, there will be two
different UCFs, (and hence two different world lines), one for the Maxwell
field and one for the gravitational field. The physically interesting
special case where the two world lines coincide will be discussed.

In this work, we seek to provide a comprehensive overview of the theory of
asymptotically shear-free NGCs, as well as their
physical applications to both flat and asymptotically-flat spacetimes. The
resulting theoretical framework unites ideas from many areas of relativistic
physics and has a crossover with several areas of mathematics, which had
previously appeared short of physical applications.

The main mathematical tool used in our description of $\mathfrak{I}^{+}$ is
the Newman--Penrose (NP), or Spin-Coefficient (SC), formalism~\cite{NPF}. Spherical functions are
expanded in spin-$s$ tensor harmonics~\cite{Spins}; in our approximations
only the $l = 0,1,2$ harmonics are retained. Basically, the detailed
calculations should be considered as expansions around the Reissner--Nordstr\"{o}m metric, which is treated as zeroth order; all other terms being
small, i.e., at least first order. We retain terms only to second order.

In Section~\ref{foundations}, we give a brief review of Penrose's conformal
null infinity $\mathfrak{I}$ along with an exposition of the NP formalism
and its application to Maxwell theory and asymptotically-flat spacetimes. There is then a
description of $\mathfrak{I}^{+}$, the stage on which most of our
calculations take place. The Bondi \textit{mass aspect} (a function on $\mathfrak{I}^{+}$) is defined by the asymptotic Weyl tensor and asymptotic shear; from it we
obtain the physical identifications of the Bondi mass and linear momentum.
Also discussed is the asymptotic symmetry group of $\mathfrak{I}^{+}$, the
Bondi--Metzner--Sachs (BMS) group~\cite{Bondi, Sachs, BMS, BMS2}. The Bondi
mass and linear momentum become basic for the physical identification of the
complex center-of-mass world line.

Section~\ref{shear-free-NGC} contains the detailed analysis of shear-free
NGCs in Minkowski spacetime. This includes the identification of the flat
space GCFs from which all regular shear-free congruences can be found. We
also show the intimate connection between the flat space GCFs, the
(homogeneous) good-cut equation, and $\mathbb{M}_{\mathbb{C}}$. As
applications, we investigate the UCF associated with
asymptotically-vanishing Maxwell fields and in particular the shear-free
congruences associated with the Li\'{e}nard--Wiechert (and complex Li\'{e}nard--Wiechert) fields. This allows us to identify a real (and complex)
center-of-charge world line, as mentioned earlier.

In Section~\ref{good-cut-eq}, we give an overview of the machinery necessary
to deal with twisting asymptotically shear-free NGCs in asymptotically-flat
spacetimes. This involves a discussion of the theory of $\mathcal{H}$-space, the construction of the good-cut equation from the asymptotic Bondi
shear and its complex four-parameter family of solutions. We point out how
the simple Minkowski space of the preceding Section~\ref{shear-free-NGC} can
be seen as a special case of the more general theory outlined here. These
results have ties to Penrose's twistor theory and the theory of
Cauchy--Riemann (CR) structures; an explanation of these crossovers is given
in Appendices~\ref{appendixA} and \ref{appendixB}.

Section~\ref{applications} provides some examples of these ideas in action.  We discuss linear perturbations off the Schwarzschild metric, Robinson--Trautman and twisting type~II algebraically special metrics, as well as asymptotically stationary spacetimes, and illustrate how the good-cut equation can be solved and the UCF determined (explicitly or implicitly) in each case.

In Section~\ref{results}, the methodology laid out in the previous Sections~\ref{shear-free-NGC}, \ref{good-cut-eq} and~\ref{applications} is applied to
the general class of asymptotically-flat spacetimes: vacuum and
Einstein--Maxwell. Here, reviewing the material of the previous section, we
use the solutions of the good-cut equation to determine all regular
asymptotically shear-free NGCs by first choosing arbitrary world lines in
the solution space and then singling out a unique one which determines the UCF (two world lines exist in
the Einstein--Maxwell case, one for the gravitational field, the other for
the Maxwell field). This identification of the unique lines comes from a
study of the transformation properties, at $\mathfrak{I}^{+}$, of the
asymptotically-defined mass and spin dipoles and the electric and magnetic
dipoles. The work of Bondi, with the identification of energy-momentum and
its evolution, allows us to make a series of surprising further physical
identifications and predictions. In addition, with a slightly different
approximation scheme, we discuss our ideas applied to the asymptotic
gravitational field with an electromagnetic dipole field as the source.

Section~\ref{gauge-invariance} contains an analysis of the gauge (or BMS)
invariance of our results.

Section~\ref{conclusion}, the Discussion/Conclusion section, begins with a
brief history of the origin of the ideas developed here, followed by
comments on alternative approaches, possible physical predictions from our
results, a summary and open questions.

Finally, we conclude with six appendices, which contain several mathematical
crossovers that were frequently used or referred to in the text: twistor theory (\ref{appendixA}); CR
structures (\ref{appendixB}); a brief exposition of the tensorial spherical
harmonics~\cite{Spins} and their Clebsch--Gordon product decompositions (\ref{appendixC}); an
overview of the metric construction on $\mathcal{H}$-space (\ref{appendixD}); the description of certain real
aspects of complex Minkowski space world lines (\ref{appendixE}); and a discussion of the `generalized good-cut equation' with an arbitrary conformal factor (\ref{appendixF}).


\subsection{Notation and definitions}
\label{N&D}

The following contains the notational conventions that will be in use
throughout the course of this review.

\begin{itemize}
\item We use the symbols `$l$', `$m$', `$n$' \dots with
several different `decorations' but always meaning a null tetrad or a null
tetrad field.

a) Though in places, e.g., in Section~\ref{spin-coefficient}, the
symbols, $l^{a}$, $m^{a}$, $n^{a}$ \ldots, i.e., with an $a,b,c \dots$
\textit{can be thought of} as the abstract representation of a null
tetrad (i.e., Penrose's abstract index notation~\cite{Spinors}), in
general, our intention is to describe vectors in a coordinate
representation.

b) The symbols, $l^{a}$, $l^{\#\,a},l^{\ast \,a}$ most often represent the
coordinate versions of different null geodesic tangent fields, e.g., one-leg
of a Bondi tetrad field or some rotated version.

c) The symbol, $\hat{l}^{a}$, (with \textit{hat}) has a \textit{very
different meaning from the others}. It is used to represent the Minkowski
components of a normalized null vector giving the null directions on an
arbitrary light cone:
\begin{equation}
\hat{l}^{a}=\frac{\sqrt{2}}{2(1+\zeta \bar{\zeta})}\left( 1+\zeta \bar{\zeta}, \zeta +\bar{\zeta}, i\bar{\zeta}-i\zeta, -1+\zeta \bar{\zeta}\right)
\equiv \left(\frac{\sqrt{2}}{2}Y_{0}^{0},\frac{1}{2}Y_{1i}^{0}\right).
\label{l.hat}
\end{equation}
As the complex stereographic coordinates $(\zeta,\bar{\zeta})$ sweep out
the sphere, the $\hat{l}^{a}$ sweeps out the entire set of directions on the
future null cone. The other members of the associated null tetrad are 
\begin{eqnarray}
\hat{m}^{a} &=& \frac{\sqrt{2}}{2(1+\zeta \bar{\zeta})}\left( 0,
1-\bar{\zeta}^{2}, -i(1+\bar{\zeta}^{2}), 2\bar{\zeta}\right) , \label{m.hat} \\
\hat{n}^{a} &=& \frac{\sqrt{2}}{2(1+\zeta \bar{\zeta})}\left( 1+\zeta 
\bar{\zeta}, -(\zeta +\bar{\zeta}), i\zeta -i\bar{\zeta}, 1-\zeta \bar{\zeta}\right).  \notag
\end{eqnarray}

\item Several different time variables ($u_{\mathrm{B}}$,
$u_{\mathrm{ret}},\tau ,s$) and derivatives with respect to them are used.

The Bondi time, $u_{\mathrm{B}}$, is closely related to the retarded time,
$u_{\mathrm{ret}}=\sqrt{2}u_{\mathrm{B}}$. The use of the retarded time, $u_{\mathrm{ret}}$, is
important in order to obtain the correct numerical factors in the
expressions for the final physical results. Derivatives with respect to these variables are
represented by
\begin{eqnarray}
\partial_{u_{\mathrm{B}}}K & \equiv & \dot{K},  \label{derivatives} \\
\partial_{u_{\mathrm{ret}}}K & \equiv & K^{\prime}=\frac{\sqrt{2}}{2}\dot{K}.  \notag
\end{eqnarray}
The $u_{\mathrm{ret}},\tau,s$, derivatives are denoted by the same prime ($^{\prime}$) since it is always applied to functions with the same
functional argument. Though we are interested in real physical spacetime,
often the time variables ($u_{\mathrm{ret}},u_{\mathrm{B}},\tau$) take
complex values close to the real ($s$ is always real). Rather than putting
on `decorations' to indicate when they are real or complex (which burdens
the expressions with an over-abundance of different symbols), we leave
reality decisions to be understood from context. In a few places where the
reality of the particular variable is manifestly first introduced (and is
basic) we decorate the symbol by a superscript (${R}$), i.e., $u_{\mathrm{B}}^{({R})}$ or $u_{\mathrm{ret}}^{({R})}$. After their introduction we revert
to the undecorated symbol.

\begin{remark} At this point we are taking the velocity of light as $c=1$
and omitting it; later, when we want the correct units to appear explicitly,
we restore the $c$. This entails, via $\tau\rightarrow c\tau$, $s\rightarrow
cs$, changing the prime derivatives to include the $c$, i.e., 
\begin{equation}
\text{K}^{\prime} \rightarrow c^{-1}\text{K}^{\prime}.
\label{c^-1}
\end{equation}
\end{remark}

\item Often the angular (or sphere) derivatives, $\eth$ and $\overline{\eth}$, are used. The notation $\eth_{(\alpha)}K$ means, apply the $\eth$ operator
to the function $K$ while holding the variable $(\alpha)$ constant.

\item The complex conjugate is represented by the overbar, e.g., $\overline{\zeta}$. When a complex variable, $\tilde{\zeta}$, is close to the complex
conjugate of $\zeta$, but independent, we use $\tilde{\zeta} \approx 
\overline{\zeta}$.

\end{itemize}

Frequently, in this work, we use terms that are not in standard use. It
seems useful for clarity to have some of these terms defined from the outset:

\begin{itemize}

\item As mentioned earlier, we use the term `generalized
light cones' to mean (real) NGCs that \textit{appear} to
have their apexes on a world line in the complexification of the
spacetime. A detailed discussion of this will be given in Sections~\ref{shear-free-NGC} and \ref{good-cut-eq}.

\item The term `complex center of mass'
(or `complex center of charge') is
frequently used. Up to the choice of constants (to give correct units) they
basically lead to the `mass-dipole plus ``$i$'' angular
momentum' (or `real electric-dipole plus ``$i$'' magnetic dipole moment'). There will be two different
types of these `complex centers of \dots';
one will be geometrically defined or \textit{intrinsic}, i.e., independent
of the choice of coordinate system, the other will be \textit{relative},
i.e., it will depend on the choice of (Bondi) coordinates. The relations
between them are non-linear and non-local.

\item A very important technical tool used throughout this work is a class
of complex analytic functions, $u_{\mathrm{B}}=G(\tau,\zeta,\overline{\zeta})$, referred to as Good-Cut Functions, (GCFs) that are closely
associated with shear-free NGCs. The details are given later. For any given
asymptotically-vanishing Maxwell field with nonvanishing total charge, the
Maxwell field itself allows one, on physical grounds, to choose a \textit{unique member} of the class referred to as the (Maxwell) Universal-Cut
Function (UCF). For vacuum asymptotically-flat spacetimes, the Weyl tensor
allows the choice of a unique member of the class referred to as the
(gravitational) UCF. For Einstein--Maxwell there will be two such functions,
though in important cases they will coincide and be referred to as UCFs.
When there is no ambiguity, in either case, they will simple be UCFs.

\item A notational irritant arises from the following situation. Very
often we expand functions on the sphere in spin-$s$ harmonics, as, e.g.,
\begin{equation*}
\chi =\chi^{0}Y_{0}+\chi^{i}Y_{1i}(\zeta,
\overline{\zeta})+\chi^{ij}Y_{2ij}(\zeta,
\overline{\zeta})+\chi^{ijk}Y_{3ijk}(\zeta, \overline{\zeta})+ \ldots ,
\end{equation*}
where the indices, $i,j,k\ldots$ \textit{represent three-dimensional Euclidean
indices}. To avoid extra notation and symbols we write scalar products and
cross-products without the use of an explicit Euclidean metric, leading to
awkward expressions like
\begin{eqnarray*}
\overrightarrow{\eta}\cdot \overrightarrow{\lambda} & \equiv & \eta^{i}\lambda^{i}\equiv \eta^{i}\lambda_{i}, \\
\mu^{k} &=& (\overrightarrow{\eta}\times \overrightarrow{\lambda})^{k}\equiv \eta^{i}\lambda^{j}\epsilon_{ijk}.
\end{eqnarray*}
This, though easy to understand and keep track of, does run into the
unpleasant fact that often the four-vector, 
\begin{equation*}
\chi^{a}=(\chi^{0},\chi^{i}),
\end{equation*}
appears as the $l=0,1$ harmonics in the harmonic expansions. Thus, care must be used when lowering or raising the relativistic index, i.e.,
$\eta_{ab}\chi^{a}=\chi_{b}=(\chi^{0},-\chi^{i})$.

\item Throughout this review (and especially in Section~\ref{results}), we will invoke comparisons between our results and those of classical electromagnetism and relativity (c.f.,~\cite{LL}).  This process rests upon our identifications of the electric and magnetic dipole and quadrupole moments in the spherical harmonic expansions of the Maxwell tensor in the Newman-Penrose formalism.  Although the identifications we make are the most natural in our framework, a numerical re-scaling is required to obtain the physical formulae in some cases: in terms of the complex dipole and quadrupole moments used for the electromagnetic field, this is given by
\begin{equation*}
D^{i}_{\mathbb{C}}=D^{i\;\mathrm{physical}}_{\mathbb{C}}, \qquad Q^{ij}_{\mathbb{C}}=\frac{\sqrt{2}}{4}Q^{ij\;\mathrm{physical}}_{\mathbb{C}}.
\end{equation*}
The conventions used here were chosen so that the numerical coefficient of $Q^{ij}_{\mathbb{C}}$ in $\phi^{0}_{0}$ was equal to one; this re-scaling can simply be viewed as choosing a different (perhaps less natural) identification for the electromagnetic quadrupole moment, or as a sort of gauge choice for our results.

\end{itemize}

\newpage

\subsection{Glossary of Symbols and Units}

In this work we make use of substantial notational machinery.  The most frequently used symbols and acronyms are gathered here for easy reference:

\renewcommand{\arraystretch}{1.15}
\ifpdf
\begin{longtable}{l p{9.5cm}}
\else
\begin{table}[htbp]
\centering
\fi
  \caption{Glossary}
  \label{glossary}
\ifpdf\\\else
  \begin{tabular}{l p{9.5cm}}
\fi
  \toprule
  \textit{Symbol/Acronym} & \textit{Definition} \\ 
  \midrule
\ifpdf
  \endfirsthead
  \multicolumn{2}{c}{\small\textbf{\tablename} \thetable{} -- \emph{Continued}}
  \\[4mm]
  \toprule
  \textit{Symbol/Acronym} & \textit{Definition} \\ 
  \midrule
  \endhead
\fi
$\mathfrak{I}^{+}$, $\mathfrak{I}_{\mathbb{C}}^{+}$ & Future null
infinity, Complex future null infinity \\ 
$\boldsymbol{I}^{+}$, $\boldsymbol{I}^{-},\boldsymbol{I}^{0}$ & Future, Past timelike infinity, Spacelike infinity \\ 
$\mathbb{M}$, $\mathbb{M}_{\mathbb{C}}$ & Minkowski space, Complex Minkowski
space \\ 
$u_{\mathrm{B}}$, $u_{\mathrm{ret}}$ & Bondi time coordinate, Retarded Bondi
time ($\sqrt{2}u_{\mathrm{B}}=u_{\mathrm{ret}}$) \\ 
$\partial_{u_{\mathrm{B}}}f=\dot{f}$ & Derivation with respect to $u_{\mathrm{B}}$ \\ 
$\partial_{u_{\mathrm{ret}}}f=f^{\prime}$ & Derivation with respect to $u_{\mathrm{ret}}$ \\ 
$r$ & Affine parameter along null geodesics \\ 
$(\zeta,\bar{\zeta})$ & $(e^{i\phi}\cot (\theta /2),\ e^{-i\phi}\cot
(\theta /2))$; stereographic coordinates on $S^{2}$ \\ 
$Y_{li...j}^{s}(\zeta,\bar{\zeta})$ & Tensorial spin-$s$ spherical harmonics
\\ 
$\eth $, $\bar{\eth}$ & $P^{1-s}\frac{\partial}{\partial \zeta}P^{s},\ \
P^{1+s}\frac{\partial}{\partial \bar{\zeta}}P^{-s}$; spin-weighted operator on
the two-sphere \\ 
$P$ & Metric function on $S^{2}$; often $P=P_{0}\equiv 1+\zeta \bar{\zeta}$
\\ 
$\eth_{(\alpha )}f$ & Application of  $\eth $-operator to $f$ while the
variable $\alpha$ is held constant \\ 
$\{l^{a},n^{a},m^{a},\bar{m}^{a}\}$ & Null tetrad system; $l^{a}n_{a}=-m^{a}\bar{m}_{a}=1$ \\ 
NGC & Null Geodesic Congruence \\ 
NP/SC & Newman--Penrose/Spin-Coefficient Formalism \\ 
$\{U,X^{A},\omega ,\xi^{A}\}$ & Metric coefficients in the Newman--Penrose
formalism \\ 
$\{\psi_{0},\psi_{1},\psi_{2},\psi_{3},\psi_{4}\}$ & Weyl tensor
components in the Newman--Penrose formalism \\ 
$\{\phi_{0},\phi_{1},\phi_{2}\}$ & Maxwell tensor components in the
Newman--Penrose formalism \\ 
$\rho$ & Complex divergence of a null geodesic congruence \\ 
$\Sigma$ & Twist of a null geodesic congruence \\ 
$\sigma $, $\sigma^{0}$ & Complex shear, Asymptotic complex shear of a NGC
\\ 
$k$ & $\frac{2G}{c^{4}}$; Gravitational constant \\ 
$\tau=s+i\lambda =T(u,\zeta,\bar{\zeta})$ & Complex auxiliary (CR)
potential function \\ 
$\partial_{\tau}f=f^{\prime}$ & Derivation with respect to $\tau$ \\ 
$\eth_{(\tau)}^{2}G(\tau,\zeta,\bar{\zeta})=\sigma^{0}(G,\zeta,\bar{\zeta})$ & Good-Cut Equation, describing asymptotically shear-free NGCs \\
$u_{\mathrm{B}}=G(\tau,\zeta,\bar{\zeta})$ & Good-Cut Function (GCF) on $\mathfrak{I}^{+}$ \\   
$L(u_{\mathrm{B}},\zeta,\bar{\zeta})=\eth_{(\tau)}G$ & Stereographic angle field for
an asymptotically shear-free NGC at $\mathfrak{I}^{+}$ \\ 
$\eth_{(u_{\mathrm{B}})}T+L\dot{T}=0$ & CR equation, describing the
embedding of $\mathfrak{I}^{+}$ into $\mathbb{C}^{2}$\\ 
$\mathcal{H}$-space & Complex four-dimensional solution space to the Good-Cut
Equation \\ 
$D_{\mathbb{C}}^{i}=D_{E}^{i}+iD_{M}^{i}={\frac12}\phi_{0}^{0i}$ & Complex electromagnetic dipole \\
$\eta^{a}(u_{\mathrm{ret}})$ & Complex center-of-charge world line, lives
in $\mathcal{H}$-space \\ 
$Q^{ij}_{\mathbb{C}}=Q^{ij}_{E}+iQ^{ij}_{M}=\frac{\sqrt{2}}{4}Q^{ij\;\mathrm{physical}}_{\mathbb{C}}$ & Complex electromagnetic quadrupole \\
$\xi^{a}(u_{\mathrm{ret}})$ & Complex center of mass world line, lives in $\mathcal{H}$-space \\
$D_{(\mathrm{grav})}^{i}=D_{(\mathrm{mass})}^{i}+ic^{-1}J^{i}$ & ~ \\
\hspace{13mm} $=-\frac{c^{2}}{6\sqrt{2}G}\psi_{1}^{0i}$ & Complex gravitational dipole \\
$Q^{ij}_{\mathrm{Grav}}=Q^{ij}_{\mathrm{Mass}}+iQ^{ij}_{\mathrm{Spin}}$ & Complex gravitational quadrupole \\ 
$u_{\mathrm{B}}=X(\tau,\zeta,\bar{\zeta})$ & Universal Cut Function (UCF) corresponding to the complex center of mass world line \\ 
$\xi^{ij}=\frac{\sqrt{2}G}{24c^{4}}Q^{ij\prime\prime}_{\mathrm{Grav}}$ & Identification between $l=2$ coefficient of the UCF and gravitational quadrupole \\
$\Psi \equiv \psi_{2}^{0}+\eth^{2}\overline{\sigma^0}+\sigma^{0} \dot{\overline{\sigma^0}}=\bar{\Psi}$ & Bondi Mass Aspect \\ 
$M_{\mathrm{B}}=-\frac{c^{2}}{2\sqrt{2}G}\Psi^{0}$ & Bondi mass \\ 
$P^{i}=-\frac{c^{3}}{6G}\Psi^{i}$ & Bondi linear three-momentum \\ 
$J^{i}=-\frac{\sqrt{2}c^{3}}{12G}\text{ Im}(\psi_{1}^{0i})$ & Vacuum linear
theory identification of angular momentum \\
\bottomrule
\ifpdf
\end{longtable}
\else
\end{tabular}
\end{table}
\fi

\clearpage

In much of what follows, we use simplified units where $c=1$.  However, in Section~\ref{results} we will revert to a notation which makes dependence upon numerical constants explicit for the sake of comparing our results with well-known quantities in classical mechanics and electromagnetism.  We therefore include the following reference table for the units of several prominent objects in our calculations to ease in verifying that correct powers of dimensional constants (e.g., $c$, $G$) appear in our final results.  Here $[\cdot]$ stands for the units of a given quantity, and
\begin{equation*}
\mathrm{L}=[\mathrm{length}], \qquad \mathrm{M}=[\mathrm{mass}], \qquad \mathrm{T}=[\mathrm{time}].
\end{equation*}

\renewcommand{\arraystretch}{1.15}
\ifpdf
\begin{longtable}{ll}
\else
\begin{table}[htbp]
\centering
\fi
  \caption{Units}
  \label{units}
\ifpdf\\\else
  \begin{tabular}{ll}
\fi
  \toprule
  \textit{Quantity} & \textit{Units} \\ 
  \midrule
\ifpdf
  \endfirsthead
  \multicolumn{2}{c}{\small\textbf{\tablename} \thetable{} -- \emph{Continued}}
  \\[4mm]
  \toprule
  \textit{Quantity} & \textit{Units} \\ 
  \midrule
  \endhead
\fi
$[G]$ & $\mathrm{L}^{3}\mathrm{M}^{-1}\mathrm{T}^{-2}$ \\
$[q]$ & $\mathrm{M}^{\frac{1}{2}}\mathrm{L}^{\frac{3}{2}}\mathrm{T}^{-1}$ \\
$[k]=[Gc^{-4}]$ & $\mathrm{M}^{-1}\mathrm{L}^{-1}\mathrm{T}^2$ \\
$[c\tau]=[cu_{\mathrm{B}}]=[G(\tau, \zeta,\bar{\zeta})]$ & $\mathrm{L}$ \\
$[\xi^{i}(\tau)]=[\eta^{i}(\tau)]=[\xi^{ij}(\tau)]$ & $\mathrm{L}$ \\
$[D^{i}_{\mathbb{C}}]$ & $\mathrm{M}^{\frac{1}{2}}\mathrm{L}^{\frac{5}{2}}\mathrm{T}^{-1}$ \\
$[D^{i}_{(\mathrm{grav})}]$ & $\mathrm{ML}$ \\
$[Q^{ij}_{\mathbb{C}}]$ & $\mathrm{M}^{\frac{1}{2}}\mathrm{L}^{\frac{7}{2}}\mathrm{T}^{-1}$ \\
$[Q^{ij}_{\mathrm{Grav}}]$ & $\mathrm{ML}^2$ \\
$[J^{i}]$ & $\mathrm{ML}^{2}\mathrm{T}^{-1}$ \\
$[\phi_{0}]=[\phi_{1}]=[\phi_{2}]$ & $\mathrm{M}^{\frac{1}{2}}\mathrm{L}^{-\frac{1}{2}}\mathrm{T}^{-1}$ \\
$[\phi^{0}_{0}]$ & $\mathrm{M}^{\frac{1}{2}}\mathrm{L}^{\frac{5}{2}}\mathrm{T}^{-1}$ \\
$[\phi^{0}_{1}]$ & $\mathrm{M}^{\frac{1}{2}}\mathrm{L}^{\frac{3}{2}}\mathrm{T}^{-1}$ \\
$[\phi^{0}_{2}]$ & $\mathrm{M}^{\frac{1}{2}}\mathrm{L}^{\frac{1}{2}}\mathrm{T}^{-1}$ \\
$[\psi_{0}]=[\psi_{1}]=[\psi_{2}]=[\psi_{3}]=[\psi_{4}]$ & $\mathrm{L}^{-2}$ \\
$[\psi^{0}_{0}]$ & $\mathrm{L}^3$ \\
$[\psi^{0}_{1}]$ & $\mathrm{L}^2$ \\
$[\psi^{0}_{2}]$ & $\mathrm{L}$ \\
$[\psi^{0}_{3}]$ & $1$ \\
$[\psi^{0}_{4}]$ & $\mathrm{L}^{-1}$ \\
\bottomrule
\ifpdf
\end{longtable}
\else
\end{tabular}
\end{table}
\fi



\newpage

\section{Foundations}
\label{foundations}

In this section, we review several of the key ideas and tools that are
indispensable in our later discussions. We keep our explanations as
concise as possible, and refrain from extensive proofs of any propositions.
The reader will be directed to the appropriate references for the
details. In large part, much of what is covered in this section should
be familiar to many workers in GR.


\subsection[Asymptotic flatness and $\mathfrak{I}^{+}$]{Asymptotic flatness and \boldmath$\mathfrak{I}^{+}$}

Ever since the work of Bondi~\cite{Bondi} illustrated the importance
of null hypersurfaces in the study of outgoing gravitational radiation, the
study of asymptotically-flat spacetimes has been one of the more important
research topics in GR. Qualitatively speaking, a spacetime can be
thought of as (future) asymptotically flat if the curvature tensor vanishes
at an appropriate rate as infinity is approached along the future-directed
null geodesics of the null hypersurfaces. The type of physical situation we
have in mind is an arbitrary compact gravitating source (perhaps with an
electric charge and current distribution), with the associated gravitational
(and electromagnetic) field. The task is to gain information about the
interior of the spacetime from the study of far-field features, multipole
moments, gravitational and electromagnetic radiation,
etc.~\cite{NewmanTod}. The arena for this study is on what is
referred to as future null infinity, $\mathfrak{I}^{+}$, the future
boundary of the spacetime. The intuitive picture of this boundary is
the set of all endpoints of future-directed null geodesics.

A precise definition of null asymptotic flatness and the boundary was
given by Penrose~\cite{Scri1, Scri2}, whose basic idea was to rescale the
spacetime metric by a conformal factor, which approaches zero
asymptotically: the zero value defining future null infinity. This process
leads to the boundary being a null hypersurface for the conformally-rescaled
metric. When this boundary can be attached to the interior of the rescaled
manifold in a regular way, then the spacetime is said to be asymptotically
flat.

As the details of this formal structure are not used here, we will rely
largely on the intuitive picture.  A thorough review of this subject can be
found in~\cite{FrauendienerLR}. However, there are a number of important
properties of $\mathfrak{I}^{+}$ arising from Penrose's construction that we
rely on~\cite{NewmanTod, Scri1, Scri2}:

(A): For both the asymptotically-flat vacuum Einstein equations and the
Einstein--Maxwell equations, $\mathfrak{I}^{+}$ is a null hypersurface of the
conformally rescaled metric.

(B): $\mathfrak{I}^{+}$ is topologically $S^{2}\times\mathbb{R}$.

(C): The Weyl tensor $C_{bcd}^{a}$ vanishes at $\mathfrak{I}^{+}$,
with the peeling theorem describing the speed of its falloff (see below).

Property (B) allows an easy visualization of the boundary,
$\mathfrak{I}^{+}$, as the past light cone of the point
$\boldsymbol{I}^{+}$, future timelike infinity. As mentioned earlier,
$\mathfrak{I}^{+}$ will be the stage for our study of asymptotically
shear-free NGCs.


\subsection{Bondi coordinates and null tetrad}

Proceeding with our examination of the properties of
$\mathfrak{I}^{+}$, we introduce, in the neighborhood of
$\mathfrak{I}^{+}$, what is known as a Bondi coordinate system:
$(u_{\mathrm{B}},r,\zeta, \bar{\zeta})$. In this system, $u_{\mathrm{B}}$, the Bondi
time, labels the null surfaces, $r$ is the affine parameter along the
null geodesics of the constant $u_{\mathrm{B}}$ surfaces and $\zeta
=e^{i\phi}\cot (\theta /2)$, the complex stereographic coordinate labelling
the null geodesics of $\mathfrak{I}^{+}$. To reach $\mathfrak{I}^{+}$, we
simply let $r\rightarrow \infty $, so that $\mathfrak{I}^{+}$ has
coordinates $(u_{\mathrm{B}},\zeta, \bar{\zeta})$. The time coordinate $u_{\mathrm{B}}$,
the topologically $\mathbb{R}$ portion of $\mathfrak{I}^{+}$, labels
`cuts' of $\mathfrak{I}^{+}$. The stereographic coordinate $\zeta$
accounts for the topological generators of the $S^{2}$ portion of
$\mathfrak{I}^{+}$, i.e., the null generators of
$\mathfrak{I}^{+}$. The choice of a Bondi coordinate system is not
unique, there being a variety of Bondi coordinate systems to choose
from. The coordinate transformations between any two, known as
Bondi--Metzner--Sachs (BMS) transformations or as the BMS group, are
discussed later in this section.

Associated with the Bondi coordinates is a (Bondi) null tetrad system,
($l^{a},n^{a},m^{a},\overline{m}^{a}$).  The first tetrad vector
$l^{a}$ is the tangent to the geodesics of the constant
$u_{\mathrm{B}}$ null surfaces given by~\cite{NewmanTod}
\begin{eqnarray}
l^{a} &=& \frac{dx^{a}}{dr}=g^{ab}\nabla_{b}u_{\mathrm{B}},  \label{Tet1} \\
l^{a}\nabla_{a}l^{b} &=& 0, \\
l^{a}\frac{\partial}{\partial x^{a}} &=& \frac{\partial}{\partial r}.
\label{Tet2}
\end{eqnarray}
The second null vector $n^{a}$ is normalized so that: 
\begin{equation}
l_{a}n^{a} = 1.
\label{Tet3}
\end{equation}
In Bondi coordinates, we have~\cite{NewmanTod} 
\begin{equation}
n^{a}\frac{\partial}{\partial x^{a}}=\frac{\partial}{\partial
  u_{\mathrm{B}}}+U\frac{\partial}{\partial r}+X^{A}\frac{\partial}{\partial
  x^{A}},
\label{Tet4}
\end{equation}
for functions $U$ and $X^{A}$ to be determined, and $A=\zeta,\overline{\zeta}$. At $\mathfrak{I}^{+}$, $n^{a}$
is tangent to the null generators of $\mathfrak{I}^{+}$.

The tetrad is completed with the choice of a complex null vector
$m^{a}$, $(m^{a}m_{a}=0)$ which is itself orthogonal to both $l_{a}$
and $n_{a}$, initially tangent to the constant $u_{\mathrm{B}}$ cuts at
$\mathfrak{I}^{+}$ and parallel propagated inward on the null
geodesics. It is normalized by 
\begin{equation}
m^{a}\bar{m}_{a}=-1.
\label{Tet5}
\end{equation}
Once more, in coordinates, we have~\cite{NewmanTod} 
\begin{equation}
m^{a}\frac{\partial}{\partial x^{a}}=\omega \frac{\partial}{\partial
  r}+\xi^{A}\frac{\partial}{\partial x^{A}},
\label{Tet6}
\end{equation}
for some $\omega$ and $\xi^{A}$ to be determined. All other scalar
products in the tetrad are to vanish.

With the tetrad thus defined, the contravariant metric of the spacetime is
given by
\begin{equation}
g^{ab}=l^{a}n^{b}+l^{b}n^{a}-m^{a}\bar{m}^{b}-m^{b}\bar{m}^{a}.
\label{Tet7}
\end{equation}
In terms of the \textit{metric coefficients} $U$, $\omega$,
$X^{A}$, and $\xi^{A}$, the metric can be written as:
\begin{eqnarray}
g^{ab} &=& \left( 
\begin{array}{ccc}
0 & 1 & 0 \\ 
1 & g^{11} & g^{1A} \\ 
0 & g^{1A} & g^{AB}
\end{array}
\right),  \label{Tet8} \\
g^{11} &=& 2(U-\omega\bar{\omega}),  \notag \\
g^{1A} &=& X^{A}-(\bar{\omega}\xi^{A}+\omega\bar{\xi}^{A}),  \notag \\
g^{AB} &=& -(\xi^{A}\bar{\xi}^{B}+\xi^{B}\bar{\xi}^{A}).  \notag
\end{eqnarray}
We thus have the spacetime metric in terms of the metric coefficients.

There remains the issue of both coordinate and tetrad freedom, i.e., local
Lorentz transformations. Most of the time we work in one arbitrary but fixed
Bondi coordinate system, though for special situations more general
coordinate systems are used. The more general transformations are given,
essentially, by choosing an arbitrary slicing of $\mathfrak{I}^{+}$, written
as $u_{\mathrm{B}}=G(s,\zeta, \bar{\zeta})$ with $s$ labelling the slices. To keep
conventional coordinate conditions unchanged requires a rescaling of
$r: r\rightarrow r^{\prime}=(\partial_{s}G)^{-1}r$. It is also
useful to be able to shift the origin of $r$ by
$r^{\prime}=r-r_{0}(u_{\mathrm{B}}, \zeta, \bar{\zeta})$ with arbitrary
$r_{0}(u_{\mathrm{B}}, \zeta, \bar{\zeta})$.

The tetrad freedom of null rotations around $n^{a}$,
performed in the neighborhood of $\mathfrak{I}^{+}$, will later play a
major role. For an arbitrary function $L(u_{\mathrm{B}}, \zeta, \bar{\zeta})$
on $\mathfrak{I}^{+}$, the null rotation about the vector
$n^{a}$~\cite{NewmanTod} is given by 
\begin{eqnarray}
l^{a} & \rightarrow & l^{\ast a}=l^{a}-\frac{\bar{L}}{r}m^{a}-\frac{L}{r}\bar{m}^{a}+0(r^{-2}),  \label{Tet10} \\
m^{a} & \rightarrow & m^{\ast \,a}=m^{a}-\frac{L}{r}n^{a}+0(r^{-2}), \\
n^{a} & \rightarrow & n^{\ast \,a}=n^{a}.
\end{eqnarray}

Eventually, by the appropriate choice of the function $L (u_{\mathrm{B}},
\zeta, \bar{\zeta})$, the new null vector, $l^{\ast a}$, can be made
into the tangent vector of an asymptotically shear-free NGC.

A second type of tetrad transformation is the rotation in the tangent $(m^{a},\overline{m}^{a})$ plane, which keeps $l^{a}$ and $n^{a}$ fixed:
\begin{equation}
m^{a}\rightarrow e^{i\lambda}m^{a}, \quad \lambda\in\mathbb{R}.
\label{Tet11}
\end{equation}
This latter transformation provides motivation for the concept of
\textit{spin weight}. A quantity $\eta_{(s)}(\zeta,\bar{\zeta})$ is
said to have spin-weight $s$ if, under the transformation,
Eq.~(\ref{Tet11}), it transforms as 
\begin{equation}
\eta \rightarrow \eta_{(s)}^{\ast}(\zeta,\bar{\zeta})=e^{is\lambda}\eta_{(s)}(\zeta,\bar{\zeta}).
\label{Tet12}
\end{equation}

An example would be to take a vector on $\mathfrak{I}^{+}$, say $\eta^{a}$,
and form the spin-weight-one quantity, $\eta_{(1)}=\eta^{a}m_{a}$.

\textbf{Comment:} For later use we note that $L (u_{\mathrm{B}}, \zeta,
\bar{\zeta})$ has spin weight, $s=1$.

For each $s$, spin-$s$ functions can be expanded in a complete basis set,
the spin-$s$ harmonics, $_{s}Y_{lm}(\zeta,\bar{\zeta})$ or spin-$s$ tensor
harmonics, $Y_{l\,i...j}^{(s)}(\zeta, \bar{\zeta})\Leftrightarrow
{}_{s}Y_{lm}(\zeta, \bar{\zeta})$ (cf.~Appendix~\ref{appendixC}).

A third tetrad transformation, the boosts, are given by 
\begin{equation}
l^{\#\,a}=Kl^{a}, \qquad n^{\#\,a}=K^{-1}n^{a}.
\label{boost}
\end{equation}
These transformations induce the idea of conformal weight, an idea similar
to spin weight. Under a boost transformation, a quantity, $\eta_{(w)}$,
will have conformal weight $w$ if 
\begin{equation}
\eta_{(w)}\rightarrow\eta_{(w)}^{\#}=K^{w}\eta_{(w)}.
\label{Tet14}
\end{equation}

Sphere derivatives of spin-weighted functions $\eta_{(s)}(\zeta,
\bar{\zeta})$ are given by the action of the operators $\eth$ and its
conjugate operator $\bar{\eth}$, defined by~\cite{Edth} 
\begin{equation}
\eth \eta_{(s)}=P^{1-s}\frac{\partial (P^{s}\eta_{(s)})}{\partial \zeta},
\label{Edth}
\end{equation}
\begin{equation}
\bar{\eth}\eta_{(s)}=P^{1+s}\frac{\partial (P^{-s}\eta_{(s)})}{\partial 
\bar{\zeta}},
\label{Edthbar}
\end{equation}
where the function $P$ is the conformal factor defining the conformal sphere
metric, 
\begin{equation*}
ds^{2}=\frac{4d\zeta d\bar{\zeta}}{P^{2}},
\end{equation*}
most often taken as the unit metric sphere by 
\begin{equation*}
P=P_{0}\equiv 1+\zeta \bar{\zeta}.
\end{equation*}


\subsection{The optical equations}

Since this work concerns NGCs and, in particular, \textit{shear-free}
and \textit{asymptotically shear-free} NGCs, it is necessary to first
define them and then study their properties.

Given a Lorentzian manifold with local coordinates, $x^{a}$, and a
NGC, i.e., a foliation by a three parameter family of null geodesics,
\begin{equation}
x^{a}=X^{a}(r,y^{w}),
\label{congruence}
\end{equation}
with $r$ the affine parametrization and the (three) $y^{w}$ labelling the
geodesics, the tangent \textit{vector field}
$l^{a}=DX^{a}\equiv\partial_{r}X^{a}$ satisfies the geodesic equation
\begin{equation*}
l^{a}\nabla_{a}l^{b}=0.
\end{equation*}

The two complex optical scalars (spin coefficients), $\rho$ and $\sigma$,
are defined by
\begin{eqnarray}
\rho &=& \nabla_{a}l_{b}\overline{m}^{a}m^{b},  \label{rho} \\
\sigma & = &  \nabla_{(a}l_{b)}m^{a}m^{b}
\notag
\end{eqnarray}
with $m^{a}$ an complex (spacelike) vector that is parallel propagated along the null geodesic field and satisfies
$m^{a}m_{a}=m^{a}l_{a}=m^{a}\overline{m}_{a}+1=0$.

The $\rho$ and $\sigma$ satisfy the \textit{optical equations} of
Sachs~\cite{Sachs}, namely, 
\begin{equation}
\frac{\partial \rho}{\partial r}=\rho^{2}+\sigma \bar{\sigma}+\Phi_{00},
\end{equation}
\begin{equation}
\frac{\partial \sigma}{\partial r}=(\rho+\bar{\rho}) \sigma +\psi_{0},
\end{equation}
\begin{eqnarray*}
\Phi_{00} &=& R_{ab}\,l^{a}l^{b}, \\
\psi_{0} &=& -C_{abcd}\,l^{a}m^{b}l^{c}m^{d},
\end{eqnarray*}
where $\Phi_{00}$ and $\Psi_{0}$ are, respectively, a Ricci and a Weyl
tensor tetrad component (see below). In flat space, i.e., with
$\Phi_{00} = \Psi_{0}=0$, excluding the degenerate case of $\rho
\overline{\rho}-\sigma \overline{\sigma}=0$, plane ($\rho=\sigma=0$) and cylindrical
fronts, the general solution is
\begin{equation}
\rho =\frac{i\Sigma -r}{r^{2}+\Sigma^{2}-\sigma^{0}\overline{\sigma}^{0}},
\label{SC21}
\end{equation}
\begin{equation}
\sigma =\frac{\sigma^{0}}{r^{2}+\Sigma^{2}-\sigma^{0}\overline{\sigma}^{0}}.
\label{SC22}
\end{equation}
The complex $\sigma^{0}$ (referred to as the asymptotic shear) and the
real $\Sigma$ (called the twist) are determined from the original
congruence, Eq.~(\ref{congruence}). Both are functions just of the
parameters, $y^{w}$. Their behavior for large $r$ is given by
\begin{eqnarray}
\rho &=& -\frac{1}{r}+\frac{i\Sigma}{r^{2}}+\frac{\Sigma^{2}}{r^{3}}-\frac{\sigma^{0}\overline{\sigma}^{0}}{r^{3}}+O(r^{-4}), \\
\sigma &=& \frac{\sigma^{0}}{r^{2}}+O(r^{-4}).
\end{eqnarray}
From this, $\sigma^{0}$ gets its name as the asymptotic shear. In
Section~\ref{shear-free-NGC}, we return to the issue of the explicit
construction of NGCs in Minkowski space and in
particular to the construction and detailed properties of
\textit{regular shear-free congruences}.

Note the important point that, in $\mathbb{M}$, the vanishing of the
asymptotic shear forces the shear to vanish. The same is not true for
asymptotically-flat spacetimes. Specifically, for future null
asymptotically-flat spaces described in a Bondi tetrad and coordinate
system, we have, from other considerations, that 
\begin{eqnarray*}
\Phi_{00} &=& O(r^{-6}), \\
\psi_{0} &=& O(r^{-5}), \\
\Sigma &=& 0,
\end{eqnarray*}
which leads to the asymptotic behavior of $\rho$ and $\sigma$, 
\begin{eqnarray*}
\rho &=& \overline{\rho}=-\frac{1}{r}+\frac{\sigma^{0}\overline{\sigma}^{0}}{r^{3}}+O(r^{-5}), \\
\sigma &=& \frac{\sigma^{0}}{r^{2}}+O(r^{-4}),
\end{eqnarray*}
with the two order symbols explicitly depending on the leading terms
in $\Phi_{00}$ and $\Psi_{0}$. The vanishing of $\sigma^{0}$
\textit{does not}, in this nonflat case, imply that $\sigma$
vanishes. This case, referred to as asymptotically shear-free, plays
the major role later. It will be returned to in greater detail in
Section~\ref{good-cut-eq}.


\subsection{The Newman--Penrose formalism}

\label{spin-coefficient}

Though the NP formalism is the basic working tool
for our analysis, this is not the appropriate venue for its detailed
exposition. Instead we will simply give an outline of the basic ideas
followed by the results found, from the application of the NP
equations, to the problem of asymptotically-flat spacetimes.

The NP version~\cite{NPF, NewmanTod, NewmanPenrose2009} of the vacuum
Einstein (or the Einstein--Maxwell) equations uses the tetrad
components
\begin{equation}
\lambda_{i}^{a}=(l^{a},n^{a},m^{a},\bar{m}^{a}),
\label{SC1}
\end{equation}
$(i=1,2,3,4)$ rather than the metric, as the basic variable. (An
alternate version, not discussed here, is to use a pair of
two-component spinors~\cite{Spinors}) The
metric, Eq.~(\ref{Tet7}), can be written compactly as 
\begin{equation}
g^{ab}=\eta^{ij}\lambda_{i}^{a}\lambda_{j}^{b},
\label{SC2}
\end{equation}
with 
\begin{equation}
\eta^{ij}=\left( 
\begin{array}{cccc}
0 & 1 & 0 & 0 \\ 
1 & 0 & 0 & 0 \\ 
0 & 0 & 0 & -1 \\ 
0 & 0 & -1 & 0
\end{array}
\right) .
\label{SC3}
\end{equation}

The complex spin coefficients, which play the role of the connection, are
determined from the Ricci rotation coefficients~\cite{NPF, NewmanTod}: 
\begin{equation}
\gamma_{jk}^{i}=\lambda_{j}^{b}\lambda_{k}^{a}\nabla_{a}\lambda
_{b}^{i}\equiv \lambda_{j}^{b}\lambda_{b;k}^{i},  \label{SC5}
\end{equation}
via the linear combinations 
\begin{equation}
\begin{array}{ccc}
\alpha = {\frac12} (\gamma_{124}-\gamma_{344}), & \lambda =-\gamma_{244}, & \kappa =\gamma_{131}, \\ 
\beta = {\frac12} (\gamma_{123}-\gamma_{343}), & \mu =-\gamma_{243}, & \rho =\gamma_{134}, \\ 
\gamma = {\frac12} (\gamma_{122}-\gamma_{342}), & \nu =-\gamma_{242}, & \sigma =\gamma_{133}, \\ 
\varepsilon = {\frac12} (\gamma_{121}-\gamma_{341}), & \pi =-\gamma_{241}, & \tau =\gamma_{132}.
\end{array}
\label{SC7}
\end{equation}

The third basic variable in the NP formalism is the Weyl tensor or,
equivalently, the following five complex tetrad components of the Weyl
tensor: 
\begin{equation}
\begin{array}{cc}
\psi_{0}=-C_{abcd}\,l^{a}m^{b}l^{c}m^{d}, & \quad
\psi_{1}=-C_{abcd}\,l^{a}n^{b}l^{c}m^{d},
\end{array}
\end{equation}

\begin{equation}
\psi_{2}=-\frac{1}{2}\left(C_{abcd}\,l^{a}n^{b}l^{c}n^{d}-C_{abcd}\,l^{a}n^{b}m^{c}\overline{m}^{d}\right),
\end{equation}

\begin{equation}
\begin{array}{cc}
\psi_{3}=C_{abcd}\,l^{a}n^{b}n^{c}\overline{m}^{d}, & \quad
\psi_{4}=C_{abcd}\,n^{a}\overline{m}^{b}n^{c}\overline{m}^{d}.
\end{array}
\end{equation}
Note that we have adopted the sign conventions of~\cite{NPF}, which differ from those in~\cite{Spinors}.

When an electromagnetic field is present, we must include the complex tetrad
components of the Maxwell field into the equations: 
\begin{eqnarray}
\phi_{0} &=& F_{ab}\,l^{a}m^{b}, \\
\phi_{1} &=& \frac{1}{2}F_{ab}\left(l^{a}n^{b}+m^{a}\overline{m}^{b}\right),  \notag \\
\phi_{2} &=& F_{ab}\,n^{a}\overline{m}^{b},  \notag
\end{eqnarray}
as well as the Ricci (or stress tensor) constructed from the three
$\phi_{i}$, e.g., $T_{ab}\,l^{a}l^{b}=\phi_{0}\overline{\phi}_{0}$, with $R_{ab}=kT_{ab}, 
k=2Gc^{-4}$.

\begin{remark}
We mention that much of the physical content and
interpretations in the present work comes from the study of the lowest
spherical harmonic coefficients in the leading terms of the far-field
expansions of the Weyl and Maxwell tensors.
\end{remark}

The NP version of the vacuum (or Einstein--Maxwell) equations consists of
three sets (or four sets) of nonlinear first-order coupled partial
differential equations for the variables: the tetrad components, the
spin coefficients, the Weyl tensor (and Maxwell field when present). Though
there is no hope that they can be solved in any general sense, many exact
solutions have been found from them. Of far more importance, large classes
of asymptotic solutions and perturbation solutions can be found. Our
interest lies in the asymptotic behavior of the asymptotically-flat
solutions. Though there are some subtle issues, integration in this class is
not difficult~\cite{NPF, NUT}. With no explanation of the integration
process, except to mention that we use the Bondi coordinate and tetrad
system of Eqs.~(\ref{Tet2}), (\ref{Tet4}), and~(\ref{Tet6}) and asymptotic
flatness ($\psi_{0}\sim O(r^{-5})$ and certain uniform smoothness conditions on sideways derivatives), we simply give the final results.

First, the radial behavior is described. The quantities with a zero
superscript, e.g., $\sigma^{0}$, $\psi_{2}^{0}$, \ldots, are
`\textit{functions of integration}', i.e., functions only of ($u_{\mathrm{B}},
\zeta, \overline{\zeta}$).

\begin{itemize}

\item The Weyl tensor:
\begin{eqnarray}
\psi_{0} &=& \psi_{0}^{0}r^{-5}+O(r^{-6}),  \label{peeling} \\
\psi_{1} &=& \psi_{1}^{0}r^{-4}+O(r^{-5}),  \notag \\
\psi_{2} &=& \psi_{2}^{0}r^{-3}+O(r^{-4}),  \notag \\
\psi_{3} &=& \psi_{3}^{0}r^{-2}+O(r^{-3}),  \notag \\
\psi_{4} &=& \psi_{4}^{0}r^{-1}+O(r^{-2}).  \notag
\end{eqnarray}

\item The Maxwell tensor: 
\begin{eqnarray}
\phi_{0} &=& \phi_{0}^{0}r^{-3}+O(r^{-4}),  \label{peeling II} \\
\phi_{1} &=& \phi_{1}^{0}r^{-2}+O(r^{-3}),  \notag \\
\phi_{2} &=& \phi_{2}^{0}r^{-1}+O(r^{-2}).  \notag
\end{eqnarray}

\item The spin coefficients and metric variables:
\begin{eqnarray}
\kappa &=& \pi =\epsilon =0, \qquad \tau =\overline{\alpha}+\beta,  \label{spincoef} \\
\rho   &=& \overline{\rho}=-r^{-1}-\sigma^{0}\overline{\sigma}^{0}r^{-3}+O(r^{-5}),  \notag \\
\sigma &=& \sigma^{0}r^{-2}+\left((\sigma^{0})^{2}\overline{\sigma}^{0}-\psi_{0}^{0}/2\right)r^{-4}+O(r^{-5}),  \notag \\
\alpha &=& \alpha^{0}r^{-1}+O(r^{-2}), \qquad \beta =\beta^{0}r^{-1}+O(r^{-2}),  \notag \\
\gamma &=& \gamma^{0}-\psi_{2}^{0}(2r^{2})^{-1}+O(r^{-3}), \qquad \lambda
=\lambda^{0}r^{-1}+O(r^{-2}),  \notag \\
\mu    &=& \mu^{0}r^{-1}+O(r^{-2}), \qquad \nu =\nu^{0}+O(r^{-1}), 
\notag
\end{eqnarray}

\begin{eqnarray*}
A       &=& \zeta\text{ or }\overline{\zeta}, \\
\xi^{A} &=& \xi^{0A}r^{-1}-\sigma^{0}\overline{\xi}^{0A}r^{-2}+\sigma^{0}\bar{\sigma}^{0}\xi^{0A}r^{-3}+O(r^{-4}), \\
\omega  &=& \omega^{0}r^{-1}-(\sigma^{0}\overline{\omega}^{0}+\psi_{1}^{0}/2)r^{-2}+O(r^{-3}), \\
X^{A}   &=& (\psi_{1}^{0}\overline{\xi}^{0A}+\overline{\psi}_{1}^{0}\xi^{0A})(6r^{3})^{-1}+O(r^{-4}), \\
U       &=& U^{0}-(\gamma^{0}+\overline{\gamma}^{0})r-(\psi_{2}^{0}+\overline{\psi}_{2}^{0})(2r)^{-1}+O(r^{-2}).
\end{eqnarray*}

\item The functions of integration are determined, using coordinate
conditions, as:
\begin{eqnarray}
\xi^{0\zeta} &=& -P, \quad \overline{\xi}^{0\zeta}=0,  \label{a} \\
\xi^{0\overline{\zeta}} &=& 0, \quad \overline{\xi}^{0\overline{\zeta}}=-P,  \label{b} \\
P            &=& 1+\zeta \overline{\zeta},  \label{c} \\
\alpha^{0}   &=& -\overline{\beta}^{0}=-\frac{\zeta}{2},  \label{d} \\
\gamma^{0}   &=& \nu^{0}=0,  \label{e} \\
\omega^{0}   &=& -\overline{\eth}\sigma^{0},  \label{f} \\
\lambda^{0}  &=& \dot{\overline{\sigma}}^{0},  \label{g} \\
\mu^{0}      &=& U^{0}=-1,  \label{h} \\
\psi_{4}^{0} &=& -\ddot{\overline{\sigma}}^{0},  \label{i} \\
\psi_{3}^{0} &=& \eth \dot{\overline{\sigma}}^{0},  \label{j} \\
\psi_{2}^{0}-\overline{\psi}_{2}^{0} &=& \overline{\eth}^{2}\sigma^{0}-\eth^{2}\overline{\sigma}^{0}+\overline{\sigma}^{0}\lambda^{0}-\sigma^{0}\overline{\lambda}^{0}.  \label{k}
\end{eqnarray}

\item The mass aspect, 
\begin{equation}
\Psi \equiv \psi_{2}^{0}+\eth^{2}\overline{\sigma}^{0}+\sigma^{0}\dot{\overline{\sigma}}^{0},
\label{Mass Aspect}
\end{equation}
satisfies the physically \textit{very important reality condition}: 
\begin{equation}
\Psi =\overline{\Psi}.  \label{reality}
\end{equation}

\item Finally, from the asymptotic Bianchi identities, we obtain the
dynamical (or evolution) relations:

\begin{eqnarray}
\dot{\psi}_{2}^{0} &=& -\eth \psi_{3}^{0}+\sigma^{0}\psi_{4}^{0}+k\phi_{2}^{0}\bar{\phi}_{2}^{0},  \label{AsyBI1} \\
\dot{\psi}_{1}^{0} &=& -\eth \psi_{2}^{0}+2\sigma^{0}\psi_{3}^{0}+2k\phi_{1}^{0}\bar{\phi}_{2}^{0},  \label{AsyBI2} \\
\dot{\psi}_{0}^{0} &=& -\eth \psi_{1}^{0}+3\sigma^{0}\psi_{2}^{0}+3k\phi_{0}^{0}\bar{\phi}_{2}^{0},  \label{AsyBI3} \\
\dot{\phi}_{1}^{0} &=& -\eth \phi_{2}^{0},  \label{AsyBI4} \\
\dot{\phi}_{0}^{0} &=& -\eth \phi_{1}^{0}+\sigma^{0}\phi_{2}^{0};
\label{AsyBI5} \\
k &=& 2Gc^{-4}.
\end{eqnarray}

\end{itemize}

\begin{remark}
These last five equations, the first of which contains the
beautiful Bondi energy-momentum loss theorem, play the fundamental role in
the dynamics of our physical quantities.
\end{remark}

\begin{remark}
Using the mass aspect, $\Psi$, with Eqs.~(\ref{i})
and~(\ref{j}), the first of the asymptotic Bianchi identities, Eq.~(\ref{AsyBI1}), can be rewritten in the concise form,
\begin{equation}
\dot{\Psi}=\dot{\sigma}\dot{\overline{\sigma}}+k\phi_{2}^{0}\overline{\phi}_{2}^{0}.
\label{conciseBI}
\end{equation}
\end{remark}

From these results, the characteristic initial problem can roughly be stated
in the following manner. At $u_{\mathrm{B}}=u_{\mathrm{B}0}$ we choose the initial
values for ($\psi_{0}^{0}, \psi_{1}^{0}, \psi_{2}^{0})$, i.e., functions
only of ($\zeta, \overline{\zeta}$). The characteristic data, the complex
Bondi shear, $\sigma^{0}(u_{\mathrm{B}}, \zeta, \overline{\zeta})$, is then
freely chosen. Since $\psi_{3}^{0}$ and $\psi_{4}^{0}$ are functions of $\sigma^{0}$, Eqs.~(\ref{g}), (\ref{i}) and~(\ref{j}) and its
derivatives, all the asymptotic variables can now be determined from
Eqs.~(\ref{AsyBI1})\,--\,(\ref{AsyBI5}).

An important consequence of the NP formalism is that it allows simple proofs
for many geometric theorems. Two important examples are the Goldberg--Sachs
theorem~\cite{GoldbergSachs} and the peeling theorem~\cite{Peeling}. The
peeling theorem is essentially given by the asymptotic behavior of the Weyl
tensor in Eq.~(\ref{peeling}) (and Eq.~(\ref{peeling II})). The
Goldberg--Sachs theorem is discussed in some detail in Section \ref{GoldbergSachs}. Both theorems are implicitly used later.

One of the immediate physical interpretations arising from the
asymptotically-flat solutions was Bondi's~\cite{Bondi} identifications, at $\mathfrak{I}^{+}$, of the interior spacetime four-momentum
(energy/momentum). Given the mass aspect, Eq.~(\ref{Mass Aspect}), 
\begin{equation*}
\Psi = \psi_{2}^{0}+\eth^{2}\overline{\sigma}^{0}+\sigma^{0}\dot{\overline{\sigma}}^{0},
\end{equation*}
and the spherical harmonic expansion 
\begin{equation}
\Psi =\Psi^{0}+\Psi^{i}Y_{1i}^{0}+\Psi^{ij}Y_{2ij}^{0}+\ldots,
\label{SC24}
\end{equation}
Bondi identified the interior mass and three-momentum with the $l=0$
and $l=1$ harmonic contributions; 
\begin{equation}
M_{\mathrm{B}}=-\frac{c^{2}}{2\sqrt{2}G}\Psi^{0},  \label{SC25}
\end{equation}
\begin{equation}
P^{i}=-\frac{c^{3}}{6G}\Psi^{i}.  \label{SC26}
\end{equation}

The evolution of these quantities, (the Bondi mass/momentum loss) is then
determined from Eq.~(\ref{conciseBI}). The details of this will be discussed in
Section~\ref{applications}.

The same clear cut asymptotic physical identification for interior angular
momentum is not as readily available. In vacuum linear theory, the angular
momentum is often taken to be 
\begin{equation}
J^{k}=-\frac{\sqrt{2}c^{3}}{12G}\text{ Im }(\psi_{1}^{0k}).
\label{SC27}
\end{equation}
However, in the nonlinear treatment, correction terms quadratic in
$\sigma^{0}$ and its derivatives are often
included~\cite{Szabados}. In the presence of a Maxwell field, this is
again modified by the addition of electromagnetic multipole
terms~\cite{PhysicalContent, AdamoNewman2}.

In our case, where we consider \textit{only quadrupole gravitational
  radiation}, the quadratic correction terms do in fact vanish and
hence Eq.~(\ref{SC27}), modified by the Maxwell terms, is correct as
it is stated.


\subsection{The Bondi--Metzner--Sachs group}

The group of coordinate transformations at $\mathfrak{I}^{+}$ that preserves
the Bondi coordinate conditions, the BMS group, is the same as the
asymptotic symmetry group that arises from approximate solutions to
Killing's equation as $\mathfrak{I}^{+}$ is approached. The BMS group has
two parts: the homogeneous Lorentz group and the supertranslation group,
which contains the Poincar\'{e} translation sub-group. Their importance to
us lies in the fact that all the physical quantities arising from our
identifications must transform appropriately under these
transformations~\cite{BMS2, PhysicalContent}.

Specifically, the BMS group is given by the supertranslations, with $\alpha
(\zeta, \overline{\zeta})$ an arbitrary regular differentiable
function on $S^{2}$: 
\begin{eqnarray}
\widehat{u}_{B}&=& u_{\mathrm{B}}+\alpha (\zeta, \overline{\zeta})  \label{supert*} \\
(\widehat{\zeta},\overline{\widehat{\zeta}})&=& (\zeta, \overline{\zeta}) 
\notag
\end{eqnarray}
and the Lorentz transformations, with $(a,b,c,d)$ the complex parameters of
$\mathrm{SL}(2,\mathbb{C})$, 
\begin{eqnarray}
\widehat{u}_{B}&=& Ku_{\mathrm{B}},  \label{Lorentz*} \\
K&=& \frac{1+\zeta \overline{\zeta}}{(a\zeta +b)(\overline{a}\overline{\zeta}+\overline{b})+(c\zeta +d)(\overline{c}\overline{\zeta}+\overline{d})}, 
\notag \\
\widehat{\zeta}&=& \frac{a\zeta +b}{c\zeta +d}, \quad ad-bc=1.  \notag
\end{eqnarray}

If $\alpha (\zeta, \overline{\zeta})$ is expanded in spherical
harmonics, 
\begin{equation}
\alpha (\zeta, \bar{\zeta})=\sum_{m,l} \alpha^{ml}Y_{lm}(\zeta, \bar{\zeta}),
\label{alpha}
\end{equation}
the $l=0,1$ terms represent the Poincar\'{e} translations, i.e.,

\begin{equation}
\alpha_{(P)}(\zeta,
\bar{\zeta})=d^{a}\hat{l}_{a}=\frac{\sqrt{2}}{2}d^{0}Y_{0}^{0}-\frac{1}{2}d^{i}Y_{1i}^{0}.
\label{alphaP}
\end{equation}
Details about the representation theory, with applications, are given later.


\subsection{Algebraically-special metrics and the Goldberg--Sachs theorem}
\label{GoldbergSachs}

Among the most studied vacuum spacetimes are those referred to as
`algebraically-special' spacetimes, i.e., vacuum spacetimes that possess
two or more coinciding principal null direction (PND) vectors. PND fields~\cite{Spinors} (in general, four locally-independent fields exist)
are defined by solutions, $L^{a}$, to the algebraic equation 
\begin{equation*}
L^{b}L_{[e}C_{a]bc[d}L_{f]}L^{c}=0, \qquad L^{a}L_{a}=0.
\end{equation*}

The Cartan--Petrov--Pirani--Penrose
classification~\cite{Petrov,Pirani,Spinors} describes the different
degeneracies (i.e., the number of coinciding PNDs):
\begin{equation*}
\begin{array}{ll}
\text{Alg.\ General}        & [1,1,1,1] \\ 
\text{Type~II}              & [2,1,1] \\ 
\text{Type~D or degenerate} & [2,2] \\ 
\text{Type~III}             & [3,1] \\ 
\text{Type~IV or N}      & [4].
\end{array}
\end{equation*}

In NP language, \textit{if} the tetrad vector $l^{a}$ is a principal
null direction, i.e., $L_{a}=l_{a}$, then automatically, 
\begin{equation*}
\psi_{0}=0.
\end{equation*}

For the algebraically-special metrics, the special cases are 
\begin{equation*}
\begin{array}{ll}
\text{Type~II}  & \psi_{0}=\psi_{1}=0 \\ 
\text{Type~III} & \psi_{0}=\psi_{1}=\psi_{2}=0 \\ 
\text{Type~IV}  & \psi_{0}=\psi_{1}=\psi_{2}=\psi_{3}=0 \\ 
\text{Type~D}   & 
\begin{array}{l}
\psi_{0}=\psi_{1}=\psi_{3}=\psi_{4}=0 \\ 
\text{with both } l^{a} \text{ and } n^{a} \text{ PNDs. }
\end{array}
\end{array}
\end{equation*}
An outstanding feature of the algebraically-special metrics is contained in
the beautiful Goldberg--Sachs theorem~\cite{GoldbergSachs}.

\begin{thm}[Goldberg--Sachs]
For a nonflat vacuum spacetime, if there is an NGC that is shear-free, i.e., there is a null vector
  field with ($\kappa=0, \sigma=0$), then the spacetime is
  algebraically special and, conversely, if a vacuum spacetime is
  algebraically special, there is an NGC with
  ($\kappa=0,\sigma=0$).
\end{thm}

In particular, this means that for all algebraically special metrics there is an everywhere shear-free NGC, and a null tetrad exists such that $\psi_{0}=\psi_{1}=0$.  The main idea of this review is an asymptotic generalization of this statement: for all asymptotically flat metrics, there exists a null tetrad such that the $l=0$ and $l=1$ harmonic coefficients of the \textit{asymptotic} Weyl tensor components $\psi_0$ and $\psi_1$ (namely, $\psi^{0i}_{0}$ and $\psi^{0i}_{1}$) vanish.  Note that this is in reality a non-trivial condition only on $\psi^{0i}_{1}$, since the other three components vanish automatically when we recall that $\psi_0$ and $\psi_1$ are spin-weight two and one respectively.


\newpage

\section{Shear-Free NGCs in Minkowski Space}
\label{shear-free-NGC}

The structure and properties of asymptotically shear-free NGCs (our
main topic) are best understood by first looking at the special case
of congruences that are shear-free everywhere (except at their
caustics). Though shear-free congruences are also found in
algebraically-special spacetimes, in this section only the shear-free
NGCs in Minkowski spacetime, $\mathbb{M}$, are discussed~\cite{AdamoNewman4}


\subsection{The flat-space good-cut equation and good-cut functions}
\label{flat-space-good-cut}

In Section~\ref{foundations}, we saw that in the NP
formalism, two of the complex spin coefficients, the \textit{optical
parameters} $\rho$ and $\sigma$ of Eqs.~(\ref{SC21}) and~(\ref{SC22}),
play a particularly important role in their description of an NGC; namely,
they carry the information of the divergence, twist and shear of the
congruence.

From Eqs.~(\ref{SC21}) and~(\ref{SC22}), the radial behavior of the optical
parameters for general \textit{shear-free} NGCs, in Minkowski space, is
given by
\begin{equation}
\rho =\frac{i\Sigma -r}{r^{2}+\Sigma^{2}}, \qquad \sigma =0,
\label{SF1}
\end{equation}
where $\Sigma$ is the twist of the congruence. A more detailed and much
deeper understanding of the shear-free congruences can be obtained by first
looking at the explicit coordinate expression, Eq.~(\ref{congruence}), for
all flat-space NGCs: 
\begin{equation}
x^{a}=u_{\mathrm{B}}(\hat{l}^{a}+\hat{n}^{a})-L\overline{\hat{m}}^{a}-\bar{L}\hat{m}^{a}+(r^{\ast}-r_{0})\hat{l}^{a},
\label{SF2}
\end{equation}
where $L(u_{\mathrm{B}},\zeta, \bar{\zeta})$ is an arbitrary complex function of the
parameters $y^{w}=(u_{\mathrm{B}},\zeta, \bar{\zeta})$; $r_{0}$, also an arbitrary
function of $(u_{\mathrm{B}},\zeta, \bar{\zeta})$, determines the origin of the affine
parameter; and $r^{\ast}$ can be chosen freely. Most frequently, to simplify
the form of $\rho $, $r_{0}$ is chosen as
\begin{equation}
r_{0}\equiv -\frac{1}{2}\left( \eth \bar{L}+\bar{\eth}L+L\dot{\bar{L}}+\bar{L}\dot{L}\right) .
\label{SF3}
\end{equation}
At this point, Eq.~(\ref{SF2}) describes an arbitrary NGC with
$(u_{\mathrm{B}},\zeta, \bar{\zeta})$ labelling the geodesics and
$r^{\ast}$ the affine distance along the individual geodesics; later
$L(u_{\mathrm{B}},\zeta, \bar{\zeta})$ will be chosen so that the
congruence is shear-free.
The tetrad ($\hat{l}^{a}, \hat{n}^{a}, \hat{m}^{a},
\overline{\hat{m}}^{a}$) is given by (\ref{l.hat}-\ref{m.hat}), see~\cite{UCF} 
%

There are several important comments to be made about
Eq.~(\ref{SF2}).  The first is that there is a simple geometric
meaning to the parameters $(u_{\mathrm{B}}, \zeta, \bar{\zeta})$: they
are the values of the Bondi coordinates of $\mathfrak{I}^{+}$, where
each geodesic of the congruence intersects $\mathfrak{I}^{+}$.
The second concerns the geometric meaning of $L$. At each point of
$\mathfrak{I}^{+}$, consider the past light cone and its sphere of null
directions. Coordinatize that sphere (of null directions) with
stereographic coordinates. The function $L(u_{\mathrm{B}},\zeta, \bar{\zeta})$
\textit{is the stereographic angle field} on $\mathfrak{I}^{+}$ that
describes the null direction of each geodesic intersecting
$\mathfrak{I}^{+}$ at the point $(u_{\mathrm{B}},\zeta, \bar{\zeta})$. The
values $L=0$ and $L=\infty$ represent, respectively, the direction
along the Bondi $l^{a}$ and $n^{a}$ vectors. This stereographic angle
field completely determines the NGC.

The twist, $\Sigma$, of the congruence can be calculated in terms of
$L(u_{\mathrm{B}},\zeta, \bar{\zeta})$ directly from Eq.~(\ref{SF2}) and the
definition of the complex divergence, Eq.~(\ref{rho}), leading to 
\begin{equation}
i\Sigma =\frac{1}{2}\left\{\eth \overline{L}+L\dot{\overline{L}}-\overline{\eth}L-\overline{L}\dot{L}\right\}.
\label{twist}
\end{equation}
We now demand that $L$ be a \textit{regular} function of its arguments
(i.e., have no infinities), or, equivalently, that all members of the NGC come
from the interior of the spacetime and not lie on $\mathfrak{I}^{+}$ itself.

It has been shown~\cite{Aronson} that the condition on the stereographic
angle field $L$ for the NGC to be shear-free is that 
\begin{equation}
\eth L+L\dot{L}=0.
\label{SF6}
\end{equation}
Our task is now to find the regular solutions of Eq.~(\ref{SF6}). The key to
doing this is via the introduction of a new complex variable $\tau$ and
complex function~\cite{Footprints, UCF}, 
\begin{equation}
\tau =T(u_{\mathrm{B}},\zeta, \bar{\zeta}).
\label{SF7}
\end{equation}
$T$ is related to $L$ by the CR equation (related to the existence
of a CR structure on $\mathfrak{I}^{+}$; see Appendix~\ref{appendixB}): 
\begin{equation}
\eth_{(u_{\mathrm{B}})}T+L\dot{T}=0.
\label{SF8}
\end{equation}

\begin{remark}
The following `gauge' freedom becomes useful later. $\tau
\rightarrow \tau^{\ast}=F(\tau)$, with $F$ analytic, leaving Eq.~(\ref{SF8}) unchanged. In other words, 
\begin{equation}
\tau^{\ast} = T^{\ast}\left(u_{\mathrm{B}},\zeta, \bar{\zeta}\right) \equiv F\left(T(u_{\mathrm{B}}, \zeta, \bar{\zeta})\right),
\label{gauge}
\end{equation}
leads to 
\begin{eqnarray*}
\eth_{(u_{\mathrm{B}})}T^{\ast} &=&F^{\prime}\eth_{(u_{\mathrm{B}})}T, \\
\dot{T}^{\ast} &=&F^{\prime}\dot{T}, \\
\eth_{(u_{\mathrm{B}})}T^{\ast}+L\dot{T}^{\ast} &=&0.
\end{eqnarray*}
\end{remark}

We assume, in the neighborhood of real $\mathfrak{I}^{+}$, i.e., near
the real $u_{\mathrm{B}}$ and $\tilde{\zeta}= \bar{\zeta}$, that $T(u_{\mathrm{B}},\zeta,
\tilde{\zeta})$ is analytic in the three arguments $(u_{\mathrm{B}},\zeta,
\tilde{\zeta})$. The inversion of Eq.~(\ref{SF7}) yields the
\textit{complex analytic cut function} 
\begin{equation}
u_{\mathrm{B}}=G(\tau, \zeta, \tilde{\zeta}).
\label{SF9}
\end{equation}
Though we are interested in real values for $u_{\mathrm{B}}$, from Eq.~(\ref{SF9}) we see that for arbitrary $\tau$ it may take complex
values. Shortly, we will also address the important issue of what values of $\tau$ are needed for real $u_{\mathrm{B}}$.

Returning to the issue of integrating the shear-free condition,
Eq.~(\ref{SF6}), using Eq.~(\ref{SF7}), we note that the derivatives
of $T$, $\eth_{(u_{\mathrm{B}})}T$ and $\dot{T}$ can be expressed in
terms of the derivatives of $G(\tau, \zeta, \bar{\zeta})$ by implicit differentiation.
The $u_{\mathrm{B}}$ derivative of $T$ is obtained by taking the $u_{\mathrm{B}}$
derivative of Eq.~(\ref{SF9}): 
\begin{equation}
1=G^{\prime}(\tau, \zeta, \bar{\zeta})\dot{T}\Rightarrow \dot{T}=\frac{1}{(G^{\prime})},
\label{SF9*}
\end{equation}
while the $\eth_{(u_{\mathrm{B}})}T$ derivative is found by applying
$\eth_{(u_{\mathrm{B}})}$ to Eq.~(\ref{SF9}),
\begin{eqnarray}
0 &=&G^{\prime}(\tau, \zeta, \bar{\zeta})\eth_{(u_{\mathrm{B}})}T+\eth_{(\tau)}G,
\label{SF9**} \\
\eth_{(u_{\mathrm{B}})}T &=&-\frac{\eth_{(\tau)}G}{G^{\prime}(\tau, \zeta, \bar{\zeta})}.  \notag
\end{eqnarray}
When Eqs.~(\ref{SF9*}) and~(\ref{SF9**}) are substituted into Eq.~(\ref{SF8}),
one finds that $L$ is given implicitly in terms of the cut function by 
\begin{eqnarray}
L(u_{\mathrm{B}},\zeta, \bar{\zeta}) &=&\eth_{(\tau)}G(\tau, \zeta, \bar{\zeta}),
\label{Ltau} \\
u_{\mathrm{B}} &=&G(\tau, \zeta, \bar{\zeta})\Leftrightarrow \tau =T(u_{\mathrm{B}},\zeta, \bar{\zeta}).
\label{G*}
\end{eqnarray}

Thus, we see that all information about the NGC can be obtained from the cut
function $G(\tau, \zeta, \bar{\zeta})$. 

By further implicit differentiation of Eq.~(\ref{Ltau}), i.e., 
\begin{eqnarray*}
\eth_{(u_{\mathrm{B}})}L(u_{\mathrm{B}},\zeta, \bar{\zeta})
&=&\eth_{(\tau)}^{2}G(\tau, \zeta, \bar{\zeta})+\eth_{(\tau)}G^{\prime}(\tau, \zeta, \bar{\zeta})\cdot \eth_{(u_{\mathrm{B}})}T, \\
\dot{L}(u_{\mathrm{B}},\zeta, \bar{\zeta}) &=&\eth_{(\tau)}G^{\prime}(\tau, \zeta, \bar{\zeta})\cdot \dot{T},
\end{eqnarray*}
using Eq.~(\ref{SF8}), the shear-free condition~(\ref{SF6}) becomes
\begin{equation}
\eth_{(\tau)}^{2}G(\tau, \zeta, \bar{\zeta})=0.
\label{SF11}
\end{equation}
This equation will be referred to as the \textit{homogeneous Good-Cut
Equation} and its solutions as flat-space \textit{Good-Cut Functions} (GCFs). In the next Section~\ref{good-cut-eq}, an inhomogeneous version, the \textit{Good-Cut Equation},
will be found for asymptotically shear-free NGCs. Its solutions will also be
referred to as GCFs.

From the properties of the $\eth^{2}$ operator, the \textit{general
  regular solution} to Eq.~(\ref{SF11}) is easily found: $G$ must contain
only $l=0$ and $l=1$ spherical harmonic contributions; thus, any
regular solution will be dependent on four arbitrary complex
parameters, $z^{a}$. If these parameters are functions of $\tau $,
i.e., $z^{a}=\xi^{a}(\tau)$, then we can express any regular solution
$G$ in terms of the complex world line $\xi^{a}(\tau)$~\cite{Footprints, UCF}: 
\begin{equation}
u_{\mathrm{B}}=G(\tau, \zeta, \bar{\zeta})=\xi^{a}(\tau)\hat{l}_{a}(\zeta, 
\bar{\zeta})\equiv \frac{\sqrt{2}\xi^{0}(\tau)}{2}-\frac{1}{2}\xi^{i}(\tau)Y_{1i}^{0}.
\label{SF12}
\end{equation}
The angle field $L(u_{\mathrm{B}},\zeta, \bar{\zeta})$ then has the form 
\begin{eqnarray}
L(u_{\mathrm{B}},\zeta, \bar{\zeta}) &=&\eth_{(\tau)}G(\tau, \zeta, \bar{\zeta})=\xi^{a}(\tau)\hat{m}_{a}(\zeta, \bar{\zeta}),
\label{angle field} \\
u_{\mathrm{B}} &=&\xi^{a}(\tau)\hat{l}_{a}(\zeta, \bar{\zeta}).
\label{parametric}
\end{eqnarray}

Thus, we have our first major result: every regular shear-free NGC in
Minkowski space is generated by the arbitrary choice of a complex world line
in what turns out to be complex Minkowski space. See Eq.~(\ref{alphaP}) for
the connection between the $l=(0,1)$ harmonics in
Eq.~(\ref{SF12}) and the Poincar\'{e} translations. We see in the
next Section~\ref{good-cut-eq} how this result generalizes to regular
asymptotically shear-free NGCs.

\begin{remark}
We point out that this construction of regular shear-free
NGCs in Minkowski space is a special example of the Kerr theorem (cf.~\cite{PenroseRindler2}). Writing Eqs.~(\ref{angle field}) and~(\ref{parametric}) as
\begin{eqnarray*}
u_{\mathrm{B}} &=& \frac{a+b\overline{\zeta}+\overline{b}\zeta +c\zeta 
\overline{\zeta}}{1+\zeta \overline{\zeta}}, \\
L &=& \frac{(\overline{b}+c\overline{\zeta})-\overline{\zeta}(a+b\overline{\zeta})}{1+\zeta \overline{\zeta}},
\end{eqnarray*}
where the ($a(\tau), b(\tau), c(\tau), d(\tau)$) are simple combinations of the $\xi^{a}(\tau)$, we then find that
\begin{eqnarray*}
L+u_{B}\overline{\zeta} &=& \overline{b}+c\overline{\zeta}, \\
u_{B}-L\zeta &=& a+b\overline{\zeta}.
\end{eqnarray*}
Noting that the right-hand side of both equations are functions only of $\tau$ and $\overline{\zeta}$, we can eliminate the $\tau$ from the two
equations, thereby constructing a function of three variables of the form
\begin{equation*}
F(L+u_{B}\overline{\zeta}, u_{B}-L\zeta,\overline{\zeta})=0.
\end{equation*}
This is a special case of the general solution to Eq.~(\ref{SF6}), which is the Kerr theorem.
\end{remark}

In addition to the construction of the angle field, $L(u_{\mathrm{B}}, \zeta,
\bar{\zeta})$, from the GCF, another quantity of great value in
applications, obtained from the GCF, is the local change in $u_{\mathrm{B}}$ as
$\tau$ changes, i.e., 
\begin{equation}
V(\tau, \zeta, \tilde{\zeta}) \equiv \partial_{\tau}G=G^{\prime}.
\label{V}
\end{equation}


\subsection{Real cuts from the complex good cuts, I}
\label{real-cuts-I}

Though our discussion of shear-free NGCs has relied, in an essential manner,
on the use of the complexification of $\mathfrak{I}^{+}$ and the complex
world lines in complex Minkowski space, it is the real structures that are
of main interest to us. We want to find the intersection of the complex GCF
with real $\mathfrak{I}^{+}$, i.e., what are the real points and real cuts
of $u_{\mathrm{B}}=G(\tau, \zeta, \overline{\zeta})$, ($\tilde{\zeta}=\overline{\zeta}$), and what are the values of $\tau$ that yield \textit{real} $u_{\mathrm{B}}$.  These reality structures were first observed in~\cite{AdamoNewman4} and recently there have been attempts to study them in the framework of holographic dualities (cf.~\cite{Adamo:2011cx} and Section~\ref{conclusion}).

To construct an associated family of real cuts from a GCF, we begin with 
\begin{equation}
u_{\mathrm{B}}=G(\tau, \zeta, \overline{\zeta})=\frac{\sqrt{2}}{2}\xi^{0}(\tau)-\frac{1}{2}\xi^{i}(\tau)Y_{1i}^{0}(\zeta, \overline{\zeta})
\label{G**}
\end{equation}
and write 
\begin{equation}
\tau =s+i\lambda
\label{s+ilambea2}
\end{equation}
with $s$ and $\lambda$ real. The cut function can then be rewritten 
\begin{eqnarray}
u_{\mathrm{B}} &=&G(\tau, \zeta, \overline{\zeta})=G(s+i\lambda, \zeta, \overline{\zeta})
\label{G_c} \\
&=&G_{R}(s,\lambda, \zeta, \overline{\zeta})+iG_{I}(s,\lambda, \zeta, \overline{\zeta}),  \notag
\end{eqnarray}
with real $G_{R}(s,\lambda, \zeta, \overline{\zeta})$ and
$G_{I}(s,\lambda, \zeta, \overline{\zeta})$. The $G_{R}(s,\lambda,
\zeta, \overline{\zeta})$ and $G_{I}(s,\lambda, \zeta, \overline{\zeta})$ are easily calculated
from $G(\tau, \zeta, \overline{\zeta})$ by 
\begin{eqnarray}
G_{R}(s,\lambda, \zeta, \overline{\zeta})
&=&\frac{1}{2}\left\{G(s+i\lambda, \zeta, \overline{\zeta})+\overline{G(s+i\lambda, \zeta, \overline{\zeta})}\right\},
\label{Gr} \\
G_{I}(s,\lambda, \zeta, \overline{\zeta})
&=&-\frac{i}{2}\left\{G(s+i\lambda, \zeta, \overline{\zeta})-\overline{G(s+i\lambda, \zeta, \overline{\zeta})}\right\}.
\label{Gi}
\end{eqnarray}

By setting 
\begin{equation}
G_{I}(s,\lambda, \zeta, \overline{\zeta})=0
\label{Gi=0}
\end{equation}
and solving for 
\begin{equation}
\lambda =\Lambda (s,\zeta, \overline{\zeta})
\label{lambda}
\end{equation}
we obtain the associated one-parameter, $s$, family of real slicings,
\begin{equation}
u_{\mathrm{B}}^{({R})}=G_{R}(s,\Lambda (s,\zeta, \overline{\zeta}), \zeta, \overline{\zeta})=\xi^{a}\left(s+i\Lambda (s,\zeta, \overline{\zeta})\right)l_{a}(\zeta, \overline{\zeta}).
\label{Gr*}
\end{equation}
Thus, the values of $\tau$ that yield real values of $u_{\mathrm{B}}$ are given
by 
\begin{equation}
\tau =s+i\Lambda (s,\zeta, \overline{\zeta}).
\label{real-tau}
\end{equation}

Perturbatively, using Eq.~(\ref{G**}) and writing $\xi^{a}(s)=\xi^{a}_{R}(s)+i\xi^{a}_{I}(s)$, we find $\lambda$ to first order:
\begin{eqnarray}
u_{\mathrm{B}} &=& \frac{\sqrt{2}}{2}\xi_{R}^{0}(s)-\frac{\sqrt{2}}{2}\xi_{I}^{0}(s)^{\prime}\lambda
-\frac{1}{2}\bigg[\xi_{R}^{i}(s)-\xi_{I}^{i}(s)^{\prime}\lambda\bigg]Y_{1i}^{0}(\zeta, \overline{\zeta})
\label{G@@} \\
&& + i\bigg[\frac{\sqrt{2}}{2}\xi_{I}^{0}(s)+\frac{\sqrt{2}}{2}\xi_{R}^{0}(s)^{\prime}\lambda\bigg]
-i\frac{1}{2}\bigg[\xi_{I}^{i}(s)+\xi_{R}^{i}(s)^{\prime}\lambda\bigg]Y_{1i}^{0}(\zeta, \overline{\zeta}),  \notag \\
u_{\mathrm{B}}^{({R})} &=& G_{R}(s,\Lambda, \zeta,\overline{\zeta})
\label{Gr@} \\
&=& \frac{\sqrt{2}}{2}\xi_{R}^{0}(s)-\frac{\sqrt{2}}{2}\xi_{I}^{0}(s)^{\prime}\lambda
-\frac{1}{2}\bigg[\xi_{R}^{i}(s)-\xi_{I}^{i}(s)^{\prime}\lambda\bigg]Y_{1i}^{0}(\zeta, \overline{\zeta}),  \notag \\
\lambda &=& \Lambda (s,\zeta,
\overline{\zeta})=-\frac{\sqrt{2}\xi_{I}^{0}(s)+\xi_{I}^{i}(s)Y_{1i}^{0}(\zeta, \overline{\zeta})}
	 {[\sqrt{2}\xi_{R}^{0}(s)^{\prime}-\xi_{R}^{i}(s)^{\prime}Y_{1i}^{0}(\zeta, \overline{\zeta})]}.
\label{Lambda@}
\end{eqnarray}

Continuing, with small values for the imaginary part of $\xi^{a}(\tau)=\xi_{R}^{a}(\tau)+i\xi_{I}^{a}(\tau)$, ($\xi_{R}^{a}(\tau),\ \xi_{I}^{a}(\tau)$ both real analytic functions) and hence small $\Lambda
(s,\zeta, \overline{\zeta})$, it is easy to see that $\Lambda
(s,\zeta, \overline{\zeta})$ (for fixed value of $s$) is a bounded smooth function
on the $(\zeta, \overline{\zeta})$ sphere, with maximum and minimum
values, $\lambda_{\max}=\Lambda
(s,\zeta_{\max}, \overline{\zeta}_{\max})$ and $\lambda_{\min}=\Lambda
(s,\zeta_{\min}, \overline{\zeta}_{\min})$. Furthermore on the ($\zeta, \overline{\zeta}$) sphere, there are a
finite line-segments worth of curves (circles) that lie between
$(\zeta_{\min},\overline{\zeta}_{\min })$ and $(\zeta_{\max}, \overline{\zeta}_{\max})$ such that $\Lambda (s,\zeta, \overline{\zeta})$ is a
monotonically increasing function on the family of curves. Hence there
will be a family of circles on the $(\zeta,\overline{\zeta})$-sphere where
the value of $\lambda$ is a constant, ranging between $\lambda_{\max}$ and $\lambda_{\min}$.

Summarizing, we have the result that in the complex $\tau$-plane there is a
ribbon or strip given by all values of $s$ and line segments parametrized
by $\lambda$ between $\lambda_{\min}$ and $\lambda_{\max}$ such that
the complex light-cones from each of the associated points, $\xi^{a}(s+i\lambda )$, all have \textit{some null
geodesics} that intersect real $\mathfrak{I}^{+}$. More specifically, for
each of the allowed values of $\tau =s+i\lambda$ there will be a circle's
worth of complex null geodesics leaving the point $\xi^{a}(s+i\lambda)$ reaching real $\mathfrak{I}^{+}$. It is the union of these null geodesics,
corresponding to the circles on the $(\zeta,\overline{\zeta})$-sphere from
the line segment, that produces the real family of cuts, Eq.~(\ref{Gr*}).

The real structure associated with a complex world line is then this
one-parameter family of slices (cuts) Eq.~(\ref{Gr*}).

\begin{remark}
We saw earlier that the shear-free angle field
was given by 
\begin{eqnarray}
L(u_{\mathrm{B}},\zeta, \bar{\zeta}) &=&\eth_{(\tau)}G(\tau, \zeta, \bar{\zeta}),
\label{A} \\
u_{\mathrm{B}} &=&G(\tau, \zeta, \bar{\zeta})\Leftrightarrow \tau =T(u_{\mathrm{B}},\zeta, \bar{\zeta}),
\label{B}
\end{eqnarray}
where real values of $u_{\mathrm{B}}$ should be used. If the real cuts, $u_{\mathrm{B}}=G_{R}\left(s,\Lambda (s,\zeta, \overline{\zeta}),\zeta, \overline{\zeta}\right)$, were used instead to calculate $L(u_{\mathrm{B}},\zeta, \bar{\zeta})$, the
results would be wrong. The restriction of $\tau$ to yield real $u_{\mathrm{B}}$,
does not commute with the application of the $\eth$ operator, i.e.,
\begin{equation*}
L(u_{\mathrm{B}},\zeta, \bar{\zeta})\neq \eth G_{R}.
\end{equation*}
The $\eth$ differentiation \textit{must be done first}, holding $\tau$
constant, before the reality of $u_{\mathrm{B}}$ is used. In other words, though we
are interested in real $\mathfrak{I}^{+}$, it is essential that we consider
its (local) complexification.
\end{remark}

There are a pair of important (dual) results that
arise from the considerations of the good cuts~\cite{AdamoNewman4, Adamo:2011cx}. From the stereographic angle
field, i.e., $L$ from Eqs.~(\ref{A}) and (\ref{B}), one can form \textit{two
different conjugate fields}, (1) the complex conjugate of $L$: 
\begin{equation}
\overline{L}=\overline{\eth}_{\overline{\tau}}\overline{G}=\overline{\xi}^{a}(\overline{\tau})\overline{\hat{m}}_{a}
\label{Lbar}
\end{equation}
and (2) the holomorphic conjugate, $\widetilde{L}$, given by 
\begin{equation}
\widetilde{L}=\overline{\eth}_{\tau}G=\xi^{a}(\tau)\overline{\hat{m}}_{a}.
\label{Ltwiddle}
\end{equation}

The two different pairs, the complex conjugate pair ($L,\overline{L}$) and
the holomorphic pair ($L,\widetilde{L}$) determine two different null vector
direction fields at $\mathfrak{I}^{+}$, the real vector field,
$l^{\ast a}$, and the complex field, $l_{C}^{\ast a}$, via
the relations
\begin{equation}
l^{a}\rightarrow l^{\ast a}=l^{a}-\frac{\bar{L}}{r}m^{a}-\frac{L}{r}\bar{m}^{a}+O(r^{-2}),
\label{Tet10*}
\end{equation}
and
\begin{equation}
l^{a}\rightarrow l_{C}^{\ast a}=l^{a}-\frac{\widetilde{L}}{r}m^{a}-\frac{L}{r}\bar{m}^{a}+O(r^{-2}).
\label{Tet10*C}
\end{equation}
Both generate, in the spacetime interior, shear-free null geodesic
congruences: the first is a real twisting shear-free congruence while the
latter is a complex twist-free congruence that consists of the light-cones from
the world line, $z^{a}=\xi^{a}(\tau)$, i.e., they focus on $\xi^{a}(\tau)$. It is this fact that they focus on the world line, $z^{a}=\xi^{a}(\tau)$ that is of most relevance to us.

The twist of the real congruence, $\Sigma
(u_{B},\zeta,\overline{\zeta})$, which comes from the complex divergence,
\begin{eqnarray}
\rho &=&-\frac{1}{r+i\Sigma } \\
2i\Sigma &=&\eth \overline{L}+L(\overline{L})^{\cdot }-\overline{\eth}L-\overline{L}\dot{L}. \\
&=&(\xi^{a}(\tau)-\overline{\xi}^{a}(\overline{\tau}))\left(
n_{a}-l_{a}\right) .  \notag
\end{eqnarray}
is proportional to the imaginary part of the complex world line and
consequently we have the real structure associated with the complex
world line coming from two (dual) places, the real cuts, Eq.~(\ref{Gr*}) and
the twist.

It is the complex point of view of the complex light-cones coming from the
complex world line that dominates our discussion.


\subsection{Approximations}

Due to the difficulties involved in the intrinsic nonlinearities and the
virtual impossibility of exactly inverting arbitrary analytic functions, it
often becomes necessary to resort to approximations. The basic
approximation will be to consider the complex world line $\xi^{a}(\tau)$
as being close to the straight line, $\xi_{0}^{a}(\tau)=\tau\delta_{0}^{a}$;
deviations from this will be considered as first order. We retain terms up
to second order, i.e., quadratic terms. Another frequently used
approximation is to terminate spherical harmonic expansions after the $l=2$
terms.

It is worthwhile to discuss some of the issues related to these
approximations.  One important issue is how to use the gauge freedom, Eq.~(\ref{gauge}), $\tau
\rightarrow \tau^{\ast}=F(\tau)$, to simplify $\xi^{a}(\tau)$ and the `velocity vector', 
\begin{equation}
v^{a}(\tau)=\xi^{a\, \prime}(\tau)\equiv \frac{d\xi^{a}}{d\tau}.
\label{v}
\end{equation}

\textbf{A Notational issue:} Given a complex analytic function (or vector)
of the complex variable $\tau$, say $G(\tau)$, then $G(\tau)$ can be
decomposed uniquely into two parts,
\begin{equation*}
G(\tau)=\mathfrak{G}_{R}(\tau)+i\mathfrak{G}_{I}(\tau),
\end{equation*}
where all the coefficients in the Taylor series for
$\mathfrak{G}_{R}(\tau)$ and $\mathfrak{G}_{I}(\tau)$ are real. With but a slight extension of
conventional notation we refer to them as real analytic functions.

With this notation, we also write
\begin{eqnarray*}
\xi^{a}(\tau) &=&\xi_{R}^{a}(\tau)+i\xi_{I}^{a}(\tau) \\
v^{a}(\tau) &=&v_{R}^{a}(\tau)+iv_{I}^{a}(\tau).
\end{eqnarray*}
By using the reparametrization of the world line, via $\tau^{\ast}=F(\tau)$, we choose $F(\tau)=\xi^{0}(\tau)$, so that (dropping the $^{\ast}$) we have
\begin{eqnarray*}
\xi^{0}(\tau) &=&\xi_{R}^{0}(\tau)=\tau,\ \ \ \xi_{I}^{0}(\tau)=0 \\
v^{0}(\tau) &=&v_{R}^{0}(\tau)=1,\ \ \ \ v_{I}^{0}(\tau)=0
\end{eqnarray*}

Finally, from the reality condition on the $u_{\mathrm{B}}$, Eqs.~(\ref{Gi}), (\ref{Gr*}) and~(\ref{lambda}) yield, with $\tau =s+i\lambda$ and $\lambda$ treated as small,
\begin{eqnarray}
u_{\mathrm{B}}^{({R})} &=&\xi_{R}^{a}(s)\hat{l}_{a}+v_{I}^{a}(s)\hat{l}_{a}\frac{\xi_{I}^{b}(s)\hat{l}_{b}}{\xi_{R}^{c\,\prime}(s)\hat{l}_{c}},
\label{realu*} \\
\lambda &=&\Lambda (s,\zeta, \overline{\zeta})=-\frac{\xi_{I}^{b}(s)\hat{l}_{b}}{\xi_{R}^{c\,\prime}(s)\hat{l}_{c}},
\label{lambda*} \\
&=&\frac{\frac{\sqrt{2}}{2}\xi_{I}^{i}(s)Y_{1i}^{0}}{1-\frac{\sqrt{2}}{2}\xi_{R}^{i\,\prime}(s)Y_{1i}^{0}}.  \notag
\end{eqnarray}
Within this slow motion approximation scheme, we have from
Eqs.~(\ref{realu*}) and~(\ref{lambda*}), 
\begin{eqnarray}
u_{\mathrm{ret}}^{({R})} &=&\sqrt{2}u_{\mathrm{B}}^{({R})}=s-\frac{1}{\sqrt{2}}\xi_{R}^{i}(s)Y_{1i}^{0}+2v_{I}^{a}(s)\hat{l}_{a}\xi_{I}^{b}(s)\hat{l}_{b},
\label{real u} \\
\lambda &\approx &\frac{\sqrt{2}}{2}\xi_{I}^{i}(s)Y_{1i}^{0}\left(1-\frac{\sqrt{2}}{2}v_{R}^{j}(s)Y_{1j}^{0}\right),
\end{eqnarray}
or, to first order, which is all that is needed,
\begin{equation*}
\lambda =\frac{\sqrt{2}}{2}\xi_{I}^{i}(s)Y_{1i}^{0}.
\end{equation*}

We then have, to linear order,
\begin{eqnarray}
\tau &=&s+i\frac{\sqrt{2}}{2}\xi_{I}^{i}(s)Y_{1i}^{0},
\label{linear lambda} \\
u_{\mathrm{ret}}^{({R})} &=&s-\frac{1}{\sqrt{2}}\xi_{R}^{i}(s)Y_{1i}^{0}.  \notag
\end{eqnarray}


\subsection{Asymptotically-vanishing Maxwell fields}
\label{AVMaxwell}

\subsubsection{A prelude}

The basic starting idea in this work is simple. It is in the generalizations
and implementations where difficulties arise.

Starting in Minkowski space in a \textit{fixed} given Lorentzian frame with
spatial origin, the electric dipole moment $\overrightarrow{D}_{E}$ is
calculated from an integral over the (localized) charge distribution. If
there is a shift, $\overrightarrow{R}$, in the origin, the dipole transforms
as
\begin{equation}
\overrightarrow{D_{E}^{\ast}}=\overrightarrow{D}_{E}-q\overrightarrow{R}.
\label{D}
\end{equation}
If $\overrightarrow{D}_{E}$ is time dependent, we obtain the center-of-charge world line by taking $\overrightarrow{D_{E}^{\ast}}=0$, i.e., from $\overrightarrow{R}=\overrightarrow{D}_{E}q^{-1}$. It is this idea that we
want to generalize and extend to gravitational fields.

The first generalization is formal and somewhat artificial: shortly it will
become quite natural. We introduce, in addition to the electric dipole
moment, the magnetic dipole moment $\overrightarrow{D}_{M}$ (also obtained
by an integral over the current distribution) and write 
\begin{equation*}
\overrightarrow{D}_{\mathbb{C}}=\overrightarrow{D}_{E}+i\overrightarrow{D}_{M}.
\end{equation*}
By allowing the displacement $\overrightarrow{R}$ to take complex values, $\overrightarrow{R}_{\mathbb{C}}$, Eq.~(\ref{D}), can be generalized to 
\begin{equation}
\overrightarrow{D_{\mathbb{C}}^{\ast}}=\overrightarrow{D}_{\mathbb{C}}-q\overrightarrow{R}_{\mathbb{C}},
\label{D_C}
\end{equation}
so that the complex center-of-charge is given by
$\overrightarrow{D_{\mathbb{C}}^{\ast}}=0$ or 
\begin{equation}
\overrightarrow{R}_{\mathbb{C}}=\overrightarrow{D}_{\mathbb{C}}q^{-1}.
\label{CCC}
\end{equation}

We emphasize that this is done in a fixed Lorentz frame and only the origin
is moved. In different Lorentz frames there will be different complex
centers of charge.

Later, directly from the general \textit{asymptotic Maxwell field itself}
(satisfying the Maxwell equations), we define the \textit{asymptotic}
complex dipole moment and give \textit{its} transformation law, including
transformations between Lorentz frames. This yields a unique complex center
of charge independent of the Lorentz frame.


\subsubsection{Asymptotically-vanishing Maxwell fields: General properties}

In this section, we describe how a complex center of charge for
asymptotically vanishing Maxwell fields in flat spacetime can be found by using the shear-free NGCs, constructed from
solutions of the homogeneous good-cut equation, to transform certain
Maxwell field components to zero. Although this serves as a good
example for our later methods in asymptotically flat spacetimes, the
reader may wish to skip ahead to Section~\ref{good-cut-eq}, where we
go directly to gravitational fields in a setting of greater generality.

Our first set of applications of shear-free NGCs comes from Maxwell theory
in Minkowski space. We review the general theory of the behavior of
asymptotically-flat or vanishing Maxwell fields assuming throughout that
there is a nonvanishing total charge, $q$. As stated in
Section~\ref{foundations}, the Maxwell field is described in terms of
its complex tetrad components, ($\phi_{0}, \phi_{1}, \phi_{2}$). In a
Bondi coordinate/tetrad system the asymptotic integration is
relatively simple~\cite{Maxwell, EDRad} resulting in the radial behavior (the
peeling theorem): 
\begin{eqnarray}
\phi_{0} &=&\frac{\phi_{0}^{0}}{r^{3}}+O(r^{-4}),
\label{Max1} \\
\phi_{1} &=&\frac{\phi_{1}^{0}}{r^{2}}+O(r^{-3}),  \notag \\
\phi_{2} &=&\frac{\phi_{2}^{0}}{r}+O(r^{-2}),  \notag
\end{eqnarray}
where the leading coefficients of $r$, ($\phi_{0}^{0}, \phi_{1}^{0},
\phi_{2}^{0}$) satisfy the evolution equations:
\begin{eqnarray}
\dot{\phi}_{0}^{0}+\eth \phi_{1}^{0} &=&0,
\label{lastMaxEqs} \\
\dot{\phi}_{1}^{0}+\eth \phi_{2}^{0} &=&0.
\label{lastmax}
\end{eqnarray}

The formal integration procedure is to take $\phi_{2}^{0}$ as an arbitrary
function of ($u_{\mathrm{B}},\zeta, \bar{\zeta}$) (the free broadcasting data), then
integrate the second, for $\phi_{1}^{0}$, with a time-independent
spin-weight $s=0$ function of integration and finally integrate the
first, for $\phi_{0}^{0}$. Using a slight modification of this, namely from the
spherical harmonic expansion, we obtain, 
\begin{eqnarray}
\phi_{0}^{0} &=&\phi_{0i}^{0}Y_{1i}^{1}+\phi_{0ij}^{0}Y_{2ij}^{1}+ \ldots,
\label{harmonic decomposition} \\
\phi_{1}^{0} &=&q+\phi_{1i}^{0}Y_{1i}^{0}+\phi_{1ij}^{0}Y_{2ij}^{0}+ \ldots, \\
\phi_{2}^{0} &=&\phi_{2i}^{0}Y_{1i}^{-1}+\phi_{2ij}^{0}Y_{2ij}^{-1}+ \ldots,
\end{eqnarray}
with the harmonic coefficients related to each other by the evolution
equations: 
\begin{eqnarray}
\phi_{0}^{0} &=&
2q\eta^{i}(u_{\mathrm{ret}})Y_{1i}^{1}+Q_{\mathbb{C}}^{ij\,
  \prime}Y_{2ij}^{1} + \ldots,
\label{ID} \\
\phi_{1}^{0} &=& q+\sqrt{2}q\eta^{i\,
  \prime}(u_{\mathrm{ret}})Y_{1i}^{0}+\frac{\sqrt{2}}{6}Q_{\mathbb{C}}^{ij\, \prime\prime}Y_{2ij}^{0}+ \ldots,  \notag \\
\phi_{2}^{0} &=& -2q\eta^{i\,
  \prime\prime}(u_{\mathrm{ret}})Y_{1i}^{-1}-\frac{1}{3}Q_{\mathbb{C}}^{ij\,
  \prime\prime\prime}Y_{2ij}^{-1}+ \ldots  \notag
\end{eqnarray}
The physical meaning of the coefficients are 
\begin{eqnarray}
q &=& \text{total electric charge}, \label{ID1} \\
q\eta^{i} &=& D_{\mathbb{C}}^{i} = \text{complex (electric \& magnetic) dipole moment} = D_{E}^{i}+iD_{M}^{i}, \notag \\
Q_{\mathbb{C}}^{ij} &=& \text{complex (electric \& magnetic) quadrupole moment}, \notag
\end{eqnarray}
etc. Recall from~\ref{N&D} that this electromagnetic quadrupole needs to be rescaled ($Q_{\mathbb{C}}^{ij}\rightarrow 2\sqrt{2}Q_{\mathbb{C}}^{ij}$) to obtain the physical quadrupole which appears in the usual expressions for Maxwell theory~\cite{LL}. For later use, the complex dipole is written as $D_{\mathbb{C}}^{i}(u_{\mathrm{ret}}) = q\eta^{i}(u_{\mathrm{ret}})$. Note that the $D_{\mathbb{C}}^{i}$
is defined relative to a given Bondi system. This is the analogue of a given
origin for the calculations of the dipole moments of Eq.~(\ref{D}).

Later in this section it will be shown that we can find a unique complex
world line, $\xi^{a}(\tau)=(\xi^{0},\xi^{i})$, (the world line
associated with a shear-free NGC), that is closely related to the
$\eta^{i}(u_{\mathrm{ret}})$. From this complex world line we can
define the \textit{intrinsic} complex dipole moment,
$D_{\mathcal{I}\mathbb{C}}^{i}=q\xi^{i}(s)$.

However, we first discuss a particular Maxwell field, $F^{ab}$, where one
of its eigenvectors is a tangent field to a shear-free NGC. This solution,
referred to as the complex Li\'{e}nard--Wiechert field is the direct
generalization of the ordinary Li\'{e}nard--Wiechert field. Though it is a
real solution in Minkowski space, \textit{it can be thought of} as arising
from a complex world line in complex Minkowski space.


\subsubsection{A coordinate and tetrad system attached to a shear-free NGC}

The parametric form of the general NGC was given
earlier by Eq.~(\ref{SF2}),
\begin{equation}
x^{a}=u_{\mathrm{B}}(\hat{l}^{a}+\hat{n}^{a})-L\overline{\hat{m}}^{a}-\bar{L}\hat{m}^{a}+(r^{\ast}-r_{0})\hat{l}^{a}.
\label{SF2*}
\end{equation}
The parameters $(u_{\mathrm{B}},\zeta, \bar{\zeta})$ labeled the
individual members of the congruence while $r^{\ast}$ was the affine
parameter along the geodesics. An alternative interpretation of the
same equation is to consider it as the coordinate transformation
between the coordinates, $x^{a}$ (or the Bondi coordinates) and the
\textit{geodesic coordinates} $(u_{\mathrm{B}},r^{\ast},\zeta,
\bar{\zeta})$. Note that while these coordinates are not Bondi coordinates,
though, in the limit, at $\mathfrak{I}^{+}$, they are. The associated
(\textit{geodesic}) tetrad is given as a function of these geodesic
coordinates, but with Minkowskian components by
Eqs.~(\ref{SF2*}). We restrict ourselves to the special case of
the coordinates and tetrad associated with the $L$ from a shear-free
NGC. Though we are dealing with a \textit{real} shear-free twisting
congruence, the congruence, as we saw, is generated by a complex
analytic world line in the complexified Minkowski space,
$z^{a}=\xi^{a}(\tau)$. The complex parameter, $\tau$, must in the end
be chosen so that the `$u_{\mathrm{B}}$' of Eq.~(\ref{G**}) is
real. The Minkowski metric and the spin coefficients associated with
this geodesic system can be calculated~\cite{UCF} in the
$(u_{\mathrm{B}},r^{\ast},\zeta, \bar{\zeta})$ frame. Unfortunately,
it must be stated parametrically, since the $\tau$ explicitly appears
via the $\xi^{a}(\tau)$ and can not be directly eliminated. (An
alternate choice of these geodesic coordinates is to use the $\tau$
instead of the $u_{\mathrm{B}}$. Unfortunately, this leads to an
analytic flat metric on the complexified Minkowski space, where the
real spacetime is hard to find.)

The use and insight given by this coordinate/tetrad system is
illustrated by its application to a special class of Maxwell
fields. We consider, as mentioned earlier, the Maxwell field where one
of its principle null vectors, $l^{\ast a}$, (an eigenvector of the
Maxwell tensor, $F_{ab}l^{\ast a}=\lambda l_{b}^{\ast}$), is a tangent
vector of a shear-free NGC. Thus, it depends on the choice of the
complex world line and is therefore referred to as the complex
Li\'{e}nard--Wiechert field. (If the world line was real it would lead
to the ordinary Li\'{e}nard--Wiechert field.) We emphasize that though
the source can formally be thought of as a charge moving on the
complex world line, the Maxwell field is a real field on real
Minkowski space. It will have a real (distributional) source at the
caustics of the congruence. Physically, its behavior is very similar
to real Li\'{e}nard--Wiechert fields, the essential difference is that
the electric dipole is now replaced by the combined electric and
magnetic dipoles. The imaginary part of the world line determines the
magnetic dipole moment.

In the spin-coefficient version of the Maxwell equations, using the geodesic
tetrad, the choice of $l^{\ast a}$ as the \textit{principle null vector}
`congruence' is just the statement that 
\begin{equation*}
\phi_{0}^{\ast}=F_{ab}l^{\ast a}m^{\ast b}=0.
\end{equation*}
This allows a very simple exact integration of the remaining Maxwell
components~\cite{Maxwell}.


\subsubsection{Complex Li\'{e}nard--Wiechert Maxwell field}

The present section, included as an illustration of the general ideas and
constructions in this work, is rather technical and complicated and can be
omitted without loss of continuity. 

The complex Li\'{e}nard--Wiechert fields (which we again emphasize
are \emph{real} Maxwell fields) are formally given by the (geodesic) tetrad
components 
of the Maxwell tensor in the null geodesic coordinate system
($u_{\mathrm{B}},r^{\ast}, \zeta, \bar{\zeta}$),
Eq.~(\ref{SF2*}). As the detailed calculations are
long~\cite{Maxwell} and take us too far afield, we only give an
outline here. The integration of the radial Maxwell equations
leads to the asymptotic behavior,
\begin{eqnarray}
\phi_{0}^{\ast} &=&0,
\label{sol.I} \\
\phi_{1}^{\ast} &=&\rho^{2}\phi_{1}^{\ast 0}, \\
\phi_{2}^{\ast} &=&\rho \phi_{2}^{\ast 0}+O(\rho^{2}),
\end{eqnarray}
with 
\begin{eqnarray}
\rho &=&-(r^{\ast}+i\Sigma )^{-1},
\label{aux} \\
2i\Sigma &=&\eth \bar{L}+L\dot{\bar{L}}-\overline{\eth}L-\bar{L}\dot{L}.  \notag
\end{eqnarray}
The $O(\rho^{2})$ expression is known in terms of ($\phi_{1}^{\ast 0},\phi
_{2}^{\ast 0}$). The function $L(u_{\mathrm{B}},\zeta, \bar{\zeta})$
is given by
\begin{eqnarray*}
L(u_{\mathrm{B}},\zeta,\bar{\zeta}) &=&\eth_{(\tau)}G(\tau,\zeta,\bar{\zeta}), \\
u_{\mathrm{B}} &=&G(\tau,\zeta,\bar{\zeta})=\xi^{a}(\tau)\hat{l}_{a}(\zeta,\bar{\zeta}),
\end{eqnarray*}
with $\xi^{a}(\tau)$ an arbitrary complex world line that determines the
shear-free congruence whose tangent vectors are the Maxwell field
eigenvectors.

\begin{remark}
In this case of the complex Li\'{e}nard--Wiechert Maxwell
field, the $\xi^{a}$ determines the intrinsic center-of-charge world line,
rather than the relative center-of-charge line.
\end{remark}

The remaining unknowns, $\phi_{1}^{\ast 0},\phi_{2}^{\ast 0}$, are
determined by the last of the Maxwell equations, 
\begin{eqnarray}
\eth \phi_{1}^{\ast 0}+L\dot{\phi}_{1}^{\ast 0}+2\dot{L}\phi_{1}^{\ast 0}
&=&0,  \label{MaxLW} \\
\eth \phi_{2}^{\ast 0}+L\dot{\phi}_{2}^{\ast 0}+\dot{L}\phi_{2}^{\ast 0}
&=&\,\dot{\phi}_{1}^{\ast 0},  \notag
\end{eqnarray}
which have been obtained from Eqs.~(\ref{AsyBI4}) and~(\ref{AsyBI5}) via the
null rotation between the Bondi and geodesic tetrads and the associated
Maxwell field transformation, namely, 
\begin{eqnarray}
l^{a} &\rightarrow &l^{\ast a}=l^{a}-\frac{\bar{L}}{r}m^{a}-\frac{L}{r}\bar{m}^{a}+O(r^{\ast -2}),
\label{NULL ROT} \\
m^{a} &\rightarrow &m^{\ast a}=m^{a}-\frac{L}{r}n^{a}, \\
n^{a} &\rightarrow &n^{\ast a}=n^{a},
\end{eqnarray}
with 
\begin{eqnarray}
\phi_{0}^{\ast 0} &=&0=\phi_{0}^{0}-2L\phi_{1}^{0}+L^{2}\phi_{2}^{0},
\label{null rot 1} \\
\phi_{1}^{\ast 0} &=&\phi_{1}^{0}-L\phi_{2}^{0},
\label{null rot 2} \\
\phi_{2}^{\ast 0} &=&\phi_{2}^{0}.
\label{null rot 3}
\end{eqnarray}

These remaining equations depend only on $L(u_{\mathrm{B}},\zeta, \bar{\zeta})$,
which, in turn, is determined by $\xi^{a}(\tau)$. In other words, the
solution is driven by the complex line, $\xi^{a}(\tau)$. As they now stand,
Eqs.~(\ref{MaxLW}) appear to be difficult to solve, partially due to the
implicit description of the $L(u_{\mathrm{B}},\zeta, \bar{\zeta})$.

Actually they are easily solved when the independent variables are changed,
via Eq.~(\ref{SF12}), from $(u_{\mathrm{B}},\zeta, \overline{\zeta})$ to the complex $(\tau ,\zeta, \overline{\zeta})$. They become, after a bit of work,

\begin{eqnarray}
\eth_{(\tau)}(V^{2}\phi_{1}^{0}) &=&0,
\label{MaxI} \\
\eth_{(\tau)}(V\phi_{2}^{0}) &=&\phi_{1}^{0\, \prime},
\label{Max2} \\
V &=&\xi^{a\prime}(\tau)\hat{l}_{a}(\zeta, \bar{\zeta}),
\end{eqnarray}
with the solution 
\begin{eqnarray}
\phi_{1}^{\ast 0} &=&\frac{q}{2}V^{-2},
\label{LWSol} \\
\phi_{2}^{\ast 0} &=&\frac{q}{2}V^{-1}\overline{\eth}_{(\tau)}\left(V^{-1}\partial_{\tau}V\right).  \notag
\end{eqnarray}
$q$ being the Coulomb charge.

Though we now have the exact solution, unfortunately it is in
complex coordinates where virtually every term depends on the complex
variable $\tau$, via $\xi^{a}(\tau)$. This is a severe impediment to a
full description and understanding of the solution in the real
Minkowski space.

In order to understand its asymptotic behavior and physical content, one
must transform it, via Eqs.~(\ref{NULL ROT})\,--\,(\ref{null rot 3}), back to a
Bondi coordinate/tetrad system. This can only be done by approximations.
After a lengthy calculation~\cite{Maxwell}, we find the Bondi peeling behavior
\begin{eqnarray}
\phi_{0} &=&r^{-3}\phi_{0}^{0}+O(r^{-4}),
\label{phiB} \\
\phi_{1} &=&r^{-2}\phi_{1}^{0}+O(r^{-3}),  \notag \\
\phi_{2} &=&r^{-1}\phi_{2}^{0}+O(r^{-2}),  \notag
\end{eqnarray}
with 
\begin{eqnarray}
\phi_{0}^{0} &=&q\left(LV^{-2}+\frac{1}{2}L^{2}V^{-1}\overline{\eth}_{(\tau)}[V^{-1}V^{\prime}]\right),
\label{phi^0*} \\
\phi_{1}^{0} &=&\frac{q}{2V^{2}}\left(1+LV\overline{\eth}_{(\tau)}[V^{-1}V^{\prime}]\right),
\label{phi^1*} \\
\phi_{2}^{0} &=&-\frac{q}{2}V^{-1}\overline{\eth}_{(\tau)}(V^{-1}V^{\prime}),
\label{phi^3*} \\
V &=&\xi^{a\,\prime}\,\hat{l}_{a}(\zeta, \bar{\zeta}).
\end{eqnarray}
Next, treating the world line, as discussed earlier, as a \textit{small
deviation from the straight line}, $\xi^{a}(\tau)=\tau \delta_{0}^{a}$,
i.e., by 
\begin{eqnarray*}
\xi^{a}(\tau) &=&\left(\tau,\xi^{i}(\tau)\right), \\
\xi^{i}(\tau) &\ll &1.
\end{eqnarray*}
The GCF and its inverse (see Section~\ref{results}) are given, to
first order, by 
\begin{eqnarray}
u_{\mathrm{ret}} &=&\sqrt{2}u_{\mathrm{B}}=\sqrt{2}G=\tau
-\frac{\sqrt{2}}{2}\xi^{i}(\tau)Y_{1i}^{0}(\zeta, \overline{\zeta}),
\label{u_r} \\
\tau &=&u_{\mathrm{ret}}+\frac{\sqrt{2}}{2}\xi^{i}(u_{\mathrm{ret}})Y_{1i}^{0}(\zeta,
\overline{\zeta}).
\end{eqnarray}

Again to first order, Eqs.~(\ref{phi^0*}),~(\ref{phi^1*}) and~(\ref{phi^3*}) yield
\begin{eqnarray}
\phi_{0}^{0} &=&2q\xi^{i}(u_{\mathrm{ret}})Y_{1i}^{1},
\label{dipole field} \\
\phi_{1}^{0} &=&q+\sqrt{2}q\xi^{i\prime}(u_{\mathrm{ret}})Y_{1i}^{0},  \notag \\
\phi_{2}^{0} &=&-2q\xi^{i\prime\prime}(u_{\mathrm{ret}})Y_{1i}^{-1},  \notag
\end{eqnarray}
the known electromagnetic dipole field, with a Coulomb charge, $q$. One then
has the physical interpretation of $q\xi^{i}(u_{\mathrm{ret}})$ as the complex dipole moment; (the electric plus `$i$' times
magnetic dipole) and $\xi^{i}(u_{\mathrm{ret}})$ is the complex center of
charge, the real part being the ordinary center of charge, while the
imaginary part is the `imaginary' magnetic center of charge. This simple
relationship between the Bondi form of the complex dipole moment, $q\xi
^{i}(u_{\mathrm{ret}})$, and the intrinsic complex center of charge, $\xi^{i}(\tau)$, is true only at linear order. The second-order relationship is
given later.

Reversing the issue, if we had instead started with an exact complex
Li\'{e}nard--Wiechert field but now given in a Bondi coordinate/tetrad system and
performed on it the transformations, Eqs.~(\ref{Tet10}) and~(\ref{null rot 1}) to the geodesic system, it would have resulted in 
\begin{equation*}
\phi_{0}^{\ast}=0.
\end{equation*}

This example was intended to show how physical meaning could be attached to
the complex world line associated with a shear-free NGC. In this case and
later in the case of asymptotically-flat spacetimes, when the GCF is
singled out by either the Maxwell field or the gravitational field, it will
be referred to it as a UCF. For either of the two cases, a flat-space
asymptotically-vanishing Maxwell field (with nonvanishing total charge) and
for a vacuum asymptotically-flat spacetime, there will be a unique UCF. In
the case of the Einstein--Maxwell fields there will, in general, be two
UCFs: one for each field.


\subsubsection{Asymptotically vanishing Maxwell fields \& shear-free NGCs}

We return now to the general asymptotically-vanishing Maxwell field,
Eqs.~(\ref{Max1}) and~(\ref{harmonic decomposition}), and its
transformation behavior under the null rotation around $n^{a}$, 
\begin{eqnarray}
l^{a} &\rightarrow &l^{\ast a}=l^{a}-\frac{\bar{L}}{r}m^{a}-\frac{L}{r}\bar{m}^{a}+0(r^{-2}),  \label{nullROT2} \\
m^{a} &\rightarrow &m^{\ast a}=m^{a}-\frac{L}{r}n^{a}+0(r^{-2}),  \notag \\
n^{a} &\rightarrow &n^{\ast a}=n^{a},  \notag
\end{eqnarray}
with $L(u_{\mathrm{B}},\zeta, \overline{\zeta})=\xi^{a}(\tau)\hat{m}_{a}$, being one of our shear-free angle fields defined by a world line, $z^{a}=\xi^{a}(\tau)$ and cut function $u_{\mathrm{B}}=\xi^{a}(\tau)\hat{l}_{a}(\zeta, \overline{\zeta})$. The leading components of the Maxwell
fields transform as 
\begin{eqnarray}
\phi_{0}^{\ast 0} &=&\phi_{0}^{0}-2L\phi_{1}^{0}+L^{2}\;\phi_{2}^{0},
\label{transformed} \\
\phi_{1}^{\ast 0} &=&\phi_{1}^{0}-L\phi_{2}^{0}, \\
\phi_{2}^{\ast 0} &=&\phi_{2}^{0}.
\end{eqnarray}

The `picture' to adopt is that the new $\phi^{\ast}$s are now given in a
tetrad defined by the complex light cone (or generalized light cone) with
origin on the complex world line. (This is obviously formal and perhaps
physically nonsensical, but mathematically quite sound, as the shear-free
congruence can be thought of as having its origin on the complex line, $\xi^{a}(\tau)$.) From the physical identifications of charge, dipole moments,
etc., of Eq.~(\ref{ID}), we can obtain the transformation law of these
physical quantities. In particular, the $l=1$ harmonic of $\phi_{0}^{0}$,
or, equivalently, the complex dipole, transforms as 
\begin{equation}
\phi_{0i}^{0\ast}=\phi_{0i}^{0}-2(L\phi_{1}^{0})|_{i}+(L^{2}\phi_{2}^{0})|_{i},
\label{dipole trans}
\end{equation}
where the notation $W|_{i}$ means \textit{extract only} the $l=1$ harmonic
from a Clebsch--Gordon expansion of $W$. A subtlety and difficulty of this
extraction process is here clarified.

\subsubsection*{\textit{The (non-)uniqueness of spherical harmonic expansions}}

An important observation, obvious but easily overlooked, concerning the
spherical harmonic expansions is that, in a certain sense, they lack
uniqueness. As this issue is significant, its clarification is important.

Assume that we have a particular spin-$s$ function on $\mathfrak{I}^{+}$,
say, $\eta_{(s)}(u_{\mathrm{B}},\zeta, \overline{\zeta})$, given in a specific
Bondi coordinate system, $(u_{\mathrm{B}},\zeta, \overline{\zeta})$, that has a
harmonic expansion given, for constant $u_{\mathrm{B}}$, by 
\begin{equation*}
\eta_{(s)}(u_{\mathrm{B}},\zeta, \overline{\zeta})=\underset{l,(ijk
  \ldots)}{\Sigma}\eta_{(s)}^{l,(ijk \ldots)}(u_{\mathrm{B}})Y_{l,(ijk \ldots)}^{(s)}
\end{equation*}
If exactly the same function was given on different cuts or slices,
say, 
\begin{equation}
u_{\mathrm{B}}=G(\tau,\zeta, \overline{\zeta}),
\label{G(s)}
\end{equation}
with 
\begin{equation*}
\eta_{(s)}^{\ast}(\tau,\zeta, \overline{\zeta})=\eta_{(s)}\left(G(\tau,\zeta,\overline{\zeta}),\zeta,\overline{\zeta}\right),
\end{equation*}
the harmonic expansion at constant $\tau$ would be different. The new
coefficients are extracted by the two-sphere integral taken at constant
$\tau$: 
\begin{equation}
\eta_{(s)}^{\ast l,(ijk\ldots )}(\tau)=\int_{S^{2}}\eta_{(s)}^{\ast}(\tau,\zeta,\overline{\zeta})\overline{Y}_{l,(ijk\ldots)}^{(s)}dS.
\label{extract}
\end{equation}
It is in this rather obvious sense that the expansions are not unique.

The transformation, Eq.~(\ref{dipole trans}), and harmonic extraction
implemented by first replacing the $u_{B}$ in all the terms of all $\phi_{0}^{0},\phi_{1}^{0},\phi_{2}^{0}$, by $u_{\mathrm{B}}=\xi^{a}(\tau)\hat{l}_{a}(\zeta,\overline{\zeta})\equiv \frac{\sqrt{2}\tau}{2}-\frac{1}{2}\xi^{i}Y_{1i}^{0}$, yields $\phi_{0i}^{0\ast}(\tau)$ with a functional
form~\cite{Adamo:2011cx}, 
\begin{equation}
\phi_{0i}^{0\ast}(\tau)=\Gamma_{i}(\phi_{0}^{0},\phi_{1}^{0},\phi_{2}^{0},\xi^{a}(\tau))=\oint_{S^{2}}[\phi_{0}^{0}-(2cL\phi_{1}^{0}-c^{2}L^{2}\phi_{2}^{0})]Y_{1i}^{-1}dS
\label{dipole trans*}
\end{equation}
is decidedly nontensorial: in fact it is very nonlocal and nonlinear.

Though it is clear that extracting $\phi_{0i}^{0\ast}(\tau)$ with this
relationship is available in principle, in practice it is impossible to do
it exactly and all examples are done with approximations: essentially using
slow motion for the complex world line.

\begin{remark}
If by some accident the Maxwell field was a complex Li\'{e}nard--Wiechert field, a world line $\xi^{a}(\tau)$ could be chosen so that
from the associated complex null cones we would have $\phi_{0}^{\ast 0}=0$.
However, though this cannot be done in general, the $l=1$ harmonics of $\phi_{0}^{\ast 0}$ can be made to vanish by the appropriate choice of the $\xi^{a}(\tau)$. This is the means by which a unique world line is chosen.
\end{remark}

\subsubsection{The complex center of charge}

The complex center of charge is defined by the vanishing of the complex
dipole moment $\phi_{0i}^{0\ast}(\tau)$; in other words, 
\begin{equation}
\Gamma_{i}(\phi_{0}^{0},\phi_{1}^{0},\phi_{2}^{0},\xi^{a})=0
\label{CCofC}
\end{equation}
determines three components of the (up to now) arbitrary complex world line, 
$\xi^{a}(\tau)$; the fourth component can be taken as $\tau$. In
practice we do this only up to second order with the use of only the $(l=0,1,2)$ harmonics. The approximation we are using is to consider the
charge $q$ as zeroth order and the dipole moments and the spatial part of
the complex world line as first order.

From Eq.~(\ref{dipole trans}), 
\begin{equation}
\phi_{0i}^{0\ast}=\Gamma_{i}\approx \phi_{0i}^{0}-2L\phi_{1}^{0}|_{i}=0
\label{dipole trans**}
\end{equation}
with the identifications, Eq.~(\ref{ID}), for $q$ and $D_{\mathbb{C}}$,
we have to first order (with $\sqrt{2}u_{B}=u_{ret}\approx \tau$),
\begin{equation}
D_{\mathbb{C}}^{i}(u_{\mathrm{ret}})=q\eta^{i}(u_{\mathrm{ret}})=q\xi^{i}(u_{\mathrm{ret}}).
\label{Bondi dipole}
\end{equation}
This is exactly the same result as we obtained earlier in Eq.~(\ref{CCC}),
via the charge and current distributions in a fixed Lorentz frame.

Carrying this calculation~\cite{Maxwell} to second order, we find the
second-order complex center of charge and the relationship between the
intrinsic complex dipole, $D_{\mathcal{I}:\mathbb{C}}^{i}$, and the complex
dipole, $D_{\mathbb{C}}^{i}$, 
\begin{eqnarray}
D_{\mathcal{I}:\mathbb{C}}^{i} &=&q\xi^{i}(s), \qquad D_{\mathbb{C}}^{i}=q\eta^{i}(s),
\label{IntrinsicE&M1} \\
\xi^{k} &=&\eta^{k}-\frac{i}{2}\eta^{i}\eta^{j\prime}\epsilon_{ijk}-\frac{\sqrt{2}}{10}q^{-1}Q_{\mathbb{C}}^{ik\prime\prime}\xi^{i},
\label{IntrinsicE&M2} \\
\eta^{k} &=&\xi^{k}+i\frac{1}{2}\xi^{i}\xi^{j\prime}\epsilon_{ijk}+\frac{\sqrt{2}}{10}q^{-1}Q_{\mathbb{C}}^{ik\prime\prime}\xi^{i}
\end{eqnarray}

The GCF, 
\begin{equation*}
u_{\mathrm{B}}=X(\tau,\zeta,\tilde{\zeta})=\xi^{a}(\tau)\hat{l}_{a}(\zeta,\tilde{\zeta}),
\end{equation*}
with this uniquely determined world line is referred to as the Maxwell UCF.

In Section~\ref{applications}, these ideas are applied to GR, with the
complex electric and magnetic dipoles being replaced by the complex
combination of the mass dipole and the angular momentum.


\newpage

\section[The Good-Cut Equation and $\mathcal{H}$-Space]{The Good-Cut Equation and \boldmath$\mathcal{H}$-Space}
\label{good-cut-eq}

In Section~\ref{shear-free-NGC}, we discussed NGCs in Minkowski
spacetime that were shear-free. In this section we consider
\textit{asymptotically} shear-free NGCs in asymptotically-flat
spacetimes. That is to say, we consider NGCs that have nonvanishing
shear in the interior of the spacetime but where, as null infinity is
approached, the shear vanishes. Whereas fully shear-free NGCs almost never occur in general asymptotically flat spacetimes, \textit{asymptotically}
shear-free congruences always exist. The case of algebraically-special spacetimes 
is the exception; they do allow one or two shear-free congruences.

We begin by reviewing the shear-free condition and follow with its
generalization to the asymptotically shear-free case. From this we derive the
generalization of the homogeneous good-cut equation to the inhomogeneous
good-cut equation. Almost all the properties of the shear-free and
asymptotically shear-free NGCs come from the study of these equations and virtually all the attributes of shear-free congruences are
shared by the asymptotically shear-free congruences. It is from the use of
these shared attributes that we will be able to extract physical
identifications and information (e.g., complex center of mass/charge, Bondi
mass, linear and angular momentum, equations of motion, etc.) from the
asymptotic gravitational fields.

Though again the use of the formal complexification of $\mathfrak{I}^{+}$,
i.e., $\mathfrak{I}_{\mathbb{C}}^{+}$, is essential for our analysis, it is
the extraction of the real structures that is important.


\subsection{Asymptotically shear-free NGCs and the good-cut equation}

We saw in Section~\ref{shear-free-NGC} that shear-free NGCs in
Minkowski space could be constructed by looking at their properties
near $\mathfrak{I}^{+}$, in one of two equivalent ways. The first was via the stereographic angle field,
$L(u_{\mathrm{B}},\zeta, \bar{\zeta})$, which gives the directions that
null rays make at their intersection with $\mathfrak{I}^{+}$. The
condition for the congruence to be shear-free was that $L$ must
satisfy 
\begin{equation}
\eth_{(u_{\mathrm{B}})}L+L\dot{L}=0.
\label{shearfree}
\end{equation}
We required solutions that were all nonsingular (regular) on the
$(\zeta, \bar{\zeta})$ sphere. (This equation has in the past most
often been solved via twistor methods~\cite{HansenNewman}.)

The second was via the complex cut function, $u_{\mathrm{B}} = G(\tau,
\zeta,\tilde{\zeta})$, that satisfied 
\begin{equation}
\eth_{(\tau)}^{2}G=0.
\label{IGC}
\end{equation}
The regular solutions were easily given by 
\begin{equation}
u_{\mathrm{B}}=G(\tau ,\zeta, \tilde{\zeta})=\xi^{a}(\tau)\hat{l}_{a}(\zeta, \tilde{\zeta})
\label{G@}
\end{equation}
with inverse function, 
\begin{equation*}
\tau =T(u_{\mathrm{B}},\zeta, \tilde{\zeta}).
\end{equation*}
They determined the $L(u_{\mathrm{B}},\zeta, \bar{\zeta})$ that satisfies
Eq.~(\ref{shearfree}) by the parametric relations 
\begin{eqnarray}
L(u_{\mathrm{B}},\zeta, \bar{\zeta}) &=&\eth_{(\tau)}G(\tau ,\zeta, \tilde{\zeta}),
\label{L} \\
u_{\mathrm{B}} &=&\xi^{a}(\tau)\hat{l}_{a}(\zeta, \tilde{\zeta}),  \notag
\end{eqnarray}
or by 
\begin{equation*}
L(u_{\mathrm{B}},\zeta, \bar{\zeta})=\eth_{(\tau)}G(\tau ,\zeta, \bar{\zeta})|_{\tau =T(u_{\mathrm{B}},\zeta, \tilde{\zeta})},
\end{equation*}
where $\xi^{a}(\tau)$ was an arbitrary complex world line in complex
Minkowski space. 

It is this pair of equations,~(\ref{shearfree}) and~(\ref{IGC}), that will
now be generalized to asymptotically-flat spacetimes.

In Section~\ref{foundations}, we saw that the asymptotic shear of the
(null geodesic) tangent vector fields, $l^{a}$, of the out-going Bondi
null surfaces was given by the free data (the Bondi shear)
$\sigma^{0}(u_{\mathrm{B}},\zeta, \bar{\zeta})$. If, near $\mathfrak{I}^{+}$, a
second NGC, with tangent vector $l^{\ast a}$, is chosen and then
described by the null rotation from $l^{a}$ to $l^{\ast a}$ around
$n^{a}$ by 
\begin{eqnarray}
l^{\ast a} &=&l^{a}+b\overline{m}^{a}+\overline{b}m^{a}+b\overline{b}n^{a},
\label{NullRot} \\
m^{\ast a} &=&m^{a}+bn^{a},  \notag \\
n^{\ast a} &=&n^{a},  \notag \\
b &=&-L/r+O(r^{-2}),  \notag
\end{eqnarray}
with $L(u_{\mathrm{B}},\zeta, \bar{\zeta})$ an arbitrary stereographic angle field,
then the asymptotic Weyl components transform as
\begin{eqnarray}
\psi_{0}^{\ast 0} &=&\psi_{0}^{0}-4L\psi_{1}^{0}+6L^{2}\psi_{2}^{0}-4L^{3}\psi_{3}^{0}+L^{4}\psi_{4}^{0},
\label{Weyl-rot}
\\
\psi_{1}^{\ast 0} &=&\psi_{1}^{0}-3L\psi_{2}^{0}+3L^{2}\psi_{3}^{0}-L^{3}\psi_{4}^{0},
\label{Weyl-rot2} \\
\psi_{2}^{\ast 0} &=&\psi_{2}^{0}-2L\psi_{3}^{0}+L^{2}\psi_{4}^{0},
\label{Weyl-rot3} \\
\psi_{3}^{\ast 0} &=&\psi_{3}^{0}-L\psi_{4}^{0},
\label{Weyl-rot4} \\
\psi_{4}^{\ast 0} &=&\psi_{4}^{0},
\label{Weyl-rot5}
\end{eqnarray}
%
and the (new) asymptotic shear of the null vector field $l^{\ast a}$ is
given by~\cite{Aronson, Footprints}
\begin{equation}
\sigma^{0\ast}=\sigma^{0}-\eth_{(u_{\mathrm{B}})}L-L\dot{L}.
\label{GC10}
\end{equation}

By requiring that the new congruence be \textit{asymptotically}
shear-free, i.e., $\sigma^{0\ast}=0$, we obtain the generalization of
Eq.~(\ref{shearfree}) for the determination of $L(u_{\mathrm{B}},\zeta,
\bar{\zeta})$, namely,
\begin{equation}
\eth_{(u_{\mathrm{B}})}L+L\dot{L}=\sigma^{0}(u_{\mathrm{B}},\zeta, \bar{\zeta}).
\label{A.shearfree}
\end{equation}
To solve this equation we again complexify $\mathfrak{I}^{+}$ to
$\mathfrak{I}_{\mathbb{C}}^{+}$ by freeing $\bar{\zeta}$ to
$\tilde{\zeta}$ and allowing $u_{\mathrm{B}}$ to take complex values close to
the real.

Again we introduce the complex potential $\tau =T(u_{\mathrm{B}},\zeta, \tilde{\zeta})$
that is related to $L$ by 
\begin{equation}
\eth_{(u_{\mathrm{B}})}T+L\dot{T}=0,
\label{CREq}
\end{equation}
with its inversion, 
\begin{equation}
u_{\mathrm{B}}=G(\tau ,\zeta, \tilde{\zeta}).
\label{Inversion}
\end{equation}
Eq.~(\ref{A.shearfree}) becomes, after the change in the
independent variable, $u_{\mathrm{B}}\Rightarrow \tau
=T(u_{\mathrm{B}},\zeta, \bar{\zeta})$, and implicit differentiation
(see Section~\ref{flat-space-good-cut} for the identical details),
\begin{equation}
\eth_{(\tau)}^{2}G=\sigma^{0}(G,\zeta, \tilde{\zeta}).
\label{GC}
\end{equation}
This, the inhomogeneous good-cut equation, is the generalization of Eq.~(\ref{IGC}).

In Section~\ref{H-space-and-good-cut-eq}, we will discuss how to construct solutions of
Eq.~(\ref{GC}) of the form, $u_{\mathrm{B}}=G(\tau ,\zeta, \tilde{\zeta})$;
however, assuming we have such a solution, it determines the angle
field $L(u_{\mathrm{B}},\zeta, \bar{\zeta})$ by the parametric relations 
\begin{eqnarray}
L(u_{\mathrm{B}},\zeta, \tilde{\zeta}) &=&\eth_{(\tau)}G,
\label{L2*} \\
u_{\mathrm{B}} &=&G(\tau ,\zeta, \tilde{\zeta}).  \notag
\end{eqnarray}
We now turn to these solutions and their properties.


\subsection[$\mathcal{H}$-space and the good-cut equation]{\boldmath$\mathcal{H}$-space and the good-cut equation}
\label{H-space-and-good-cut-eq}

Eq.~(\ref{GC}), written in earlier literature as 
\begin{equation}
\eth^{2}Z=\sigma^{0}(Z,\zeta, \tilde{\zeta}),
\label{GC2}
\end{equation}
is a well-known and well-studied partial differential equation, often
referred to as the ``good-cut equation''~\cite{HansenNewman,PropHspace}. For
sufficiently regular $\sigma^{0}(u_{\mathrm{B}},\zeta, \bar{\zeta})$ (which is
assumed here) it has been proven~\cite{PropHspace} that the solutions are
determined by points in a complex four-dimensional space, $z^{a}$, referred
to as $\mathcal{H}$-space, i.e., solutions are given as 
\begin{equation}
u_{\mathrm{B}}=Z(z^{a},\zeta, \tilde{\zeta}).
\label{solution}
\end{equation}

Later in this section, by choosing an arbitrary complex analytic world line
in $\mathcal{H}$-space, $z^{a}=\xi^{a}(\tau)$, we describe how to
construct the shear-free angle field, $L(u_{\mathrm{B}},\zeta, \tilde{\zeta})$.
First, however, we discuss properties and the origin of Eq.~(\ref{solution}).

Roughly or intuitively one can see how the four complex parameters enter the
solution from the following argument. We can write Eq.~(\ref{GC2})
as the integral equation
\begin{equation}
Z=z^{a}\hat{l}_{a}(\zeta, \tilde{\zeta})+\mathop{\displaystyle \oint} \sigma^{0}(Z,\eta ,\tilde{\eta})K_{0,-2}^{+}(\eta ,\tilde{\eta}, \zeta, \tilde{\zeta})dS_{\eta}
\end{equation}
with 
\begin{eqnarray*}
K_{0,-2}^{+}(\zeta, \tilde{\zeta},\eta ,\tilde{\eta}) &\equiv &-\frac{1}{4\pi}\frac{(1+\tilde{\zeta}\eta)^{2}(\eta -\zeta )}{(1+\zeta \tilde{\zeta})(1+\eta \tilde{\eta})(\tilde{\eta}-\tilde{\zeta})}, \\
dS_{\eta} &=&4i\frac{d\eta \wedge d\tilde{\eta}}{(1+\eta \tilde{\eta})^{2}},
\end{eqnarray*}
where $z^{a}\hat{l}_{a}(\zeta, \tilde{\zeta})$ is the kernel of the $\eth^{2}$
operator (the solution to the homogeneous good-cut equation) and
$K_{0,-2}^{+}(\zeta, \tilde{\zeta}, \eta, \tilde{\eta})$ is the Green's
function for the $\eth^{2}$ operator~\cite{Greens}. By iterating this
equation, with the kernel being the zeroth iterate, i.e., 
\begin{eqnarray}
Z_{n}(\zeta, \tilde{\zeta}) &=&z^{a}\hat{l}_{a}(\zeta, \tilde{\zeta})+\int_{S^{2}}K_{0,-2}^{+}(\zeta, \tilde{\zeta},\eta ,\tilde{\eta})\sigma
(Z_{n-1},\eta ,\tilde{\eta})dS_{\eta},
\label{I} \\
Z_{0}(\zeta, \tilde{\zeta}) &=&z^{a}\hat{l}_{a}(\zeta, \tilde{\zeta}),
\end{eqnarray}
one easily sees how the four $z^{a}$ enter the solution. Basically,
the $z^{a}$ come from the solution to the homogeneous equation.

It should be noted again that the $z^{a}\hat{l}_{a}(\zeta, \tilde{\zeta})$ is
composed of the $l=(0,1)$ harmonics, 
\begin{equation}
z^{a}\hat{l}_{a}(\zeta, \tilde{\zeta})=\frac{1}{\sqrt{2}}z^{0}-\frac{1}{2}z^{i}Y_{1i}^{0}(\zeta, \tilde{\zeta}).
\label{homogeneous}
\end{equation}
Furthermore, the integral term does not contribute to these lowest harmonics.
This means that solutions can be written 
\begin{equation}
u_{\mathrm{B}}=Z(z^{a},\zeta, \tilde{\zeta})\equiv z^{a}\hat{l}_{a}(\zeta, \tilde{\zeta})+Z_{l\geq 2}(z^{a},\zeta, \tilde{\zeta}),
\label{inhom}
\end{equation}
with $Z_{l\geq 2}$ containing spherical harmonics $l=2$ and higher.  

We note that using this form of the solution implies that we have set
stringent coordinate conditions on the $\mathcal{H}$-space by
requiring that the first four spherical harmonic coefficients be the
four $\mathcal{H}$-space coordinates.  Arbitrary coordinates would
just mean that these four coefficients were arbitrary functions of
other coordinates.  How these special coordinates change under the BMS
group is discussed later.

\begin{remark}
It is of considerable interest that on $\mathcal{H}$-space
there is a natural quadratic complex metric -- as constructed in Appendix~\ref{appendixD} -- that is given by the surprising relationship~\cite{Hspace, PropHspace} 
\begin{eqnarray}
ds_{(\mathcal{H})}^{2} &=&g_{(\mathcal{H})ab}\,dz^{a}dz^{b}\equiv \left(\frac{1}{8\pi}\int_{S^{2}}\frac{dS}{(dZ)^{2}}\right)^{-1},
\label{H metric} \\
dZ &\equiv &\nabla_{a}Z\;dz^{a}, \\
dS &=&4i\frac{d\zeta \wedge d\tilde{\zeta}}{(1+\zeta \tilde{\zeta})^{2}}.
\end{eqnarray}
%
Remarkably this turns out to be a Ricci-flat metric with a
nonvanishing anti-self-dual Weyl tensor and vanishing self-dual Weyl tenor, i.e., it is intrinsically a
complex anti-self-dual vacuum metric.  For vanishing Bondi shear, $\mathcal{H}$-space
reduces to complex Minkowski space (i.e.,
$g_{(\mathcal{H})ab}|_{\sigma^{0}=0}=\eta_{ab}$).
\end{remark}


\subsubsection{Solutions to the shear-free equation}

Returning to the issue of the solutions to the shear-free condition, i.e.,
Eq.~(\ref{A.shearfree}), $L(u_{\mathrm{B}},\zeta, \tilde{\zeta})$, we see that they
are easily constructed from the solutions to the good-cut equation,
$u_{\mathrm{B}} = Z(z^{a},\zeta, \tilde{\zeta})$. By choosing an arbitrary complex
world line in the $\mathcal{H}$-space, i.e., 
\begin{equation}
z^{a}=\xi^{a}(\tau),
\label{world-line}
\end{equation}
we write the GCF as 
\begin{equation}
u_{\mathrm{B}}=G(\tau ,\zeta, \tilde{\zeta})\equiv Z(\xi^{a}(\tau),\zeta, \tilde{\zeta}),
\label{GCF}
\end{equation}
or, from Eq.~(\ref{inhom}), 
\begin{equation}
u_{\mathrm{B}}=G(\tau ,\zeta,
\tilde{\zeta})=\frac{1}{\sqrt{2}}\xi^{0}(\tau)-\frac{1}{2}\xi^{i}(\tau)Y_{1i}^{0}(\zeta,
\tilde{\zeta})+\xi^{ij}(\tau)Y_{2ij}^{0}(\zeta, \tilde{\zeta})+ \ldots .
\label{UCF***}
\end{equation}
This leads immediately, via Eqs.~(\ref{L2*}) and~(\ref{UCF***}), to the
parametric description of the shear-free stereographic angle field
$L(u_{\mathrm{B}}, \zeta, \tilde{\zeta})$, as well as the Bondi shear
$\sigma^{0}(u_{\mathrm{B}},\zeta, \tilde{\zeta})$:
\begin{eqnarray}
u_{\mathrm{B}} &=&
\frac{1}{\sqrt{2}}\xi^{0}(\tau)-\frac{1}{2}\xi^{i}(\tau)Y_{1i}^{0}(\zeta,
\tilde{\zeta})+\xi^{ij}(\tau)Y_{2ij}^{0}(\zeta, \tilde{\zeta})+ \dots,
\label{u_c} \\
L(u_{\mathrm{B}},\zeta, \tilde{\zeta}) &=&
\xi^{i}(\tau)Y_{1i}^{1}(\zeta,\tilde{\zeta})-6\xi^{ij}(\tau)Y_{2ij}^{1}(\zeta,
\tilde{\zeta})+ \dots,
\label{L**} \\
\sigma^{0}(u_{\mathrm{B}},\zeta, \tilde{\zeta}) &=& 24\xi^{ij}(\tau)Y_{2ij}^{2}+ \dots .
\label{sigma***}
\end{eqnarray}
We denote the inverse to Eq.~(\ref{UCF***}) by 
\begin{equation}
\tau =T(u_{\mathrm{B}},\zeta, \tilde{\zeta}),
\label{T}
\end{equation}
and refer to the complex world line $\xi^{a}(\tau)$ as the `virtual' source of the congruence.
The asymptotic twist of the asymptotically shear-free NGC is exactly as in
the flat-space case, 
\begin{equation}
i\Sigma =\frac{1}{2}\left\{\eth\overline{L}+L\dot{\overline{L}}-\overline{\eth}L-\overline{L}\dot{L}\right\}.
\label{twist*}
\end{equation}
As in the flat-space case, the derived quantity
\begin{equation}
V(\tau,\zeta, \tilde{\zeta}) \equiv \partial_{\tau}G=G^{\prime}
\end{equation}
plays a large role in applications. (In the case of the
Robinson--Trautman metrics~\cite{RobinsonTrautman, RTmetrics} $V$ is the basic variable for the construction of the metric.)  

Using the gauge freedom, $\tau \rightarrow \tau^{\ast}=\Phi(\tau)$, as in the Minkowski-space case, we impose the simple condition
\begin{equation}
\xi^{0}=\tau.
\label{tau}
\end{equation}

\textbf{A Brief Summary:} The description and analysis of the
asymptotically shear-free NGCs in asymptotically-flat spacetimes is
remarkably similar to that of the flat-space regular shear-free NGCs. We
have seen that all \textit{regular} shear-free NGCs in Minkowski space and
asymptotically-flat spaces are generated by solutions to the good-cut
equation, with each solution determined by the choice of an arbitrary
complex analytic world line in complex Minkowski space or $\mathcal{H}$-space. The basic
governing variables are the complex GCF, $u_{\mathrm{B}} = G(\tau
,\zeta, \tilde{\zeta})$, and the stereographic angle field on
$\mathfrak{I}_{\mathbb{C}}^{+}$, $L(u_{\mathrm{B}},\zeta,
\tilde{\zeta})$, restricted to real $\mathfrak{I}^{+}$. In every
sense, the flat-space case can be considered as a special case of the
asymptotically-flat case.

In Sections~\ref{applications} and~\ref{results}, we will show that in every asymptotically flat
spacetime a special complex-world line (along with its associated NGC and GCF) can be singled out using physical considerations.  This special GCF
is referred to as the (gravitational) UCF, and is denoted by
\begin{equation}
u_{\mathrm{B}}=X(\tau ,\zeta, \tilde{\zeta}).
\label{UCF}
\end{equation}


\subsection{Real cuts from the complex good cuts, II}

The construction of real structures from the complex structures, i.e.,
finding the complex values of $\tau$ that yield real values of
$u_{\mathrm{B}}$ and the associated real cuts, is virtually identical
to the flat-space construction of Section~\ref{shear-free-NGC}. The real structure associated
with the complex Minkowski space complex world lines is easily extended to
the $\mathcal{H}$-space world lines associated with asymptotically flat
spacetimes. The only difference is that we start with the GCF 
\begin{equation}
u_{\mathrm{B}}=G(\tau, \zeta, \overline{\zeta})=\xi^{a}(\tau)\hat{l}_{a}(\zeta, \bar{\zeta})+G_{l\geq 2}(\tau, \zeta, \overline{\zeta})
\label{G}
\end{equation}
rather than the flat-space 
\begin{equation*}
u_{\mathrm{B}}=G(\tau, \zeta, \overline{\zeta})=\xi^{a}(\tau)\hat{l}_{a}(\zeta, \bar{\zeta}).
\end{equation*}

Again assuming that the Bondi shear is sufficiently small and the $\mathcal{H}$-space complex world line is not too far from the
``real'', the solution to the good-cut
equation~(\ref{GC}), i.e., Eq.~(\ref{G}), with $\tau =s+i\lambda$, is
decomposed into real and imaginary parts,
%
\begin{equation}
G(\tau,\zeta,\overline{\zeta})=\frac{1}{2}\left(G(s+i\lambda ,\zeta,\overline{\zeta})+\overline{G}(s-i\lambda ,\zeta,\overline{\zeta})\right)+\frac{1}{2}\left(G(s+i\lambda ,\zeta,\overline{\zeta})-\overline{G}(s-i\lambda ,\zeta,\overline{\zeta})\right).
\label{decomposition}
\end{equation}
Setting the imaginary part to zero and solving for $\lambda$ we obtain an
expression of the form,
\begin{equation*}
\lambda =\Lambda (s,\zeta,\overline{\zeta}).
\end{equation*}
As in the flat case, for fixed $s=s_{0}$, $\Lambda$ has values on a line
segment bounded between some $\lambda_{\min}$ and $\lambda_{\max}$ .
The allowed values of $\tau$ are again on a ribbon in the $\tau $-plane (i.e., region which is topologically $\mathbb{R}\times I$ for an interval $I$);
all values of $s$ and allowed values on the $\lambda $-line segments.

Each level curve of the function $\lambda =\Lambda (s_{0},\zeta,\overline{\zeta})=$ constant on the $(\zeta,\overline{\zeta})$-sphere (closed curves or isolated points) determines a specific subset of the null
directions and associated null geodesics on the light-cone of the complex
point $\xi^{a}(s_{0}+i\Lambda (s_{0},\zeta,\overline{\zeta}))$ that
intersect the real $\mathfrak{I}^{+}$. These geodesics will be referred
to as \textit{`real' geodesics}. As $\lambda$ moves over all allowed values of its segment, we obtain the set of $\mathcal{H}$-space points, $\xi^{a}(s_{0}+i\Lambda (s_{0},\zeta,\overline{\zeta}))$ and their collection of
`real' geodesics. From Eq.~(\ref{decomposition}), these
`real' geodesics intersect $\mathfrak{I}^{+}$ on the real cut
\begin{equation}
u^{(R)}_{\mathrm{B}}=G_{R}(s_{0}+i\Lambda (s_{0},\zeta,\overline{\zeta}),\zeta,\overline{\zeta}).
\label{real slicing}
\end{equation}

As $s$ varies we obtain a one-parameter family of cuts. If these cuts do
not intersect with each other we say that the complex world line $\xi^{a}(\tau)$ is by definition `timelike.' This occurs when the time component of
the real part of the complex velocity vector, $v^{a}(\tau)=\mathrm{d}\xi^{a}(\tau)/\mathrm{d}\tau $, is sufficiently large.

\subsection{Summary of Real Structures}

To put the ideas of this section into perspective we collect the claims.

\begin{itemize}
\item In Minkowski space, the future directed light-cones emanating from a
real timelike world line, $x^{a}=\xi^{a}(s)$, intersect future null
infinity, $\mathfrak{I}^{+}$, on a one-parameter family of spherical
non-intersecting cuts.

\item The complex light-cones emanating from a timelike complex analytic
curve in complex Minkowski space, $z^{a}=\xi^{a}(\tau)$ parametrized by
the complex parameter $\tau =s+i\lambda $, has for each fixed value of $s$ and $\lambda$ a limited set of null geodesics that reach real $\mathfrak{I}^{+}$. However, for a ribbon in the complex $\tau $-plane
(i.e., a region topologically $\mathbb{R}\times I$, with $s\in \mathbb{R}$
and $\lambda \in I=[\lambda_{\min},\lambda_{\max}]$), there will be many
null geodesics intersecting $\mathfrak{I}^{+}$. Such null geodesics were
referred to as \textit{`real' geodesics}
. More specifically, for a fixed $s$, there is a specific range of $\lambda\ values$ such that all the real null geodesics intersect $\mathfrak{I}^{+}$ in a full cut, leading to a one-parameter family of real
(distorted sphere) slicings of $\mathfrak{I}^{+}$. The ribbon is the
generalization of the real world line and the slicings are the analogues of
the spherical slicings. When the ribbon shrinks to a line it degenerates to
the real case. We can consider the ribbon as a generalized world line and
the `real' null geodesics from
a constant $s$ portion of the ribbon as a generalized light-cone.

\item For the case of asymptotically flat spacetimes, the real light-cones
from interior points are replaced by the virtual light-cones generated by
the asymptotically shear-free NGCs. These cones emanate from a complex
virtual world line $z^{a}=\xi^{a}(\tau)$ in the associated $\mathcal{H}$-space. As in the case of complex Minkowski space, there is a ribbon in
the $\tau $-plane where the `real' null
geodesics originate from. The `real' null geodesics coming from a cross-section
of the strip at fixed $s$ (as in the complex Minkowski case), intersect $\mathfrak{I}^{+}$ in a cut; the collection of cuts yielding a one-parameter
family. The situation is exactly the same as in the complex
Minkowski space case except that the spherical harmonic decomposition of
these cuts is in general more complicated.
\end{itemize}

\subsubsection*{\textit{Example: the (charged) Kerr metric}}
Considering the Kerr or the charged Kerr metrics (or even more generally any
asymptotically flat stationary metric), we have immediately that the Bondi
shear $\sigma^{0}$ vanishes and hence the associated $\mathcal{H}$-space
is complex Minkowski space (cf.~\cite{AdamoNewman3, Adamo:2010ey} and Appendix~\ref{appendixF}). From the stationarity and a real origin shift
and rotation, the complex world line can be put into the form
\begin{equation}
\xi^{a}(\tau)=(\tau,0,0,ia),
\label{Kerr world-line}
\end{equation}
with $a$ being the Kerr parameter. The complex cut function is then
\begin{eqnarray}
u_{\mathrm{B}} &=&\xi^{a}(\tau)\hat{l}_{a}(\zeta,\bar{\zeta})  \label{Kerr cut} \\
&=&\frac{\tau}{\sqrt{2}}-\frac{i}{2}aY_{1,3}^{0}(\zeta,\overline{\zeta}),
\notag \\
Y_{1,3}^{0}(\zeta,\overline{\zeta}) &=&-\sqrt{2}\frac{1-\zeta \overline{\zeta}}{1+\zeta \overline{\zeta}},  \notag
\end{eqnarray}
so that the angle fields of \ref{Lbar}-\ref{Ltwiddle} are 
\begin{eqnarray*}
L &=&\sqrt{2}ia\frac{\overline{\zeta}}{1+\zeta \overline{\zeta}}, \\
\overline{L} &=&-\sqrt{2}ia\frac{\zeta}{1+\zeta \overline{\zeta}}, \\
\widetilde{L} &=&\sqrt{2}ia\frac{\zeta}{1+\zeta \overline{\zeta}}.
\end{eqnarray*}

Using $\tau =s+i\lambda$ in Eq.~(\ref{Kerr cut}), the reality
condition $u_{\mathrm{B}}=u^{(R)}_{\mathrm{B}}$ on the cut function is
that
\begin{equation*}
\lambda =\Lambda (s,\zeta,\bar{\zeta})=\frac{\sqrt{2}}{2}aY_{1,3}^{0}(\zeta
,\overline{\zeta}),
\end{equation*}
so that on the $\tau $-ribbon, $\lambda$ ranges between $\pm$ $\sqrt{2}$ and the real slices from the ribbon becomes simply $u_{\mathrm{B}}=s/\sqrt{2}$.

Though we are certainly not making the claim that one can in reality `observe' these
complex world lines that arise from (asymptotically) shear-free congruences, we nevertheless claim that they can be observed in a different sense.  In the next two sections our goal will be to show that, just as a complex center of charge world line in $\mathbb{M}_{\mathbb{C}}$ can be selected, so too can a complex center of mass world line be singled out in $\mathcal{H}$-space.  As we will see, some surprising physical identifications arise from this program, and it is in this sense which the footprints of these complex world lines can be observed.


\newpage

\section{Simple Applications}
\label{applications}

In this section we give four simple examples of the use of shear-free and
asymptotically shear-free NGCs in GR. The first is for asymptotically-linearized perturbations off the Schwarzschild metric, while the next two are
from the class of algebraically-special metrics, namely the
Robinson--Trautman metric and the type~II twisting metrics; the fourth is for
asymptotically static/stationary metrics.


\subsection{Linearized off Schwarzschild}

As a first example, we describe how the shear-free NGCs are applied in
linear perturbations off the Schwarzschild metric. The ideas used here are
intended to clarify the more complicated issues in the full nonlinear
asymptotic theory. We will see that these linear perturbations greatly
resemble our results from Section~\ref{AVMaxwell} on the determination of the
intrinsic center of charge in Maxwell theory, when there were small
deviations from the Coulomb field.

We begin with the Schwarzschild spacetime, treating the Schwarzschild mass, 
$M_{\mathrm{Sch}}\equiv M_{\mathrm{B}}$, as a zeroth-order quantity, and integrate the
linearized Bianchi identities for the linear Weyl tensor corrections. Though
we could go on and find the linearized connection and metric, we stop just
with the Weyl tensor. The radial behavior is given by the peeling theorem,
so that we can start with the linearized asymptotic Bianchi
identities, Eqs.~(\ref{AsyBI1})\,--\,(\ref{AsyBI3}).

Our main variables for the investigation are the asymptotic Weyl tensor
components and the Bondi shear, $\sigma^{0}$, with their related
differential equations, i.e., the asymptotic Bianchi identities, Eq.~(\ref{AsyBI1}), (\ref{AsyBI2}) and (\ref{reality}). Assuming the gravitational
radiation is weak, we treat $\sigma^{0}$ and $\dot{\sigma}^{0}$ as small.
Keeping only linear terms in the Bianchi identities, the equations for $\psi_{1}^{0}$ and $\Psi$ (the mass aspect) become
\begin{eqnarray}
\dot{\psi}_{1}^{0}+ \eth \Psi &=& \eth^{3}\overline{\sigma}^{0},
\label{2*} \\
\dot{\Psi}&=& 0,
\label{2**} \\
\Psi &=& \overline{\Psi},
\label{2***} \\
\Psi & \equiv & \psi_{2}^{0}+\eth^{2}\overline{\sigma}^{0}.
\label{2****}
\end{eqnarray}
The $\psi_{1}^{0}$ is small (first order), while the
\begin{equation}
\Psi =\Psi^{0}+\Psi^{i}Y_{1i}^{0}+\Psi^{ij}Y_{2ij}^{0}+ \ldots
\label{MassAspect***}
\end{equation}
has the zeroth-order Schwarzschild mass plus first-order terms
\begin{eqnarray}
\Psi^{0} &=&-\frac{2\sqrt{2}G}{c^{2}}M_{\mathrm{Sch}}+\delta \Psi^{0},
\label{PSI0}
\\
\Psi^{i} &=&-\frac{6G}{c^{3}}P^{i}.
\label{PSI1}
\end{eqnarray}

In \textit{linear theory}, the complex (mass) dipole moment, 
\begin{equation}
D_{(\mathrm{grav})}^{i}=D_{(\mathrm{mass})}^{i}+ic^{-1}J^{i}
\label{D_c}
\end{equation}
is given~\cite{Szabados}, on a particular Bondi cut with a Bondi tetrad (up
to dimensional constants), by the $l=1$ harmonic components of
$\psi_{1}^{0}$, i.e., from the $\psi_{1}^{0\,i}$ in the expansion 
\begin{equation}
\psi_{1}^{0}=\psi_{1}^{0i}Y_{1i}^{1}+\psi_{1}^{0ij}Y_{2ij}^{1}+ \ldots
\label{expansion1}
\end{equation}
For a different cut and different tetrad, one needs the transformation law
to the new $\psi_{1}^{\ast 0}$ and new $\psi_{1}^{\ast 0\,i}$. Under
the tetrad transformation (a null rotation around $n^{a}$) to the
asymptotically shear-free vector field, $l^{\ast a}$, Eq.~(\ref{nullROT2}),
\begin{equation*}
l^{a}\rightarrow l^{\ast a}=l^{a}-\frac{\bar{L}}{r}m^{a}-\frac{L}{r}\bar{m}^{a}+O(r^{-2}),
\end{equation*}
with, from Eqs.~(\ref{u_c}) and~(\ref{L**}),
\begin{eqnarray}
u_{\mathrm{B}} &=& \xi^{a}(\tau)\hat{l}_{a}(\zeta, \overline{\zeta})+\xi^{ij}(\tau)Y_{2ij}^{0}(\zeta, \overline{\zeta})+ \ldots \\
&=& \frac{1}{\sqrt{2}}\xi^{0}(\tau)-\frac{1}{2}\xi^{i}(\tau)Y_{1i}^{0}(\zeta,
\overline{\zeta})+\xi^{ij}(\tau)Y_{2ij}^{0}(\zeta, \overline{\zeta})+ \ldots \\
L(u_{\mathrm{B}},\zeta, \overline{\zeta}) &=& \xi^{i}(\tau)Y_{1i}^{1}(\zeta, \overline{\zeta})-6\xi^{ij}(\tau)Y_{2ij}^{1}(\zeta, \overline{\zeta})+ \ldots 
\end{eqnarray}
the linearized transformation is given by~\cite{Aronson} 
\begin{equation}
\psi_{1}^{0\ast}=\psi_{1}^{0}-3L\Psi.
\label{psi trans 2}
\end{equation}
The extraction of the $l=1$ part of $\psi_{1}^{0\ast}$ should, in
principle, be taken on the new cut given by $u_{\mathrm{B}}=\xi^{a}(\tau)\hat{l}_{a}(\zeta, \overline{\zeta})+\xi^{ij}(\tau)Y_{2ij}^{0}(\zeta, \overline{\zeta})+\dots$ with constant $\tau$. However, because of 
the linearization, the extraction can be taken on the $u_{\mathrm{B}}$ constant
cuts. Following the same line of reasoning that led to the definition
of center of charge, we demand the vanishing of the $l=1$ part of
$\psi_{1}^{0\ast}$.

This leads immediately to 
\begin{equation}
\psi_{1}^{0}|_{l=1}=3L\Psi |_{l=1},
\label{2**}
\end{equation}
or, using the decomposition into real and imaginary parts,
$\psi_{1}^{0i}=\psi_{1R}^{0i}+i\psi_{1I}^{0i}$ and
$\xi^{i}(u_{\mathrm{ret}})=\xi_{R}^{i}(u_{\mathrm{ret}})+i\xi_{I}^{i}(u_{\mathrm{ret}})$,

\begin{eqnarray}
\psi_{1R}^{0i} &=& -\frac{6\sqrt{2}G}{c^{2}}M_{\mathrm{Sch}}\xi_{R}^{i}(u_{\mathrm{ret}}),
\\
\psi_{1I}^{0i} &=& -\frac{6\sqrt{2}G}{c^{2}}M_{\mathrm{Sch}}\xi_{I}^{i}(u_{\mathrm{ret}}).
\end{eqnarray}

Identifying~\cite{Szabados, KerrNewman} the (intrinsic) \textit{angular
momentum}, either from the conventional linear identification or from the
Kerr metric, as 
\begin{equation}
J^{i}=S^{i}=M_{\mathrm{Sch}}c\xi_{I}^{i}
\label{J^i}
\end{equation}
and the mass dipole as 
\begin{equation}
D_{(\mathrm{mass})}^{i}=M_{\mathrm{Sch}}\xi_{R}^{i},
\label{D_mass}
\end{equation}
we have
\begin{equation}
\psi_{1}^{0i}=-\frac{6\sqrt{2}G}{c^{2}}D_{(\mathrm{grav})}^{i}=-\frac{6\sqrt{2}G}{c^{2}}(D_{(\mathrm{mass})}^{i}+ic^{-1}J^{i}).
\label{D-c*}
\end{equation}
By inserting Eq.~(\ref{D-c*}) into Eq.~(\ref{2*}), taking, respectively, the
real and imaginary parts, using Eq.~(\ref{PSI1}) and the reality of
$\Psi$,
we find 
\begin{equation}
P^{i}=M_{\mathrm{Sch}}\xi_{R}^{i\,\prime}\equiv M_{\mathrm{Sch}}v_{R}^{i},
\label{P^i}
\end{equation}
the kinematic expression of linear momentum and 
\begin{equation}
J^{i\,\prime}=0,
\label{cons.A.M.}
\end{equation}
the conservation of angular momentum.

Finally, from the $l=(0,1)$ parts of Eq.~(\ref{2**}), we have, at this
approximation, that the mass and linear momentum remain constant,
i.e., $M=M_{\mathrm{Sch}}=M_{\mathrm{B}}$ and $\delta \Psi^{0}=0$. Thus, we obtain the
trivial equations of motion for the center of mass, 
\begin{equation}
M_{\mathrm{Sch}}\xi_{R}^{i\, \prime\prime}=0.
\label{EqsOf Motion}
\end{equation}

The linearization off Schwarzschild, with our identifications, lead to a
stationary spinning spacetime object with the standard classical mechanics
kinematic and dynamic description. It was the linearization that let to such simplifications, and in Section~\ref{results}, when nonlinear terms are included (in similar calculations), much
more interesting and surprising physical results are found.


\subsection{The Robinson--Trautman metrics}

The algebraically-special type~II Robinson--Trautman (RT) metrics are
expressed in conventional RT coordinates, ($\tau, r, \zeta,
\overline{\zeta}$), $\tau$ now real, by~\cite{RobinsonTrautman}
\begin{equation}
ds^{2}=2 \left( K-\frac{V^{\prime}}{V}r+\frac{\psi_{2}^{0}}{r} \right) d\tau^{2}+2d\tau
dr-r^{2}\frac{2d\zeta d\overline{\zeta}}{V^{2}P_{0}^{2}},
\label{RTMmetric}
\end{equation}
%
with
\begin{eqnarray}
K &=& 2V^{2}P_{0}^{2}\partial_{\overline{\zeta}}\partial_{\zeta}\log VP_{0},
\label{K} \\
P_{0} &=& 1+\zeta \overline{\zeta},
\label{P} \\
\psi_{2}^{0} &=& \psi_{2}^{0}(\tau).
\label{psi2}
\end{eqnarray}
The unknowns are the Weyl component $\psi_{2}^{0}$ (closely related to the
Bondi mass), which is a function only of (real) $\tau$ and the variable, $V(\tau, \zeta, \overline{\zeta})$, both of which are variables in the RT equation
(see below). There remains the freedom 
\begin{equation}
\tau \rightarrow \tau^{\ast}=g(\tau),
\end{equation}
which often is chosen so that $\psi_{2}^{0}(\tau)$~=~constant. However, we
make a different choice. In the spherical harmonic expansion of $V$, 
\begin{equation}
V=v^{a}\hat{l}_{a}(\zeta, \overline{\zeta})+v^{ij}Y_{2ij}^{0}+ \ldots,
\label{R-T expansion}
\end{equation}
the $\tau$ is chosen by normalizing the four-vector, $v^{a}$, to one,
i.e., $v^{a}v_{a}=1$. The final field equation, the RT equation, is 
\begin{equation}
\psi_{2}^{0\,\prime}-3\psi_{2}^{0}\frac{V^{\prime}}{V^{3}}-V^{3}\left(\eth_{(\tau)}^{2}\overline{\eth}_{(\tau)}^{2}V-V^{-1}\overline{\eth}_{(\tau)}^{2}V\cdot \eth_{(\tau)}^{2}V\right) =0.
\label{R-T eq}
\end{equation}
These spacetimes, via the Goldberg--Sachs theorem, possess a degenerate
shear-free PND field, $l^{a}$, that is surface-forming,
(i.e., twist free). Using the tetrad constructed from $l^{a}$ we have that
the Weyl components are of the form 
\begin{eqnarray*}
\psi_{0} &=&\psi_{1}=0, \\
\psi_{2} &\neq &0.
\end{eqnarray*}
Furthermore, the metric contains a `real timelike world line,
$x^{a}=\xi^{a}(\tau)$,' with normalized velocity vector
$v^{a}=\xi^{a\, \prime}$. All of these properties allow us to identify
the RT metrics as being analogous to the real Li\'{e}nard--Wiechert
solutions of the Maxwell equations. 

Assuming for the moment that we have integrated the RT equation and
know $V=V(\tau ,\zeta, \overline{\zeta})$, then, by the integral 
\begin{equation}
u_{\mathrm{B}}=\int V(\tau,\zeta, \overline{\zeta})d\tau \equiv X_{\text{RT}}(\tau,
\zeta, \bar{\zeta}),
\label{UCF.R-T}
\end{equation}
the UCF for the RT metrics has been found. The freedom of adding $\alpha
(\zeta, \overline{\zeta})$ to the integral is just the supertranslation
freedom in the choice of a Bondi coordinate system. From
$X_{\mathrm{RT}}(\tau, \zeta, \bar{\zeta})$ a variety of information can be obtained: the
Bondi shear, $\sigma^{0}$, is given parametrically by 
\begin{eqnarray}
\sigma^{0}(u_{\mathrm{B}},\zeta, \bar{\zeta})
&=&\eth_{(\tau)}^{2}X_{\mathrm{RT}}(\tau ,\zeta, \bar{\zeta}),
\label{shearR-T} \\
u_{\mathrm{B}} &=&X_{\mathrm{RT}}(\tau ,\zeta, \bar{\zeta}),  \notag
\end{eqnarray}
as well as the angle field $L$ by
\begin{eqnarray}
L(u_{\mathrm{B}},\zeta, \bar{\zeta}) &=&\eth_{(\tau)}X_{\mathrm{RT}}(\tau, \zeta, \bar{\zeta}),
\label{L.R-T} \\
u_{\mathrm{B}} &=&X_{\mathrm{RT}}(\tau, \zeta, \bar{\zeta}).  \notag
\end{eqnarray}
In turn, from this information the RT metric (in the neighborhood of $\mathfrak{I}^{+}$) can, in principle, be re-expressed in terms of the Bondi
coordinate system, though in practice one must revert to approximations. These
approximate calculations lead, via the Bondi mass aspect evolution equation,
to both Bondi mass loss and to equations of motion for the world line, $x^{a}=\xi^{a}(\tau)$. An alternate approximation for the mass loss and
equations of motion is to insert the spherical harmonic expansion of $V$
into the RT equation and look at the lowest harmonic terms. We omit further
details aside from mentioning that we come back to these calculations
in a more general context in Section~\ref{results}.


\subsection{Type II twisting metrics}

It was pointed out in the previous section that the RT metrics are the
general relativistic analogues of the (real) Li\'{e}nard--Wiechert Maxwell
fields. The type~II algebraically-special twisting metrics are the
gravitational analogues of the \textit{complex} Li\'{e}nard--Wiechert Maxwell
fields described earlier. Unfortunately they are far more complicated than
the RT metrics. In spite of the large literature and much effort there
are very few known solutions and much still to be
learned~\cite{RTmetrics, RTmetrics2, Lind}.  We give a very brief
description of them, emphasizing only the items of relevance to us.

A null tetrad system (and null geodesic coordinates) can be adopted for the
type~II metrics so that the Weyl tetrad components are 
\begin{eqnarray*}
\psi_{0} &=&\psi_{1}=0, \\
\psi_{2} &\neq &0.
\end{eqnarray*}
It follows from the Goldberg--Sachs theorem that the degenerate principal
null congruence is geodesic and shear-free. Thus, from the earlier
discussions it follows that there is a \textit{unique} angle field, $L(u_{\mathrm{B}},\zeta, \overline{\zeta})$. As with the complex Li\'{e}nard--Wiechert Maxwell fields, the type~II metrics and Weyl tensors are
given in terms of the angle field, $L(u_{\mathrm{B}},\zeta, \overline{\zeta}) $. In fact, the entire metric and the field equations (the asymptotic
Bianchi identities) can be written in terms of $L$ and a Weyl tensor
component (essentially the Bondi mass). Since $L(u_{\mathrm{B}},\zeta, 
\overline{\zeta})$ describes a unique shear-free NGC, it can be written
parametrically in terms of a \textit{unique} GCF, namely the UCF $X_{(\text{type~II})}(\tau,\zeta, \overline{\zeta})$. So, we have that 
\begin{eqnarray*}
L(u_{\mathrm{B}},\zeta, \overline{\zeta}) &=&\eth_{(\tau)}X_{(\text{type~II})}, \\
u_{\mathrm{B}} &=&X_{(\text{type~II})}(\tau ,\zeta, \overline{\zeta}).
\end{eqnarray*}
Since $X_{(\text{type~II})}(\tau ,\zeta, \overline{\zeta})$ can be expanded in
spherical harmonics, the $l=(0,1)$ harmonics can be identified with a
(unique) complex world line in $\mathcal{H}$-space. The asymptotic
Bianchi identities then yield both kinematic equations (for angular momentum
and the Bondi linear momentum) and equations of motion for the world line,
analogous to those obtained for the Schwarzschild perturbation and the RT
metrics. As a kinematic example, the imaginary part of the world line is
identified as the intrinsic spin, the same identification as in the Kerr
metric,
\begin{equation}
S^{i}=M_{\mathrm{B}}c\xi_{I}^{i}.
\label{Identify.S}
\end{equation}
In Section~\ref{results}, a version of these results will
be derived in a far more general context.

Recently, the type II Einstein-Maxwell equations were studied using a slow-motion perturbation expansion around the Reissner-N\"{o}rdstrom metric, keeping spherical harmonic contributions up to $l=2$.  It was found that the above-mentioned world-line coincides in this case with that given by the Abraham-Lorentz-Dirac equation, prompting us to consider such spacetimes as `type II particles' in the same way that one can refer to Reissner-N\"{o}rdstrom-Schwarzschild or Kerr-Newman `particles'~\cite{Newman:2011im}.


\subsection{Asymptotically static and stationary spacetimes}


By defining \textit{asymptotically static or stationary} spacetimes as
those asymptotically-flat spacetimes where the asymptotic variables are
`time' independent, i.e., $u_{\mathrm{B}}$ independent, we can look at our procedure
for transforming to the complex center of mass (or complex center of
charge). This example, though very special, has the huge advantage in that
it can be done exactly, without the use of perturbations~\cite{AdamoNewman3}.

Imposing time independence on the asymptotic Bianchi identities,
Eqs.~(\ref{AsyBI1})\,--\,(\ref{AsyBI3}), 
\begin{eqnarray*}
\dot{\psi}_{2}^{0} &=&-\eth \psi_{3}^{0}+\sigma^{0}\psi_{4}^{0}, \\
\dot{\psi}_{1}^{0} &=&-\eth \psi_{2}^{0}+2\sigma^{0}\psi_{3}^{0}, \\
\dot{\psi}_{0}^{0} &=&-\eth \psi_{1}^{0}+3\sigma^{0}\psi_{2}^{0},
\end{eqnarray*}
and reality condition
\begin{equation*}
\Psi \equiv \psi_{2}^{0}+\eth^{2}\overline{\sigma}+\sigma\dot{\overline{\sigma}}=\overline{\Psi},
\end{equation*}
we have, using Eqs.~(\ref{i}) and~(\ref{j}) with $\dot{\sigma}^{0}=0$, that
\begin{eqnarray}
\psi_{3}^{0} &=&\psi_{4}^{0}=0,
\label{psi0.1} \\
\eth \psi_{2}^{0} &=&0,
\label{ethpsi2} \\
\eth \psi_{1}^{0} &=&3\sigma^{0}\psi_{2}^{0},
\label{ethpsi1} \\
\Psi &\equiv &\psi_{2}^{0}+\eth^{2}\overline{\sigma}=\overline{\psi}_{2}^{0}+\overline{\eth}^{2}\sigma =\overline{\Psi}.
\label{reality*}
\end{eqnarray}

From Eq.~(\ref{reality*}), we find (after a simple calculation) that the
imaginary part of $\psi_{2}^{0}$ is determined by the `magnetic'~\cite{NewmanTod} part of the Bondi shear (spin-weight $s=2$) and thus must contain
harmonics only of $l\geq 2$. But from Eq.~(\ref{ethpsi2}), we find that $\psi_{2}^{0}$ contains only the $l=0$ harmonic. From this it follows that the
`magnetic' part of the shear must vanish. The remaining part of the shear,
i.e., the `electric' part, which by assumption is time independent, can be
made to vanish by a supertranslation, via the Sachs theorem:
\begin{eqnarray}
\widehat{u}_{B} &=& u_{\mathrm{B}}+\alpha (\zeta, \overline{\zeta}),
\label{sachs} \\
\widehat{\sigma}(\zeta, \overline{\zeta}) &=& \sigma (\zeta, \overline{\zeta})+\eth^{2}\alpha (\zeta, \overline{\zeta}).  \notag
\end{eqnarray}
In this Bondi frame, (i.e., frame with a vanishing shear),
Eq.~(\ref{ethpsi1}), implies that 
\begin{eqnarray}
\psi_{1}^{0} &=& \psi_{1}^{0i}Y_{1i}^{1},
\label{identifications} \\
\psi_{1}^{0i} &=& -\frac{6\sqrt{2}G}{c^{2}}D_{(\mathrm{grav})}^{i}=-\frac{6\sqrt{2}G}{c^{2}}(D_{(\mathrm{mass})}^{i}+ic^{-1}J^{i}),
\label{ident*}
\end{eqnarray}
using the conventionally accepted physical identification of the complex
gravitational dipole. (Since the shear vanishes, this agrees with probably
all the various attempted identifications.)

From the mass identification, $\psi_{2}^{0}$ becomes 
\begin{equation}
\psi_{2}^{0}=-\frac{2\sqrt{2}G}{c^{2}}M_{\mathrm{B}}.
\label{Msch}
\end{equation}
Since the Bondi shear is zero, the asymptotically shear-free congruences are
determined by the same GCFs as in flat spaces, i.e., we have
\begin{eqnarray}
L(u_{\mathrm{B}},\zeta, \bar{\zeta}) &=&\eth_{(\tau)}G(\tau,\zeta, \bar{\zeta})=\xi^{a}(\tau)\hat{m}_{a}(\zeta, \bar{\zeta}),
\label{angle-field*} \\
u_{\mathrm{B}} &=&\xi^{a}(\tau)\hat{l}_{a}(\zeta, \bar{\zeta}).
\end{eqnarray}

Our procedure for the identification of the complex center of mass, namely
setting $\psi_{1}^{\ast 0}=0$ in the transformation, Eq.~(\ref{Weyl-rot2}), 
\begin{equation*}
\psi_{1}^{\ast 0}=\psi_{1}^{0}-3L\psi_{2}^{0}+3L^{2}\psi_{3}^{0}-L^{3}\psi_{4}^{0}
\end{equation*}
leads, after using Eqs.~(\ref{identifications}),~(\ref{psi0.1}) and~(\ref{angle-field*}), to 
\begin{eqnarray}
\psi_{1}^{0} &=& 3L\psi_{2}^{0},
\label{D_cgrav} \\
\psi_{1}^{0i} &=& -\frac{6\sqrt{2}G}{c^{2}}D_{(\mathrm{grav})}^{i},  \notag \\
D_{(\mathrm{grav})}^{i} &=& M_{\mathrm{B}}\xi^{i}.  \notag
\end{eqnarray}

From the time independence, $\xi^{i}$, the spatial part of the world line
is a constant vector. By a (real) spatial Poincar\'{e} transformation (from
the BMS group), the real part of $\xi^{i}$ can be made to vanish, while by
ordinary rotation the imaginary part of $\xi^{i}$ can be made to point in
the three-direction.  Using the the gauge freedom in the choice of $\tau$ we set $\xi^{0}(\tau)=\tau$.
Then pulling all these items together, we have for
the complex world line, the UCF, $L(u_{\mathrm{B}},\zeta, \bar{\zeta})$ and the
angular momentum, $J^{i}$:
\begin{eqnarray}
\xi^{a}(\tau) &=&(\tau ,0,0,i\xi^{3})
\label{Identifications*} \\
u_{\mathrm{B}} &=& X(\tau,\zeta, \bar{\zeta})=\xi^{a}(\tau)\hat{l}_{a}(\zeta,\bar{\zeta})\equiv\frac{\tau}{\sqrt{2}}-\frac{i}{2}\xi
^{3}Y_{1,3}^{0},  \notag \\
L(u_{\mathrm{B}},\zeta, \bar{\zeta}) &=& i\xi_{I}^{3}Y_{1,3}^{1},  \notag \\
J^{i} &=& S^{i}=M_{\mathrm{B}}c\xi^{3}\delta_{3}^{i}=M_{\mathrm{B}}c(0,0,\xi^{3})=M_{\mathrm{B}}c\xi_{I}^{i}.
\notag
\end{eqnarray}
Thus, we have the complex center of mass on the complex world line, $z^{a}=\xi^{a}(\tau)$.

These results for the lower multipole moments, i.e., $l=0,1$, are \textit{identical to those of the Kerr metric} presented earlier! The higher moments are still present
(appearing in higher $r^{-1}$ terms in the Weyl tensor) and are not affected
by these results.


\newpage

\section{Main Results}
\label{results}

We saw in Sections~\ref{shear-free-NGC} and \ref{good-cut-eq} how
shear-free and asymptotically shear-free NGCs determine arbitrary
complex analytic world lines in the auxiliary complex $\mathcal{H}$-space
(or complex Minkowski space). In the examples from Sections~\ref{shear-free-NGC}
and~\ref{applications}, we saw how, in each of the cases, one could
pick out a special GCF, referred to as the UCF, and the associated complex
world line by a transformation to the complex center of mass or charge
by requiring that the complex dipoles vanish. In the present section
we consider the same problem, but now perturbatively for the general
situation of asymptotically-flat spacetimes satisfying either the
vacuum Einstein or the Einstein--Maxwell equations in the neighborhood
of future null infinity. Since the calculations are relatively
long and complicated, we give the basic ideas in outline form
and then present the final results for Einstein--Maxwell spacetimes without detailed steps.

We begin with the Reissner--Nordstr\"{o}m metric, considering both the
mass and the charge as zeroth-order quantities, and perturb from it. The
perturbation data is considered to be first order and the perturbations
themselves are general in the class of analytic asymptotically-flat
spacetimes. Though our considerations are for arbitrary mass and charge
distributions in the interior, we look at the fields in the
neighborhood of $\mathfrak{I}^{+}$. The calculations are carried to
second order in the perturbation data. Throughout we use expansions in
spherical harmonics and their tensor harmonic versions, but terminate
the expansions after $l=2$. Clebsch--Gordon expansions are frequently
used; see Appendix~\ref{appendixC}.


\subsection{A brief summary -- Before continuing}

Very briefly, for the purpose of organizing the many strands so far
developed, we summarize our procedure for finding the complex center of
mass. We begin with the gravitational radiation data, the Bondi shear,
$\sigma^{0}(u_{\mathrm{B}}, \zeta, \bar{\zeta})$ and solve the good-cut equation, 
\begin{equation*}
\eth^{2}Z=\sigma^{0}(Z,\zeta, \bar{\zeta}),
\end{equation*}
with solution $u_{\mathrm{B}}=Z(z^{a},\zeta, \bar{\zeta})$ and the four complex
parameters $z^{a}$ defining the solution space. Next we consider an
arbitrary complex world line in the solution space,
$z^{a}=\xi^{a}(\tau)=(\xi^{0}(\tau),\xi^{i}(\tau))$, so that
$u_{\mathrm{B}}=Z(\xi^{a}(\tau),\zeta, \bar{\zeta})=G(\tau ,\zeta,
\bar{\zeta})$, a GCF, which can be expanded in spherical harmonics as
\begin{eqnarray}
u_{\mathrm{B}} &=&G(\tau, \zeta, \bar{\zeta})=\xi^{a}(\tau)\hat{l}_{a}(\zeta, \bar{\zeta})+\xi^{ij}(\tau)Y_{2ij}^{0}+\ldots
\label{AsyGCF} \\
&=&\frac{\xi^{0}(\tau)}{\sqrt{2}}-\frac{1}{2}\xi^{i}(\tau)Y_{1i}^{0}+\xi^{ij}(\tau)Y_{2ij}^{0}+\ldots  \notag
\end{eqnarray}
Assuming slow motion and the gauge condition $\xi^{0}(\tau)=\tau$
(see Section~\ref{good-cut-eq}), we have
\begin{equation}
u_{\mathrm{B}}=\frac{\tau}{\sqrt{2}}-\frac{1}{2}\xi^{i}(\tau)Y_{1i}^{0}+\xi^{ij}(\tau)Y_{2ij}^{0}+ \ldots
\label{G2}
\end{equation}
(Though the world line is arbitrary, the quadrupole term, $\xi^{ij}(\tau)$,
and higher harmonics, are determined by both the Bondi shear and the world line.)

The inverse function, 
\begin{eqnarray}
\tau &=& T(u_{\mathrm{ret}},\zeta, \bar{\zeta}),  \label{u_ret} \\
u_{\mathrm{ret}} &=&\sqrt{2}u_{\mathrm{B}},  \notag
\end{eqnarray}
can be found by the following iteration process~\cite{UCF}: writing
Eq.~(\ref{G2}) as 
\begin{equation}
\tau =u_{\mathrm{ret}}+F(\tau ,\zeta, \bar{\zeta}),
\label{iteration}
\end{equation}
with 
\begin{equation}
F(\tau ,\zeta,
\bar{\zeta})=\frac{\sqrt{2}}{2}\xi^{i}(\tau)Y_{1i}^{0}(\zeta,
\bar{\zeta})-\sqrt{2}\xi^{ij}(\tau)Y_{1ij}^{0}(\zeta, \bar{\zeta})+ \ldots,
\end{equation}
the iteration relationship, with the zeroth-order iterate,
$\tau_{0}=u_{\mathrm{ret}}$, is
\begin{equation}
\tau_{n}=u_{\mathrm{ret}}+F(\tau_{n-1},\zeta, \bar{\zeta}).
\label{iteration*}
\end{equation}
To second order, this is
\begin{equation*}
\tau =T(u_{\mathrm{ret}},\zeta,
\bar{\zeta})=u_{\mathrm{ret}}+F\left(u_{\mathrm{ret}}+F(u_{\mathrm{ret}},\zeta,
\overline{\zeta}),\zeta,\overline{\zeta}\right) \approx u_{\mathrm{ret}}+F+F\partial_{u_{\mathrm{ret}}}F,
\end{equation*}
%
but for most of our calculations, all
that is needed is the first iterate, given by 
\begin{equation}
\tau =T(u_{\mathrm{ret}},\zeta,\bar{\zeta})=u_{\mathrm{ret}}+\frac{\sqrt{2}}{2}\xi^{i}(u_{\mathrm{ret}})Y_{1i}^{0}(\zeta,\bar{\zeta})-\sqrt{2}\xi^{ij}(u_{\mathrm{ret}})Y_{1ij}^{0}(\zeta,\bar{\zeta}).
\label{first It}
\end{equation}
This relationship is, in principle, an important one.

We also have the linearized \textit{reality} relations -- easily found
earlier or from Eq.~(\ref{first It}):
\begin{eqnarray}
\tau &=&s+i\lambda ,
\label{1*} \\
\lambda &=&\Lambda (s,\zeta,\overline{\zeta})=\frac{\sqrt{2}}{2}\xi_{I}^{i}(s)Y_{1i}^{0}-\sqrt{2}\xi_{I}^{ij}(s)Y_{2ij}^{0},
\label{two*} \\
\tau &=&s+i\left(\frac{\sqrt{2}}{2}\xi_{I}^{i}(s)Y_{1i}^{0}-\sqrt{2}\xi_{I}^{ij}(s)Y_{2ij}^{0}\right),
\label{two**} \\
u_{\mathrm{ret}}^{(R)} &=&\sqrt{2}G_{R}(s,\zeta,\overline{\zeta})=\sqrt{2}u_{\mathrm{B}}^{(R)}=s-\frac{\sqrt{2}}{2}\xi_{R}^{i}(s)Y_{1i}^{0}+\sqrt{2}\xi_{R}^{ij}(s)Y_{2ij}^{0}.
\label{3*}
\end{eqnarray}

The associated angle field, $L$, and the Bondi shear, $\sigma^{0}$, are
given parametrically by

\begin{eqnarray}
L(u_{\mathrm{B}},\zeta, \bar{\zeta}) &=&\eth_{(\tau)}G(\tau,\zeta, \bar{\zeta})
\label{L***} \\
&=&\xi^{i}(\tau)Y_{1i}^{1}-6\xi^{ij}(\tau)Y_{2ij}^{1}+\ldots  \notag
\end{eqnarray}
and
\begin{eqnarray}
\sigma^{0}(u_{\mathrm{B}},\zeta, \bar{\zeta}) &=&\eth_{(\tau)}^{2}G(\tau, \zeta, \bar{\zeta}),
\label{sigma****} \\
&=&24\xi^{ij}(\tau)Y_{2ij}^{2}+\ldots ,  \notag
\end{eqnarray}
using the inverse to $u_{\mathrm{B}}=G(\tau,\zeta,\bar{\zeta})$, Eq.~(\ref{first It}). The asymptotically shear-free NGC is given by performing the null
rotation 
\begin{eqnarray}
l^{\ast a} &=&l^{a}+b\overline{m}^{a}+\overline{b}m^{a}+b\overline{b}n^{a},
\label{null-rot*} \\
m^{\ast a} &=&m^{a}+bn^{a},  \notag \\
n^{\ast a} &=&n^{a},  \notag \\
b &=&-L/r+O(r^{-2}).  \notag
\end{eqnarray}

As stated in (\ref{Weyl-rot})-(\ref{Weyl-rot5}), under (\ref{null-rot*}) the transformed asymptotic Weyl tensor
becomes
\begin{eqnarray}
\psi_{0}^{*0} &=&\psi_{0}^{0}-4L\psi_{1}^{0}+6L^{2}\psi_{2}^{0}-4L^{3}\psi_{3}^{0}+L^{4}\psi_{4}^{0},
\label{Weyl-rot*} \\
\psi_{1}^{*0} &=&\psi_{1}^{0}-3L\psi_{2}^{0}+3L^{2}\psi_{3}^{0}-L^{3}\psi_{4}^{0},
\label{Weyl-rot2*} \\
\psi_{2}^{*0} &=&\psi_{2}^{0}-2L\psi_{3}^{0}+L^{2}\psi_{4}^{0},
\label{Weyl-rot3*} \\
\psi_{3}^{*0} &=&\psi_{3}^{0}-L\psi_{4}^{0},
\label{Weyl-rot4*} \\
\psi_{4}^{*0} &=&\psi_{4}^{0}.
\label{Weyl-rot5*}
\end{eqnarray}

The procedure for finding the complex center of mass is centered on Eq.~(\ref{Weyl-rot2*}), where we search for and set to zero the $l=1$ harmonic in 
$\psi_{1}^{\ast 0}$ on a $\tau$~=~constant slice. This determines the
complex center-of-mass world line and singles out a particular GCF referred
to as the UCF,
\begin{equation}
u_{\mathrm{B}}=X(\tau,\zeta,\bar{\zeta})=G(\tau,\zeta,\bar{\zeta}),
\label{complexUCF}
\end{equation}
with the real version,
\begin{equation}
u_{\mathrm{ret}}^{(R)}=X_{R}(s,\zeta, \bar{\zeta})=G_{R}(s,\zeta, \bar{\zeta}),
\label{realUCF}
\end{equation}
for the gravitational field in the general asymptotically-flat case.

For the case of the Einstein--Maxwell fields, in general there will be two
complex world lines and two associated UCFs, one for the center of charge, the other for the center
of mass. For later use we note that the
gravitational world line will be denoted by $\xi^{a}$, while the
electromagnetic world line by $\eta^{a}$. Later we consider the special
case when the two world lines and the two UCFs coincide, i.e., $\xi^{a}=\eta^{a}$.

From the assumption that $\sigma^{0}$ and $L$ are first order and, from
Eqs.~(\ref{j}) and (\ref{k}) (e.g., $\psi_{3}^{0}=\eth \dot{\overline{\sigma}}^{0}$), Eq.~(\ref{Weyl-rot2*}), \textit{to second order}, is 
\begin{equation}
\psi_{1}^{\ast 0}=\psi_{1}^{0}-3L(\Psi -\eth^{2}\bar{\sigma}^{0}),
\label{psi1*.2}
\end{equation}
where $\psi_{2}^{0}$ has been replaced by the mass aspect (\ref{Mass Aspect}): $\Psi \approx \psi_{2}^{0}+\eth^{2}\bar{\sigma}^{0}$.

Using the spherical harmonic expansions (see Eqs.~(\ref{L***})
and~(\ref{sigma****})), 
\begin{eqnarray}
\Psi &=&\Psi^{0}+\Psi^{i}Y_{1i}^{0}+\Psi^{ij}Y_{2ij}^{0}+\ldots, \\
\psi_{1}^{0} &=&\psi_{1}^{0i}Y_{1i}^{1}+\psi_{1}^{0ij}Y_{2ij}^{1}+ \ldots, \\
\psi_{1}^{\ast 0} &=&\psi_{1}^{\ast 0i}Y_{1i}^{1}+\psi_{1}^{\ast
0ij}Y_{2ij}^{1}+ \ldots, \\
L(u_{\mathrm{B}},\zeta, \bar{\zeta}) &=&\xi^{i}(\tau)Y_{1i}^{1}-6\xi^{ij}(\tau
)Y_{2ij}^{1}+\ldots, \\
\sigma^{0}(u_{\mathrm{B}},\zeta, \bar{\zeta}) &=&24\xi^{ij}(\tau)Y_{2ij}^{2}+\ldots
\end{eqnarray}
and remembering that $\Psi^{0}$ is zeroth order, Eq.~(\ref{psi1*.2}),
becomes
\begin{equation*}
\psi_{1}^{\ast 0} =\psi_{1}^{0} -3[\xi^{i}(\tau)Y_{1i}^{1}-6\xi^{ij}(\tau)Y_{2ij}^{1}][\Psi^{0}+\Psi^{i}Y_{1i}^{0}+\{\Psi^{ij}-24\overline{\xi}^{ij}(\tau)\}Y_{2ij}^{0}]
\end{equation*}
or, re-arranging and performing the relevant Clebsh-Gordon expansions,
\begin{multline}
\label{INFO}
\psi^{0}_{1} = \psi^{*0}_{1}+3\Psi^{0}\xi^{i}Y^{1}_{1i}+\frac{3\sqrt{2}i}{2}\xi^{k}\Psi^{j}\epsilon_{kji}Y^{1}_{1i}-\frac{108}{5}\xi^{ik}\Psi^{k}Y^{1}_{1i} -\frac{18}{5}\xi^{k}(\Psi^{ik}-24\bar{\xi}^{ik})Y^{1}_{1i}\\
-\frac{216\sqrt{2}i}{5}\xi^{kj}(\Psi^{kl}-24\bar{\xi}^{kl})\epsilon_{jli}Y^{1}_{1i} + \; l\geq 2 \, \mbox{harmonic contributions}.
\end{multline}
%
Note that though Eq.~(\ref{INFO}) depends
initially on both $\tau$ and $u_{\mathrm{ret}}$, with $\tau =T(u_{\mathrm{ret}},\zeta,\overline{\zeta})$, we will eventually replace all the $u_{\mathrm{B}}$ (or $u_{\mathrm{ret}}$) by their expressions in terms of $\tau$, using (\ref{G2}). The transformation equation is then a function only of $\tau$ and ($\zeta,\bar{\zeta}$), at least to the given order in our perturbative framework.

This equation, though complicated and unattractive, is our main source of
information concerning the complex center-of-mass world line. The
information is extracted in the following way: Considering only the $l=1$ harmonics at constant $\tau$ in (\ref{INFO}), we set the $l=1$ harmonics
of $\psi_{1}^{\ast 0}$ (with constant $\tau $) to zero (i.e., $\psi_{1}^{\ast 0i}=0$). The three resulting relations are used to \textit{determine} the three spatial components, $\xi^{k}(\tau)$, of $\xi
^{a}(\tau)$ (with $\xi^{0}=\tau$).  This fixes the complex center of mass in terms of $\psi_{1}^{0i},\
\Psi^{0}$, $\Psi^{i}$, and other data which is readily interpreted physically. Alternatively it allow us to express $\psi_{1}^{0i}$ in terms of the $\xi^{a}(\tau)$.

Extracting this information takes a bit of effort.


\subsection{The complex center-of-mass world line}

Before trying to determine the $l=1$ harmonics of (\ref{INFO}), several comments and repetitions (for emphasis) are in
order:

\begin{enumerate}
\item As previously noted, Eq.~(\ref{INFO}) is a
function of both $\tau$ (via the $\xi^{i}$, $\xi^{ij}$)
and $u_{\mathrm{ret}}$ (via the $\psi_{1}^{0i}$ and $\Psi $). The extraction of the $l=1$ part of $\psi_{1}^{\ast 0}$ must be
taken on the constant $\tau$ cuts. In other words $u_{\mathrm{ret}}$ must
be eliminated by using (\ref{G2}).

\item This elimination of $u_{\mathrm{B}}$ (or $u_{\mathrm{ret}}$) is done in the
linear terms via the expansion: 
\begin{eqnarray}
\eta (u_{\mathrm{ret}}) &=&\eta \left( \tau -\frac{\sqrt{2}}{2}\xi_{R}^{i}(\tau)Y_{1i}^{0}+\sqrt{2}\xi_{R}^{ij}(\tau)Y_{2ij}^{0}\right)
\label{taylor} \\
&\approx &\eta (\tau)-\frac{\sqrt{2}}{2}\eta (s)^{\prime}\left[ \xi_{R}^{i}(s)Y_{1i}^{0}-2\xi_{R}^{ij}(s)Y_{2ij}^{0}\right]  \notag \\
u_{\mathrm{ret}} &=&\tau -\frac{\sqrt{2}}{2}\xi^{i}(\tau)Y_{1i}^{0}+\sqrt{2}\xi^{ij}(\tau)Y_{2ij}^{0}+\ldots  \notag
\end{eqnarray}

In the nonlinear terms we can simply use 
\begin{equation*}
u_{\mathrm{ret}}=\tau .
\end{equation*}

\item In the Clebsch--Gordon expansions of the harmonic products, though we
need both the $l=1$ and $l=2$ terms in the calculation, we keep at the end
only the $l=1$ terms for the $\psi_{1}^{0i}$. (Note that there are no $l=0$
terms since $\psi_{1}^{\ast 0}$ is spin weight $s=1$.)


\item For completeness, we have included into the calculations Maxwell
fields with both a complex dipole (electric and magnetic), $D_{\mathbb{C}}^{i}=q\eta^{i}=q(\eta_{R}^{i}+i\eta_{I}^{i})$ and complex
quadrupole (electric and magnetic) fields $Q_{\mathbb{C}}^{kj}=Q_{E}^{kj}+iQ_{M}^{kj}$.
\end{enumerate}

We begin by focusing on the $l=1$ portion of the right-hand-side of~(\ref{INFO}) in the complex center of mass frame.  Assuming that $\psi^{*0i}_{1}=0$, we see that all remaining terms on this side of the equation are non-linear, so we can simply make the replacement $u_{\mathrm{ret}}\rightarrow\tau$.  On the left-hand side of the equation, extracting the $l=1$ component of $\psi^{0}_{1}$ on a constant $\tau$ slice is more complicated though; using (\ref{taylor}), we have that
\begin{equation}
\label{psitrans1}
\psi^{0}_{1}(u_{\mathrm{ret}})=\psi^{0}_{1}(\tau)-\frac{\sqrt{2}}{2}\psi^{0}_{1}(\tau)'\xi^{i}(\tau)Y^{1}_{1i}+\sqrt{2}\psi^{0}_{1}(\tau)'\xi^{ij}(\tau)Y^{1}_{2ij}.
\end{equation}
Using the Bianchi identity (\ref{AsyBI2}) and inserting the proper factor of $\sqrt{2}$ to account for the retarded Bondi time, we have that (to second order)
\begin{equation*}
\psi^{0\prime}_{1}=-\frac{\sqrt{2}}{2}\eth \Psi +\frac{\sqrt{2}}{2}\eth^{3}\bar{\sigma}^{0}+\sqrt{2}k\phi^{0}_{1}\bar{\phi}^{0}_{2}.
\end{equation*}
As $\psi^{0\prime}_{1}$ enters (\ref{psitrans1}) only in non-linear terms, we only need to extract the linear portion of this Bianchi identity.  Recalling (suppressing factors of $c$ for the time being) that
\begin{equation*}
\phi^{0}_{1}=q+\sqrt{2}q\eta^{i\prime}Y^{0}_{1i}+\frac{\sqrt{2}}{6}Q^{ij\prime\prime}_{\mathbb{C}}Y^{0}_{2ij}, \qquad \phi^{0}_{2}=-2q\eta^{i\prime\prime}Y^{1}_{1i}-\frac{1}{3}Q^{ij\prime\prime\prime}_{\mathbb{C}}Y^{1}_{2ij},
\end{equation*}
we readily determine that
\begin{equation}
\label{psitrans2}
\psi^{0\prime}_{1}=\sqrt{2}\Psi^{i}Y^{1}_{1i}+3\sqrt{2}(\Psi^{ij}-24\bar{\xi}^{ij})Y^{1}_{2ij}-2\sqrt{2}kq^{2}\bar{\eta}^{i\prime\prime}Y^{1}_{1i}-\frac{\sqrt{2}kq}{3}\bar{Q}^{ij\prime\prime\prime}_{\mathbb{C}}Y^{1}_{2ij}.
\end{equation}

Feeding~(\ref{psitrans2}) into Eq.~(\ref{psitrans1}) and performing the relevant Clebsch--Gordon expansions, we find:
\begin{multline}
\label{psitrans3}
\psi^{0}_{1}(u_{\mathrm{ret}})|_{l=1}= \psi^{0i}_{1}-\frac{\sqrt{2}i}{2}\Psi^{j}\xi^{k}\epsilon_{jki}+\frac{864}{5}\xi^{j}\bar{\xi}^{ji}-\frac{12}{5}\Psi^{j}\xi^{ji} +\frac{3456\sqrt{2}i}{5}\xi^{lj}\bar{\xi}^{lk}\epsilon_{jki}\\
+\sqrt{2}ikq^{2}\xi^{k}\bar{\eta}^{j\prime\prime}\epsilon_{jki} +\frac{24}{5}kq^{2}\xi^{ji}\bar{\eta}^{j\prime\prime}+\frac{2}{5}kq\xi^{j}\bar{Q}^{ji\prime\prime\prime}_{\mathbb{C}}+\frac{24\sqrt{2}i}{5}kq \xi^{lj}\bar{Q}^{lk\prime\prime\prime}_{\mathbb{C}}\epsilon_{jki}.
\end{multline}
Here, we have implicitly used the fact that, to our level of approximation, $\Psi^{ij}=-24\bar{\xi}^{ij}$. 

We can now incorporate this into~(\ref{INFO}) to obtain the full complex center of mass equation as a function of $\tau$.  The $l=0,1$ components of the mass aspect are replaced by the expressions
\begin{equation*}
\Psi^{0}=-\frac{2\sqrt{2}G}{c^2}M_{\mathrm{B}}, \qquad \Psi^{i}= -\frac{6G}{c^3}P^{i}, \qquad \Psi^{ij}=-24\bar{\xi}^{ij};
\end{equation*}
we insert $k=2Gc^{-4}$ and the appropriate factors of $c$ elsewhere,
at which point (\ref{INFO}) can be re-expressed in a manner that
determines the complex center of mass, with all terms being functions
of $\tau$:
\begin{multline}
\label{main eq}
\psi^{0i}_{1}= -\frac{6\sqrt{2}G}{c^2}M_{\mathrm{B}}\xi^{i}+\frac{6\sqrt{2}i}{c^3}GP^{k}\xi^{j}\epsilon_{kji}-\frac{576G}{5c^3}P^{k}\xi^{ki}+\frac{6912\sqrt{2}i}{5}\xi^{lj}\bar{\xi}^{lk}\epsilon_{jki} \\
-\frac{2\sqrt{2}i}{c^6}Gq^{2}\xi^{k}\bar{\eta}^{j\prime\prime}\epsilon_{jki}-\frac{48G}{5c^6}q^{2}\xi^{ji}\bar{\eta}^{j\prime\prime}-\frac{4G}{5c^7}q\xi^{j}\bar{Q}^{ji\prime\prime\prime}_{\mathbb{C}}-\frac{16\sqrt{2}i}{5c^7}Gq\xi^{lj}\bar{Q}^{lk\prime\prime\prime}_{\mathbb{C}}\epsilon_{jki}.
\end{multline}
Note that the linear term
\begin{equation*}
\psi_{1}^{0i}=-\frac{6\sqrt{2}G}{c^{2}}M_{B}\xi^{i}=3\Psi^{0}\xi^{i}
\end{equation*}
coincides with the earlier results in the stationary case, Eq.~(\ref{D_cgrav}).

Now, we recall our identification for the complex gravitational dipole,
\begin{equation}
\label{complex grav.dipole}
\psi^{0i}_{1}(\tau)=-\frac{6\sqrt{2} G}{c^2}\left(D^{i}_{(\mathrm{mass})}+ic^{-1}J^{i}\right),
\end{equation}
as well as the identification between the $l=2$ harmonic coefficient of the UCF and the gravitational quadrupole:
\begin{equation}
\label{gravquad}
\xi^{ij}=\frac{\sqrt{2}G}{24c^4}Q^{ij\prime\prime}_{\mathrm{Grav}}=\frac{\sqrt{2}G}{24c^4}(Q^{ij\prime\prime}_{\mathrm{Mass}}+iQ^{ij\prime\prime}_{\mathrm{Spin}}).
\end{equation}
Feeding these into~(\ref{main eq}), we can separate out the real and imaginary parts via~(\ref{complex grav.dipole}) to obtain expressions for the mass dipole and angular momentum, our primary results:
\begin{multline}
\label{D}
D^{i}_{(\mathrm{mass})}= M_{\mathrm{B}}\xi^{i}_{R}-c^{-1}P^{k}\xi^{j}_{I}\epsilon_{jki}+\frac{4G}{5c^5}P^{k}Q^{ki\prime\prime}_{\mathrm{Mass}}+\frac{2G}{5c^6}Q^{lj\prime\prime}_{\mathrm{Spin}}Q^{lk\prime\prime}_{\mathrm{Mass}}\epsilon_{jki} \\
+\frac{q^2}{3c^4}\left(\xi^{k}_{R}\eta^{j\prime\prime}_{I}-\xi^{k}_{I}\eta^{j\prime\prime}_{R}\right)\epsilon_{jki}+\frac{Gq^2}{15c^8}\left(\eta^{j\prime\prime}_{R}Q^{ji\prime\prime}_{\mathrm{Mass}}+\eta^{j\prime\prime}_{I}Q^{ji\prime\prime}_{\mathrm{Spin}}\right)+\frac{\sqrt{2}q}{15c^5}\left(\xi^{j}_{R}Q^{ji\prime\prime\prime}_{E}+\xi^{j}_{I}Q^{ji\prime\prime\prime}_{M}\right) \\
+\frac{\sqrt{2}Gq}{45c^9}\left(Q^{lj\prime\prime}_{\mathrm{Mass}}Q^{lk\prime\prime\prime}_{M}-Q^{lj\prime\prime}_{\mathrm{Spin}}Q^{lk\prime\prime\prime}_{E}\right)\epsilon_{jki}
\end{multline}
\begin{multline}
\label{J}
J^{i}=cM_{\mathrm{B}}\xi^{i}_{I}+\xi^{j}_{R}P^{k}\epsilon_{jki}+\frac{4G}{5c^4}P^{k}Q^{ki\prime\prime}_{\mathrm{Spin}}+\frac{q^2}{3c^3}\left(\xi^{k}_{R}\eta^{j\prime\prime}_{R}+\xi^{k}_{I}\eta^{j\prime\prime}_{I}\right)\epsilon_{jki}+\frac{\sqrt{2}q}{15c^4}\left(\xi^{j}_{I}Q^{ji\prime\prime\prime}_{E}-\xi^{j}_{R}Q^{ji\prime\prime\prime}_{M}\right) \\
+\frac{Gq^2}{15c^7}\left(\eta^{j\prime\prime}_{R}Q^{ji\prime\prime}_{\mathrm{Spin}}-\eta^{j\prime\prime}_{I}Q^{ji\prime\prime}_{\mathrm{Mass}}\right)+\frac{\sqrt{2}Gq}{45c^8}\left(Q^{lj\prime\prime}_{\mathrm{Mass}}Q^{lk\prime\prime\prime}_{E}+Q^{lj\prime\prime}_{\mathrm{Spin}}Q^{lk\prime\prime\prime}_{M}\right)\epsilon_{jki}.
\end{multline}

Though these results are discussed at greater length later, we point out that
Eqs.~(\ref{D}) and (\ref{J}) already contains terms of obvious physical interest. 
Note that the first two items in $J^{i}$ are the spin, 
\begin{equation}
\overrightarrow{S}=cM_{\mathrm{B}}\overrightarrow{\xi}_{I}
\label{spin}
\end{equation}
(identified via the special case of the Kerr--Newman metric) and the orbital
angular momentum
\begin{equation}
\overrightarrow{L}=\overrightarrow{\xi}_{R}\times\overrightarrow{P}.
\label{orbital}
\end{equation}
The mass dipole $D_{(\mathrm{mass})}^{i}$ has the conventional term
$M_{\mathrm{B}}\overrightarrow{\xi}_{R}$ and a momentum-spin coupling
term (which appears to be new):
\begin{equation}
D_{(\mathrm{mass})}=M_{\mathrm{B}}\overrightarrow{\xi}_{R}+\frac{1}{c^{2}M_{\mathrm{B}}}\overrightarrow{P}\times\overrightarrow{S} +\cdots
\label{mass dipole*}
\end{equation}

We will see shortly that there is also a great deal of physical content to be
found in the nonlinear terms of Eq.~(\ref{main eq}).


\subsection{The evolution of the complex center of mass}
\label{evolution-complex-center-of-mass}

The evolution of the mass dipole and the angular momentum, defined
from the $\psi_{1}^{0i}$, Eq.~(\ref{main eq}) and Eqs.~(\ref{D})
with (\ref{J}), is determined via the Bianchi identity 
\begin{equation}
\label{combianchi}
\dot{\psi}_{1}^{0}=-\eth \psi_{2}^{0}+2\sigma^{0}\psi_{3}^{0}+2k\phi_{1}^{0}\bar{\phi}_{2}^{0}.
\end{equation}
This relationship allows us the determine -- kinematically -- the Bondi
momentum in terms of the dipole and the complex world line.

By extracting the $l=1$ harmonic from Eq.~(\ref{combianchi}), a
process which involves several Clebsch--Gordon expansions, we find
\begin{multline}
\label{psi1,0'}
\psi_{1}^{0i\prime} = \sqrt{2}c\Psi^{i}+\frac{1728\sqrt{2}i}{5}\xi^{kj}\bar{\xi}^{kl\prime}-\frac{4\sqrt{2}G}{c^5}q^{2}\bar{\eta}^{i\prime\prime}-\frac{4\sqrt{2}i}{c^6}Gq^{2}\eta^{j\prime}\bar{\eta}^{k\prime\prime}\epsilon_{kji} \\
+\frac{8G}{5c^7}q\bar{\eta}^{j\prime\prime}Q^{ji\prime\prime}_{\mathbb{C}}-\frac{8G}{5c^7}q\eta^{j\prime}\bar{Q}^{ji\prime\prime\prime}_{\mathbb{C}}-\frac{8\sqrt{2}i}{15c^8}GQ^{kj\prime\prime}_{\mathbb{C}}\bar{Q}^{kl\prime\prime\prime}_{\mathbb{C}}\epsilon_{lji}.
\end{multline}
Using our various identifications for the complex gravitational
dipole, the Bondi momentum, and the complex gravitational quadrupole,
(\ref{psi1,0'}) can be written as
\begin{multline}
\label{D'&J'}
\left(D_{(\mathrm{mass})}^{i}+ic^{-1}J^{i}\right)^{\prime} = P^{i}-\frac{12i}{5c^6}GQ^{kj\prime\prime}_{\mathrm{Grav}}\bar{Q}^{kl\prime\prime\prime}_{\mathrm{Grav}}\epsilon_{jli}+\frac{2q^2}{3c^3}\bar{\eta}^{i\prime\prime}+\frac{2i}{3c^4}q^{2}\eta^{j\prime}\bar{\eta}^{k\prime\prime}\epsilon_{kji} \\
-\frac{2\sqrt{2}}{15c^5}q\bar{\eta}^{j\prime\prime}Q^{ji\prime\prime}_{\mathbb{C}}+\frac{2\sqrt{2}}{15c^5}q\eta^{j\prime}\bar{Q}^{ji\prime\prime\prime}_{\mathbb{C}}+\frac{4i}{45c^6}Q^{kj\prime\prime}_{\mathbb{C}}\bar{Q}^{kl\prime\prime\prime}_{\mathbb{C}}\epsilon_{lji},
\end{multline}
or in terms of real and imaginary parts:
\begin{multline}
\label{D'}
D^{i\prime}_{(\mathrm{mass})} = P^{i} +\frac{2q^2}{3c^3}\eta^{i\prime\prime}_{R}+\frac{2q^2}{3c^4}\left(\eta^{j\prime}_{R}\eta^{k\prime}_{I}\right)^{\prime}\epsilon_{kji}-\frac{12G}{5c^6}\left(Q^{kj\prime\prime}_{\mathrm{Mass}}Q^{kl\prime\prime}_{\mathrm{Spin}}\right)^{\prime}\epsilon_{jli} \\
+\frac{2\sqrt{2}q}{15c^5}\left(\eta^{j\prime}_{R}Q^{ji\prime\prime\prime}_{E}-\eta^{j\prime\prime}_{R}Q^{ji\prime\prime}_{E}+\eta^{j\prime}_{I}Q^{ji\prime\prime\prime}_{M}-\eta^{j\prime\prime}_{I}Q^{ji\prime\prime}_{M}\right)+\frac{4}{45c^6}\left(Q^{kj\prime\prime}_{E}Q^{kl\prime\prime}_{M}\right)^{\prime}\epsilon_{lji},
\end{multline}
\begin{multline}
\label{J'}
J^{i\prime} = -\frac{2q^2}{3c^2}\eta^{i\prime\prime}_{I}+\frac{2q^2}{3c^3}\left(\eta^{j\prime}_{R}\eta^{k\prime\prime}_{R}+\eta^{j\prime}_{I}\eta^{k\prime\prime}_{I}\right)\epsilon_{kji}-\frac{12G}{5c^5}\left(Q^{kj\prime\prime}_{\mathrm{Mass}}Q^{kl\prime\prime\prime}_{\mathrm{Mass}}+Q^{kj\prime\prime}_{\mathrm{Spin}}Q^{kl\prime\prime\prime}_{\mathrm{Spin}}\right)\epsilon_{jli} \\
+\frac{2\sqrt{2}q}{15c^4}\left(\eta^{j\prime}_{I}Q^{ji\prime\prime}_{E}-\eta^{j\prime}_{R}Q^{ji\prime\prime}_{M}\right)^{\prime}+\frac{4}{45c^5}\left(Q^{kj\prime\prime}_{E}Q^{kl\prime\prime\prime}_{E}+Q^{kj\prime\prime}_{M}Q^{kl\prime\prime\prime}_{M}\right)\epsilon_{lji}.
\end{multline}

Eq.~(\ref{J'}), which is the conservation of angular
momentum, has several things to note.  As there are two terms
appearing as total derivatives (the first and fourth), it \emph{might}
be more natural to include them in an alternative definition of
angular momentum~\cite{Adamo:2011cx}:
\begin{equation}
\label{altangmom}
J^{i}_{T}=J^{i}+\frac{2q^{2}}{3c^2}\eta^{i\prime}_{I}-\frac{2\sqrt{2}q}{15c^4}\left(\eta^{j\prime}_{I}Q^{ji\prime\prime}_{E}-\eta^{j\prime}_{R}Q^{ji\prime\prime}_{M}\right).
\end{equation}
This results in an alternative flux law for angular momentum
conservation,
\begin{eqnarray}
J^{i\prime}_{T} & = & (\mathrm{Flux})^{i}_{T}, \nonumber \\
(\mathrm{Flux})^{i}_{T} & = & \frac{2q^2}{3c^3}\left(\eta^{j\prime}_{R}\eta^{k\prime\prime}_{R}+\eta^{j\prime}_{I}\eta^{k\prime\prime}_{I}\right)\epsilon_{kji}-\frac{12G}{5c^5}\left(Q^{kj\prime\prime}_{\mathrm{Mass}}Q^{kl\prime\prime\prime}_{\mathrm{Mass}}+Q^{kj\prime\prime}_{\mathrm{Spin}}Q^{kl\prime\prime\prime}_{\mathrm{Spin}}\right)\epsilon_{jli} \label{altflux} \\
& & +\frac{4}{45c^5}\left(Q^{kj\prime\prime}_{E}Q^{kl\prime\prime\prime}_{E}+Q^{kj\prime\prime}_{M}Q^{kl\prime\prime\prime}_{M}\right)\epsilon_{lji}, \nonumber
\end{eqnarray}
whose terms appear to agree with the known angular momentum flux due to gravitational quadrupole and electromagnetic dipole and quadrupole radiation~\cite{LL}.

As for the evolution equation for the mass dipole (\ref{D'}), we can
obtain an expression for the Bondi linear momentum by taking the
derivative (with respect to retarded Bondi time) of equation (\ref{D})
to eliminate $D^{i}_{(\mathrm{mass})}$ and find:
\begin{equation}
\label{P+}
P^{i} = M_{\mathrm{B}}\xi^{i\prime}_{R}-\frac{2q^2}{3c^3}\eta^{i\prime\prime}_{R}+\mathfrak{P}^{i}_{1}+\mathfrak{P}^{i}_{2}+\mathfrak{P}^{i}_{3},
\end{equation}
where $\mathfrak{P}^{i}_{1}$, $\mathfrak{P}^{i}_{2}$ and
$\mathfrak{P}^{i}_{3}$ are non-linear terms representing
dipole-dipole, dipole-quadrupole and quadrupole-quadrupole coupling
respectively,
\begin{eqnarray*}
\mathfrak{P}^{i}_{1} & = & \frac{q^2}{3c^4}\left(3\eta^{k\prime\prime}_{R}\xi^{j}_{I}-\eta^{j\prime\prime}_{I}\xi^{k}_{r}-2\eta^{k\prime}_{R}\eta^{j\prime}_{I}\right)^{\prime}\epsilon_{jki}-\frac{M_{\mathrm{B}}}{c}\left(\xi^{k\prime\prime}_{R}\xi^{j}_{I}\right)^{\prime}\epsilon_{jki}, \\
\mathfrak{P}^{i}_{2} & = & \frac{4G}{5c^5}M_{\mathrm{B}}\left(\xi^{k\prime}_{R}Q^{ki\prime\prime}_{\mathrm{Mass}}\right)^{\prime}+\frac{2\sqrt{2}q}{15c^5}\left(\eta^{j\prime\prime}_{R}Q^{ji\prime\prime}-\eta^{j\prime}_{R}Q^{ji\prime\prime\prime}_{E}+\eta^{j\prime\prime}_{I}Q^{ji\prime\prime}_{M}-\eta^{j\prime}_{I}Q^{ji\prime\prime\prime}_{M}\right) \\
& & +\frac{\sqrt{2}q}{15c^5}\left(\xi^{j}_{R}Q^{ji\prime\prime\prime}_{E}+\xi^{j}_{I}Q^{ji\prime\prime\prime}_{M}\right)^{\prime}+\frac{Gq^2}{15c^8}\left(\eta^{j\prime\prime}_{I}Q^{ji\prime\prime}_{\mathrm{Spin}}-7\eta^{j\prime\prime}_{R}Q^{ji\prime\prime}_{\mathrm{Mass}}\right)', \\
\mathfrak{P}^{i}_{3} & = & \frac{2G}{c^6}\left(Q^{kj\prime\prime}_{\mathrm{Mass}}Q^{kl\prime\prime}_{\mathrm{Spin}}\right)^{\prime}\epsilon_{jli}+\frac{4}{45c^6}\left(Q^{kj\prime\prime}_{E}Q^{kl\prime\prime}_{M}\right)^{\prime}\epsilon_{jli}+\frac{\sqrt{2}Gq}{45c^9}\left(Q^{lj\prime\prime}_{\mathrm{Mass}}Q^{lk\prime\prime\prime}_{M}-Q^{lj\prime\prime}_{\mathrm{Spin}}Q^{lk\prime\prime\prime}_{E}\right)^{\prime}\epsilon_{jki}.
\end{eqnarray*}

%

\begin{remark}
In the calculation leading to (\ref{P+}), non-linear terms with $P^{i}$ (or its derivatives) were replaced by the linear expression $P^{i}\approx M_{\mathrm{B}}\xi_{R}^{i\prime}-\frac{2}{3}c^{-3}q^{2}\eta_{R}^{i\prime \prime}$.  Additionally, we have neglected the time derivatives of the Bondi mass, $M_{\mathrm{B}}^{\prime}$; this is because, as we shall see momentarily, these derivatives are themselves second order quantities and hence give a vanishing contribution to equation (\ref{P+}) at our level of approximation.
\end{remark}

\subsubsection*{\textit{Physical Content}}

\begin{itemize}
\item The first term of $P^{i}$ is the standard Newtonian kinematic
expression for the linear momentum, $M\ \overrightarrow{v}$.

\item The second term, $-\frac{2}{3}c^{-3}q^{2}\eta_{R}^{i\,\prime \prime}$, which is a contribution from the second derivative of the electric dipole
moment, $q\eta_{R}^{i}$, plays a special role for the case when the complex
center of mass coincides with the complex center of charge, $\eta^{a}=\xi
^{a}$. In this case, the second term is exactly the contribution to the
momentum that yields the classical radiation reaction force of classical
electrodynamics~\cite{LL}.

\item Many of the remaining terms in $P^{i}$, though apparently second
order, are really of higher order when the dynamics are considered. Others
involve quadrupole interactions, which contain high powers of $c^{-1}$.

\item In the expression for $J^{i}$ we have already identified, in
earlier discussions, the first two terms $M_{\mathrm{B}}c\xi_{I}^{j}$ and $M_{\mathrm{B}}\xi_{R}^{k\,\prime}\xi_{R}^{i}\epsilon_{ikj}$ as the
intrinsic spin angular momentum and the orbital angular momentum. The
further terms, a spin-spin, spin-quadrupole and quadrupole-quadrupole
interaction terms, are considerably smaller.

\item As mentioned earlier, in Eq.~(\ref{J'}) we see that there are
five flux terms, the second is from the gravitational quadrupole flux, the
third and fifth are from the classical electromagnetic dipole and
electromagnetic quadrupole flux, while the fourth come from
dipole-quadrupole coupling. \textit{The Maxwell dipole part is identical to
that derived from pure Maxwell theory}~\cite{LL}. We emphasize that this
angular momentum flux law has little to do directly with the chosen
definition of angular momentum. The imaginary part of the Bianchi
identity (\ref{combianchi}),\ with the reality condition $\Psi =\overline{\Psi}$, \emph{is} the angular momentum conservation law. How to identify the
different terms, i.e., identifying the time derivative of the angular
momentum and the flux terms, comes from different arguments. The
identification of the Maxwell contribution to total angular momentum and the
flux contain certain arbitrary assignments: some terms on the left-hand side
of the equation, i.e., terms with a time derivative, could have been moved
onto the right-hand side and been called `flux' terms. However, our
assignments were governed by the question of what terms appeared most
naturally to be on different sides. The first term appears to be a new
prediction.

\item The angular momentum conservation law can be considered as the
evolution equation for the imaginary part of the complex world line, i.e., $\xi_{I}^{i}(u_{\mathrm{ret}})$. The evolution for the real part is found
from the Bondi energy-momentum loss equation.

\item In the special case where the complex centers of mass and charge
coincide, $\eta^{a}=\xi^{a}$, we have a rather attractive identification:
since now the magnetic dipole moment is given by $D_{M}^{i}=q\xi_{I}^{i}$
and the spin by $S^{i}=M_{\mathrm{B}}c\xi_{I}^{i}$, we have that the
gyromagnetic ratio is 
\begin{equation}
\frac{|S^{i}|}{|D_{M}^{i}|}=\frac{M_{\mathrm{B}}c}{q}
\label{g=2}
\end{equation}
leading to the Dirac value of $g$, i.e., $g=2$.

\end{itemize}


\subsection{The evolution of the Bondi energy-momentum}
\label{evolution-Bondi-energy-momentum}

Finally, to obtain the equations of motion, we substitute the kinematic
expression for $P^{i}$ into the Bondi evolution equation, the Bianchi
identity, Eq.~(\ref{AsyBI1}); 
\begin{equation*}
\dot{\psi}_{2}^{0}=-\eth \psi_{3}^{0}+\sigma^{0}\psi_{4}^{0}+k\phi_{2}^{0}\bar{\phi}_{2}^{0},
\end{equation*}
or its much more useful and attractive (real) form
\begin{equation}
\label{evolution}
\Psi^{\prime}=\frac{\sqrt{2}}{c}\sigma^{0\prime}\overline{\sigma}^{0\prime}+\frac{\sqrt{2}k}{c}\phi^{0}_{2}\bar{\phi}^{0}_{2}.
\end{equation}

\begin{remark}
The Bondi mass, $M_{\mathrm{B}}=-\frac{c^{2}}{2\sqrt{2}G}\Psi^{0}$, and the original mass of the Reissner--Nordstr\"{o}m (Schwarzschild) unperturbed metric, $M_{\mathrm{RN}}=-\frac{c^{2}}{2\sqrt{2}G}\psi_{2}^{0\,0}$, i.e., the $l=0$ harmonic of $\psi_{2}^{0}$, differ by a quadratic term in the shear, the $l=0$ part of $\sigma \dot{\overline{\sigma}}$. This suggests that the observed mass of an object is partially determined by its time-dependent quadrupole moment.
\end{remark}

Upon extracting the $l=0$ harmonic portion of Eq.~(\ref{evolution}) as well as inserting our various physical identifications for the objects involved, we obtain the Bondi mass loss theorem:
\begin{multline}
\label{M'}
M_{\mathrm{B}}^{\prime} = -\frac{G}{5c^7}\left(Q^{jk\prime\prime\prime}_{\mathrm{Mass}}Q^{jk\prime\prime\prime}_{\mathrm{Mass}}+Q^{jk\prime\prime\prime}_{\mathrm{Spin}}Q^{jk\prime\prime\prime}_{\mathrm{Spin}}\right)- \frac{4q^2}{3c^5}\left(\eta^{k\prime\prime}_{E}\eta^{k\prime\prime}_{E}+\eta^{k\prime\prime}_{M}\eta^{k\prime\prime}_{M}\right) \\
-\frac{4}{45c^7}\left(Q^{jk\prime\prime\prime}_{E}Q^{jk\prime\prime\prime}_{E}+Q^{jk\prime\prime\prime}_{M}Q^{jk\prime\prime\prime}_{M}\right).
\end{multline} 
This mass/energy loss equation contains the classical energy loss due to
electric and magnetic dipole radiation and electric and magnetic quadrupole ($Q_{E}^{ij},Q_{M}^{ij}$) radiation. (Note that agreement with the physical quadrupole radiation is recovered after making the aforementioned rescaling $Q^{ij}_{\mathbb{C}}\rightarrow 2\sqrt{2}Q^{ij}_{\mathbb{C}}$.) The gravitational energy loss is the
conventional quadrupole loss by the identification (\ref{gravquad}) of $\xi^{ij}$ with
the gravitational quadrupole moment $Q_{\mathrm{Grav}}^{ij}$.

The momentum loss equation, from the $l=1$ part of
Eq.~(\ref{evolution}), is then identified with the recoil force
due to momentum radiation:
\begin{equation*}
P^{i\prime}=F_{\mathrm{recoil}}^{i},
\end{equation*}
where 
\begin{multline}
\label{recoil}
F_{\mathrm{recoil}}^{i}= \frac{2G}{15c^6}\left(Q^{jl\prime\prime\prime}_{\mathrm{Spin}}Q^{kj\prime\prime\prime}_{\mathrm{Mass}}-Q^{jl\prime\prime\prime}_{\mathrm{Mass}}Q^{kj\prime\prime\prime}_{\mathrm{Spin}}\right)\epsilon_{kli}+\frac{2q^2}{3c^4}\left(\eta^{j\prime\prime}_{R}\eta^{k\prime\prime}_{I}-\eta^{j\prime\prime}_{I}\eta^{k\prime\prime}_{R}\right)\epsilon_{kji} \\
-\frac{4\sqrt{2}q}{15c^5}\left(\eta^{j\prime\prime}_{R}Q^{ji\prime\prime\prime}_{E}+\eta^{j\prime\prime}_{I}Q^{ji\prime\prime\prime}_{M}\right)-\frac{4}{135c^6}\left(Q^{jl\prime\prime\prime}_{M}Q^{kj\prime\prime\prime}_{E}-Q^{jl\prime\prime\prime}_{E}Q^{kj\prime\prime\prime}_{M}\right).
\end{multline}
Finally, we can substitute in the $P^i$ from (\ref{P+}) to obtain Newton's second law of motion:
%
\begin{equation}
M_{\mathrm{B}}\xi_{R}^{i\,\prime\prime}=F^{i},
\label{newton2nd}
\end{equation}
with
\begin{eqnarray}
F^{i} &=&-M_{\mathrm{B}}^{\prime}\xi_{R}^{i\,\prime}+\frac{2}{3}c^{-3}q^{2}\eta_{R}^{i\,\prime\prime\prime}+F_{\mathrm{recoil}}^{i}-\Xi^{i\,\prime},
\label{F^i} \\
\Xi &=&\mathfrak{P}_{1}+\mathfrak{P}_{2}+\mathfrak{P}_{3}
\end{eqnarray}

\subsubsection*{\textit{Physical Content}}

There are several things to observe and comment on concerning Eqs.~(\ref{newton2nd}) and~(\ref{F^i}):

\begin{itemize}

\item If the complex world line associated with the Maxwell center
of charge coincides with the complex center of mass, i.e., if
$\eta^{i}= \xi^{i}$, the term 
\begin{equation}
\frac{2}{3}c^{-3}q^{2}\xi_{R}^{i\,\prime \prime \prime}
\label{rad.react}
\end{equation}
becomes the classical electrodynamic radiation reaction force.

\item This result follows directly from the Einstein--Maxwell
equations. There was no model building other than requiring that the two
complex world lines coincide. Furthermore, there was no mass renormalization;
the mass was simply the conventional Bondi mass as seen at infinity. The
problem of the runaway solutions, though not solved here, is converted to
the stability of the Einstein--Maxwell equations with the `coinciding'
condition on the two world lines. If the two world lines do not coincide,
i.e., the Maxwell world line forms independent data, then there is no
problem of unstable behavior. This suggests a resolution to the problem of
the unstable solutions: one should treat the source as a structured object,
not a point, and centers of mass and charge as independent quantities.

\item The $F_{\mathrm{recoil}}^{i}$ is the recoil force from momentum
radiation.

\item The $\Xi^{i\, \prime}=-F_{\mathrm{RR}}^{i}$ can be interpreted as
the gravitational radiation reaction.

\item The first term in $F^{i}$, i.e.,
  $-M_{\mathrm{B}}^{\prime}\xi_{R}^{i\,\prime}$, is identical to a term in
  the classical Lorentz--Dirac equations of motion. Again it is nice to see it appearing, but with the use
of the mass loss equation it is in reality third order.

\end{itemize}


\subsection{Other related results}

The ideas involved in the identification, at future null infinity, of
interior physical quantities that were developed in the proceeding sections
can also be applied to a variety of different perturbation schemes. 
Bramson, Adamo and Newman~\cite{Bramson, AdamoNewman1, AdamoNewman2} have
investigated how gravitational perturbations originating solely from a
Maxwell radiation field can be carried through again using the asymptotic
Bianchi identities to obtain, in a different context, the same
identifications: a complex center-of-mass/charge world line, energy and
momentum loss, as well as an angular momentum flux law that agrees exactly
with the predictions of classical electromagnetic field theory. This scheme
yields (up to the order of the perturbation) an approximation for the metric
in the interior of the perturbed spacetime.

We briefly describe this procedure. One initially chooses as a background an
exact solution of the Einstein equations; three cases were studied: flat
Minkowski spacetime, the Schwarzschild spacetime with a `small' mass and
the Schwarzschild spacetime with a finite, `zero order', mass. For such
backgrounds, the set of spin coefficients is known and fixed. On this
background the Maxwell equations were integrated to obtain the desired
electromagnetic field that acts as the gravitational perturbation. Bramson
has done this for a pure electric dipole solution~\cite{Bramson, AdamoNewman1}
on the Minkowski background. Recent work has used an electric and magnetic
dipole field with a Coulomb charge~\cite{AdamoNewman2}. The resulting
Maxwell field, in each case, is then inserted into the asymptotic Bianchi
identities, which, in turn, determine the behavior of the perturbed
asymptotic Weyl tensor, i.e., the Maxwell field induces nontrivial changes
to the gravitational field. Treating the Maxwell field as first order, the
calculations were done to second order, as was done earlier in this review.

Using the just obtained Weyl tensor terms, one can proceed to the
integration of the spin-coefficient equations and the second-order
metric tensor. For example, one finds that the dipole Maxwell field
induces a second-order Bondi shear, $\sigma^{0}$. (This in principle
would lead to a fourth-order gravitational energy loss, which in our
approximation is ignored.)

Returning to the point of view of this section, 
the perturbed Weyl tensor can now be used to obtain the same physical
identifications described earlier, i.e., by employing a null rotation to set 
$\psi_{1}^{0\ast i}=0$, equations of motion and asymptotic physical
quantities, (e.g., center of mass and charge, kinematic expressions for
momentum and angular momentum, etc.) for the interior of the system could be
found. Although we will not repeat these calculations here, we present a
few of the results. Though the calculations are similar to the earlier ones,
they differ in two ways: there is no first-order freely given Bondi shear
and the perturbation term orders are different.

For instance, the perturbations induced by a Coulomb charge and general
electromagnetic dipole Maxwell field in a Schwarzschild background lead to
energy, momentum, and angular momentum flux relations~\cite{AdamoNewman2}:
\begin{eqnarray}
M_{\mathrm{B}}^{\prime} &=& -\frac{2}{3c^{5}}\left( D_{E}^{i\,
  \prime\prime}D_{E}^{i\prime \prime}+D_{M}^{i\prime \prime}D_{M}^{i\,
  \prime\prime}\right) ,
\label{classical results} \\
P^{i\prime} &=& \frac{1}{3c^{4}}D_{E}^{k\, \prime\prime}D_{M}^{j\, \prime\prime}\epsilon_{kji},  \notag \\
J^{k\prime} &=& \frac{2}{3c^{3}}\left( D_{E}^{i\,
  \prime\prime}D_{E}^{j\, \prime}+D_{M}^{i\, \prime\prime}D_{M}^{j\, \prime}\right) \epsilon_{ijk},  \notag
\end{eqnarray}
all of which agree exactly with predictions from classical field theory~\cite{LL}. 

The familiarity of such results is an exhibit in favor of the physical
identification methods described in this review, i.e., they are a
confirmation of the consistency of the identification scheme.


\newpage

\section{Gauge (BMS) Invariance}
\label{gauge-invariance}


The issue of gauge invariance, the understanding of which is not obvious or
easy, must now be addressed. The claim is that the work described here is
in fact gauge (or BMS) invariant.

First of all we have, $\mathfrak{I}_{\mathbb{C}}^{+}$, or its real part, $\mathfrak{I}^{+}$. On $\mathfrak{I}_{\mathbb{C}}^{+}$, for each choice of
spacetime interior and solution of the Einstein--Maxwell equations, we have
its UCF, either in its complex version, $u_{\mathrm{B}}=X(\tau,\zeta,\tilde{\zeta})$, or its real version, Eq.~(\ref{realUCF}). The geometric picture of
the UCF is a one-parameter family of slicings (complex or real) of $\mathfrak{I}_{\mathbb{C}}^{+}$ or $\mathfrak{I}^{+}$. This is a geometric
construct that has a different appearance or description in different Bondi
coordinate systems. It is this difference that we must investigate. We
concentrate on the complex version.

Under the action of the supertranslation, Eq.~(\ref{supert*}), we have: 
\begin{equation}
X(\tau,\zeta,\tilde{\zeta})\rightarrow X^{*}(\tau,\zeta,\tilde{\zeta})=X(\tau,\zeta,\tilde{\zeta})+\alpha (\zeta,\tilde{\zeta}),
\label{Xhat}
\end{equation}
with $\alpha (\zeta,\tilde{\zeta})$ an arbitrary complex smooth function on
(complexified) $S^{2}$. Its effect is to add on a constant to each spherical
harmonic coefficient of $X$. The special case of translations, with 
\begin{equation}
\alpha (\zeta, \tilde{\zeta})=d^{a}\hat{l}_{a}(\zeta, \tilde{\zeta}),
\label{alphaTrans}
\end{equation}
simply adds to the $l=(0,1)$ harmonic components the complex constants
$d^{a}$, so, via Eq.~(\ref{AsyGCF}), we have the (complex)
Poincar\'{e} translations,
\begin{equation}
\xi^{a}\rightarrow \xi^{a*}=\xi^{a}+d^{a}.
\label{translation*}
\end{equation}

The action of the homogeneous Lorentz transformations,
Eq.~(\ref{Lorentz*}), 
\begin{eqnarray}
u^{*}_{\mathrm{B}} &=&Ku_{\mathrm{B}},  \label{Lorentz**} \\
K &=&\frac{1+\zeta \overline{\zeta}}{(a\zeta +b)(\overline{a}\overline{\zeta}
+\overline{b})+(c\zeta +d)(\overline{c}\overline{\zeta}+\overline{d})}, \nonumber \\
\zeta^{*} &=&\frac{a\zeta +b}{c\zeta +d}; \quad ad-bc=1.
 \label{FLT} \\
e^{i\lambda} &=&\frac{c\zeta +d}{\overline{c}\overline{\zeta}+\overline{d}} \nonumber
\end{eqnarray}
is considerably more complicated. It leads to 
\begin{equation}
X^{*}(\tau,\zeta^{*},\tilde{\zeta}^{*})=KX(\tau, \zeta, 
\tilde{\zeta}).
\label{KX}
\end{equation}

Before discussing the relevant effects of the Lorentz transformations on our
considerations we first digress and describe an important technical issue
concerning representation of the homogeneous Lorentz group.

The representation theory of the Lorentz group, developed and described by
Gelfand, Graev and Vilenkin~\cite{GGV} used homogeneous functions of two
complex variables (homogeneous of degrees, $n_{1}-1$ and $n_{2}-1$) as the
representation space. Here we summarize these ideas via an equivalent
method~\cite{Harmonics1,Reps} using spin-weighted functions on the
sphere as the representation spaces. In the notation of Gelfand, Graev and Vilenkin,
representations are labelled by the two numbers $(n_{1},n_{2})$ or by $(s,w)$,
with $(n_{1},n_{2})=(w-s+1,w+s+1)$. The `$s$' is the same `$s$' as in the
spin weighted functions and `$w$' is the conformal weight~\cite{NewmanTod}
(sometimes called `boost weight'). The different representations are written
as $D_{(n_{1},n_{2})}$. The special case of irreducible unitary
representations, which occur when $(n_{1},n_{2})$ are not integers, plays no
role for us and will not be discussed. We consider only the case when
$(n_{1},n_{2})$ are integers so that the $(s,w)$ take integer or half
integer values. If $n_{1}$ and $n_{2}$ are both positive integers or both negative
integers, we have, respectively, the positive or negative integer
representations. The representation space, for each $(s,w)$, are the
functions on the sphere, $\eta_{(s,w)}(\zeta ,\overline{\zeta})$, that can
be expanded in spin-weighted spherical
harmonics, $_{s}Y_{lm}(\zeta, \overline{\zeta})$, so that 
\begin{equation}
\eta_{(s,w)}(\zeta, \overline{\zeta})=\sum_{l=s}^{\infty}\eta_{(lm)}\ _{s}Y_{lm}(\zeta, \overline{\zeta}).
\label{expansion}
\end{equation}

Under the action of the Lorentz group (\ref{Lorentz**})\,--\,(\ref{FLT}), they transform as
\begin{equation}
\widehat{\eta}_{(s,w)}(\widehat{\zeta},\overline{\widehat{\zeta}})=e^{is\lambda}K^{w}\eta_{(s,w)}(\zeta,
\overline{\zeta}).
\label{transforms}
\end{equation}
These representations, in general, are neither irreducible nor totally
reducible. For us the important point is that many of these representations
possess an invariant finite-dimensional subspace which (often) corresponds
to the usual finite dimensional tensor representation space. Under the
transformation, Eq.~(\ref{transforms}), the finite number of coefficients in
these subspaces transform among themselves. \textit{It is this fact
  which we heavily utilize}.  More specifically we have two related
situations: (1)~when the $(n_{1},n_{2})$ are both positive integers (or $w\geq |s|$) there will be finite dimensional invariant subspaces, $D_{(n_{1},n_{2})}^{+}$, which are spanned by the basis vectors 
$_{s}Y_{lm}(\zeta, \overline{\zeta})$, with $l$ given in the range,
$|s|\leq l \leq w$. All the finite dimensional representations can be
obtained in this manner. And (2)~when the $(-n_{1},-n_{2})$ are both
negative integers (i.e., we have a negative integer representation)
there will be an \textit{infinite} dimensional invariant subspace,
$D_{(-n_{1},-n_{2})}^{-}$, described elsewhere~\cite{Harmonics1}. One,
however, can obtain a \textit{finite} dimensional representation for
each negative integer case by the following construction: One forms
the factor space, $D_{(-n_{1},-n_{2})}/D_{(-n_{1},-n_{2})}^{-}$. This space is
isomorphic to one of the finite dimensional spaces associated with the
positive integers. The explicit form of the isomorphism, which is not
needed here, is given in Held et al.~\cite{Harmonics1,GGV}. 

Of major interest for us is not so much the invariant subspaces but instead
their interactions with their compliments (the full vector space modulo the
invariant subspace). Under the action of the Lorentz transformations applied
to a general vector in the representation space, the components of the
invariant subspaces remain in the invariant subspace but in addition
components of the complement move into the invariant subspace. On the
other hand, the components of the invariant subspaces do not move into the
complement subspace: the transformed components of the compliment
involve only the original compliment components. The transformation thus has
a non-trivial Jordan form. 

Rather than give the full description of these invariant subspaces we
confine ourselves to the few cases of relevance to us.

\subsubsection*{\textit{I. The Good-Cut Function}}
Though our interest is primarily in the negative integer representations, we first address the positive integer case of the $s=0$ and $w=1$, [$(n_{1},n_{2})=(2,2)$], representation. The harmonics, $l=(0,1)$ form the invariant subspace. The cut 
function,
$X(\tau, \zeta, \overline{\zeta})$, for each fix values of $\tau$,
lies in this space. 

We write the GCF as
\begin{eqnarray}
u_{\mathrm{B}} &=&X(\xi^{a}(\tau),\zeta, \overline{\zeta}),  \label{GCF expanded} \\
&=&\xi^{a}\hat{l}_{a}(\zeta, \overline{\zeta})+\sum_{l,|m|\leq l} H^{lm}(\xi^{a})Y_{lm}(\zeta ,\overline{\zeta}).  \nonumber
\end{eqnarray}

After the Lorentz transformation, the geometric slicings have not changed
but their description in terms of $(u_{\mathrm{B}},\zeta,\overline{\zeta})$ has changed
to that of $(u^{*}_{\mathrm{B}},\zeta^{*},\overline{\zeta}^{*})$. This leads to
\begin{eqnarray}
u^{*}_{\mathrm{B}} &=&KX=X^{*},  \label{X'=KX} \\
&=&\xi^{a*}\hat{l}_{a}^{*}(\zeta^{*}, \overline{\zeta}^{*})+\sum_{l,|m|\leq l} H^{lm\,*}(\xi^{a*})Y_{lm}(\zeta^{*}, \overline{\zeta}^{*}).  \nonumber
\end{eqnarray}

Using the transformation properties of the invariant subspace and its
compliment we see that the coordinate transformation must have the form:
\begin{equation}
\xi^{a*}=F^{a}(\xi^{b},H^{lm}(\xi^{b}),\ldots),
\label{chi'}
\end{equation}
in other words it moves the higher harmonic coefficients down to the
$l=(0,1)$ coefficients. The higher harmonic coefficients transform among
themselves;
\begin{equation*}
H^{lm\,*}(\xi^{a*})=F^{lm}(\ldots,H^{l^{*}m^{*}}(\xi^{b}),\dots).
\end{equation*}

Treating the $\xi^{a}$ and $\xi^{a*}$ as functions of $\tau $, we
have
\begin{equation}
\label{V&V'}
V=\partial_{\tau}X=v^{a}\frac{\partial X}{\partial\xi^{a}}=v^{a}\hat{l}_{a}+v^{a}\sum_{l,|m|\leq l}\frac{\partial H^{lm}(\xi^{b})}{\partial\xi^{a}}Y_{lm}(\zeta,\overline{\zeta}),
\end{equation}
where
\begin{equation*}
v^{a}=\frac{d\xi^{a}}{d\tau}, \qquad v^{a*}=\frac{\partial F^{a}}{\partial\xi^{b}}\frac{d\xi^{b}}{d\tau}=F^{a},_{b}v^{b}.
\end{equation*}
It then follows that $V$ transforms as
\begin{multline}
V^{*} = KV = v^{a*}\left(\hat{l}_{a}^{*}+\sum_{l,|m|\leq l}H^{lm\,*},_{a}\ Y_{lm}(\zeta^{*},\overline{\zeta}^{*})\right) \\
= v^{b}F^{a},_{b}\left(\hat{l}_{a}^{*}+\sum_{l,|m|\leq l}H^{lm\,*},_{a}\ Y_{lm}(\zeta^{*},\overline{\zeta}^{*})\right).
\end{multline}
%
Our $\mathcal{H}$-space coordinates, $z^{a}=\xi^{a}$, and their $\tau $-derivatives, $v^{a}$, \textit{are} the $l=(0,1)$ coefficients of harmonic expansions of the $X$ and $V$ respectively.  We have shown that a Lorentz transformation induces a specific coordinate transformation (and associated vector transformation) on these coefficients.

\subsubsection*{\textit{II. The Mass Aspect}}
A second important example concerns the Bondi mass aspect, (where we have
introduced the $Y_{0}^{0}$ for simplicity of treatment of numerical factors)
\begin{equation}
\Psi \equiv \Psi_{(0,-3)}
=\Psi^{0}Y_{0}^{0}+\Psi^{i}Y_{1i}^{0}+\Psi^{ij}Y_{2ij}^{0}+\ldots
\label{0-3}
\end{equation}
This is a $s=0$ and $w=-3$, $[(n_{1},n_{2})=(-2,-2)]$ quantity. The
invariant subspace has a basis set of the harmonics with $l\geq 2$. The
factor space is \textit{isomorphic} to the \textit{finite dimensional}
positive integer space $(n_{1},n_{2})=(2,2)$ and hence the harmonic
coefficients of $l=(0,1$) lie in this finite dimensional representation
space. From this isomorphism we know that functions of the form, $\Psi^{0}Y_{0}^{0}+\Psi^{i}Y_{1i}^{0}$, whose four coefficients are proportional to the Bondi four-momentum 
\begin{equation}
P^{a}=(Mc,P^{i}),
\label{Bondi P}
\end{equation}
form a Lorentzian four-vector. Note that we have rescaled the $\Psi^{0}$ in
Eq.~(\ref{0-3}) by the $Y^{0}_{0}$, differing from that of Eq.~(\ref{SC24}) in order to give the spherical harmonic coefficients
immediate physical meaning without the use of the factors in equations
Eq.~(\ref{SC25}) and Eq.~(\ref{SC26}).

This is the justification for calling the $l=(0,1)$ harmonics of the mass
aspect a Lorentzian four-vector. (Technically, the Bondi four-momentum is a
co-factor but we have allowed ourselves a slight notational irregularity.)

\subsubsection*{\textit{III. The Complex Dipole Moment}}
The Weyl tensor component, $\psi_{1}^{0}$, has $s=1$ and $w=-3$, $[(n_{1},n_{2})=(-3,-1)]$. The associated finite dimensional factor
space is isomorphic to the finite part of the
$s=-1, w=1, [(n_{1},n_{2})=(3,1)]$ representation. We have that 
\begin{equation}
\psi_{1}^{0}\equiv
\psi_{1(1,-3)}^{0}=\psi_{1}^{0i}Y_{1i}^{1}+\psi_{1}^{0ij}Y_{2ij}^{1}+\ldots
\label{1,-1}
\end{equation}
leads to the finite-dimensional representation space
\begin{equation}
\text{finite space}=\text{span}\left\{\psi_{1i}^{0}Y_{1i}^{-1}\right\}.
\label{finite rep space}
\end{equation}

What finite tensor transformation this corresponds to is a
slightly more complicated question than that of the previous examples of Lorentzian
vectors. In fact, it corresponds to the Lorentz transformations applied to
(complex) self-dual antisymmetric two-index tensors~\cite{PhysicalContent}.
We clarify this with an example from Maxwell theory: from a given \textbf{E}
and \textbf{B}, the Maxwell tensor, $F^{ab}$, and then its self-dual
version can be constructed: 
\begin{equation*}
W^{ab+}=F^{ab}+iF^{\ast ab}.
\end{equation*}
A Lorentz transformation applied to the tensor, $W^{ab+}$, is
equivalent~\cite{LL} to the same transformation applied to 
\begin{equation}
\psi_{1}^{0\ i}\leftrightarrow (\mathbf{E}+i\mathbf{B)}^{i}.
\label{E+iB}
\end{equation}

This allows us to assign Lorentzian invariant physical meaning
to our identifications of the complex mass
dipole moment and angular momentum vector, $D_{(\mathrm{mass})}^{i}+iJ^{i}$.

\subsubsection*{\textit{IV. General Invariants}}
Our last example is a general discussion of how to construct Lorentzian invariants from the representation spaces. Though we will
confine our remarks to just the cases of $s=0$, it is easy to extend them
to non-vanishing $s$ by having the two functions have respectively
spin-weight $s$ and $-s$.

Consider pairs of conformally weighted functions ($s=0$), $W_{(w)},
G_{(-w-2)}$, with weights respectively, $(w,-w-2)$. They are considered
to be in dual spaces. Our claim is that the integrals of the form
\begin{equation}
I=\int G_{(-w-2)}W_{(w)}d\Omega   \label{Invariants}
\end{equation}
are Lorentz invariants.

We first point out that under the fractional linear transformation, $\zeta^{*} \leftrightarrow \zeta$, given by Eq.~(\ref{FLT}), the area
element on the sphere
\begin{equation}
d\Omega =\frac{4i\,d\zeta\wedge d\overline{\zeta}}{(1+\zeta \overline{\zeta})^{2}}
\label{sphere area element}
\end{equation}
transforms as~\cite{Harmonics1}
\begin{equation}
d\Omega^{*}=K^{2}d\Omega .
\label{area transf}
\end{equation}
This leads immediately to
\begin{eqnarray}
I &=&\int G_{(-w-2)}^{*}W_{(w)}^{*}d\Omega^{*},
\label{proof} \\
&=&\int K^{-w-2}G_{(-w-2)}K^{w}W_{(w)}K^{2}d\Omega ,  \nonumber \\
&=&\int G_{(-w-2)}W_{(w)}d\Omega ,  \nonumber
\end{eqnarray}
the claimed result.

There are several immediate simple applications of Eq.~(\ref{Invariants}). By choosing an arbitrary $w=-2,s=0$ function, say $G_{(-2)}(\zeta, \overline{\zeta})$ and $W_{(0)}=1$, we immediately have a Lorentzian
scalar,
\begin{equation}
N\equiv \int G_{(-2)}d\Omega =\int G_{(-2)}^{\prime}d\Omega^{\prime}.
\label{scalar 1}
\end{equation}

If this is made more specific by choosing $G_{(-2)}=V^{-2}$, we have the
remarkable result (proved in Appendix~\ref{appendixD}) that this scalar
yields the $\mathcal{H}$-space metric norm of the ``velocity'' $v^a$, via
\begin{equation}
8\pi (g_{ab}v^{a}v^{b})^{-1}\equiv \int V^{-2}d\Omega .
\label{H*}
\end{equation}

A simple variant of this arises by taking the derivative of (\ref{H*}) with
respect to $v^{a}$, and multiplying by an arbitrary vector, $w^{a}$
leading to
\begin{equation}
8\pi w^{a}v_{a}(g_{cd}v^{c}v^{d})^{-2}\equiv \int V^{-3}w^{a}Z,_{a}d\Omega .
\label{v_a}
\end{equation}

Many other versions can easily be found.


\newpage

\section{Discussion/Conclusion}
\label{conclusion}

\subsection{History/background}
\label{background}

The work reported in this document has had a very long gestation period. It
began in 1965~\cite{KerrNewman} with the publication of a paper where a
complex coordinate transformation was performed on the
Schwarzschild/Reissner--Nordstr\"om solutions. This, in \textit{a
  precise sense}, moved the `point source' onto a complex world line
in a complexified spacetime. It thereby led to a derivation of the
spinning and the charged-spinning particle metrics. How and why this
procedure worked was considered to be rather mysterious and a great
deal of effort by a variety of people went into trying to unravel
it. In the end, the use of the complex coordinate transformation for
the derivation of these metrics appeared as if
it was simply an accident; that is, a trick with no immediate significance.
Nevertheless, the idea of a complex world line, appearing in a natural
manner, was an intriguing thought, which frequently returned. Some years
later, working on an apparently unrelated subject, we studied and found the
condition for a regular NGC, in asymptotically-flat
spacetime, to have a vanishing asymptotic shear~\cite{Aronson}. This led to the realization that a regular NGC was generated by a complex world line, though originally there was no relationship between the two complex world lines. This condition (our
previously discussed shear-free condition, Eq.~(\ref{A.shearfree}))
was eventually shown to be closely related to Penrose's asymptotic twistor theory, and in the
flat-space case it led to the Kerr theorem and totally shear-free NGCs. From
a very different point of view, searching for asymptotically shear-free
complex null surfaces, \textit{the good-cut equation} was found with its
four-complex parameter solution space, leading to the theory of $\mathcal{H}$-space.

Years later, the different strands came together. The shear-free condition
was found to be closely related to the good-cut equation; namely, that one
equation could be transformed into the other. The major surprise came when
we discovered that the \textit{regular} solutions of either equation were 
generated by complex world lines in an
auxiliary space~\cite{Footprints}. These complex world lines were
interpreted as being complex analytic curves in the associated $\mathcal{H}$-space. The deeper meaning of this remains a major question still to be
fully resolved; it is this issue which is \textit{partially} addressed in
the present work.

At first, these complex world lines were associated with the
spinning, charged and uncharged particle metrics -- type D algebraically
special metrics, but now can be seen as just special cases of the
asymptotically flat solutions. Since these metrics were algebraically
special, among the many possible asymptotically shear-free NGCs there was
(at least) one \textit{totally shear-free} (rather than asymptotically
shear-free) congruence coming from the Goldberg--Sachs theorem. Their
associated world lines were the ones first discovered in 1965 (coming,
by accident, from the complex coordinate transformation), and became the complex
center-of-mass world line (which coincided with the complex center of charge
in the charged case.). This observation was the clue for how to search for
the generalization of the special world line associated with
algebraically-special metrics and thus, in general, how to look for the
special world line (and congruence) to be identified with the complex center
of mass for arbitrary asymptotically flat spacetimes.

For the algebraically-special metrics, the null tetrad system at
$\mathfrak{I}^{+}$ with one leg being the tangent null vector to the
shear-free congruence leads to the vanishing of the asymptotic Weyl
tensor component, i.e., $\psi_{0}^{\ast}=\psi_{1}^{\ast}=0$. For the
general case, no tetrad exists with that property but one can always
find a null tetrad with one leg being tangent to an asymptotically shear-free congruence so that the $l=1$
harmonics of $\psi_{1}^{0\ast}$ vanish. It is precisely that choice of
tetrad that led to our definition of the complex center of mass.


\subsection{Other choices for physical identification}

The question of whether our definition of the complex center of mass is the
best possible definition, or even a reasonable one, is not easy to answer. We
did try to establish a criteria for choosing such a definition: (i)~it should
predict already known physical laws or reasonable new laws, (ii)~it should
have a clear geometric foundation and a logical consistency and
(iii)~it should agree with special cases, mainly the
algebraically-special metrics or analogies with flat-space Maxwell
theory. We did try out several other possible choices~\cite{PhysicalContent} and
found them all failing. This clearly does not rule out others that we
did not think of, but at the present our choice appears to be both
natural and effective in making contact with physical phenomena.
There still remains the mystery of the
physical meaning of $\mathcal{H}$-space or why it leads to such reasonable
physical results. Possible resolutions appear to lie in the duality between
the complex world lines and real but twisting NGCs~\cite{Adamo:2011cx}.


\subsection{Predictions}

Our equations of motion are simultaneously satisfactory and unsatisfactory:
they yield the equations of motion for an isolated object with a great deal
of internal structure (time-dependent multipoles with the emission of
gravitational and electromagnetic radiation) in the form of Newton's second
law. In addition, they contain a definition of angular momentum with an
angular momentum flux. The dipole part of the angular momentum flux agrees
with classical electromagnetic theory. Unfortunately, there appears to be no immediate
way to study or describe interacting particles in this manner.

However, there are some areas where these ideas might be tested, though the effects
would be very small. For example, earlier we saw that there was a
contribution to the Bondi mass (an addition to the Reissner--Nordstr\"om mass)
from the quadrupole moment,

\begin{equation}
M_{\mathrm{B}}-M_{RN}\equiv \Delta
M=-\frac{1}{5}Gc^{-6}\mathrm{\ Re}\ Q_{\mathrm{Grav}}^{ij\,
  \prime\prime}\overline{Q}_{\mathrm{Grav}}^{ij\, \prime\prime\prime}.
\label{deltaM}
\end{equation}
There were predicted contributions to both the momentum and angular momentum
flux from the gravitational and electromagnetic quadrupole radiation as well
as new terms in the angular momentum flux equation, i.e., the
charge/magnetic-dipole coupling term
\begin{equation*}
\frac{2}{3}q^{2}c^{-2}\eta_{I}^{i\,\prime}=\frac{2}{3}qc^{-2}D_{M}^{i\,\prime}.
\end{equation*}
This interpretation of this term is slightly ambiguous in that it could be
identified, in the flux equation, as either a contribution to the angular
momentum flux (as we are suggesting now) or as a contribution to the angular
momentum itself. See Eq.~(\ref{J'}).

There are other unfamiliar terms that can be thought of as predictions of
this theoretical construct, though how to possibly measure them is not at all clear.


\subsection{Summary of results}

\begin{enumerate}

\item From the asymptotic Weyl and Maxwell tensors, with their transformation
properties, we were able (via the asymptotically shear-free NGC) to
obtain two complex world lines -- a complex `center of mass' and
`complex center of charge' in the auxiliary $\mathcal{H}$-space. When
`viewed' from a Bondi coordinate and tetrad system, this led to an
expression for the real center of mass of the gravitating system and a
kinematic expression for the total angular momentum (including intrinsic
spin and orbital angular momentum), as seen from null infinity. It is
interesting to observe that the \textit{kinematical} expressions for the
classical linear momentum and angular momentum came directly from the
gravitational \textit{dynamical} laws (Bianchi identities) for the evolution
of the Weyl tensor.

\item From the real parts of one of the asymptotic Bianchi identities,
Eq.~(\ref{AsyBI1}), we found the standard kinematic expression for the
Bondi linear momentum, $P=M\xi_{R}^{\prime}+\dots$, with the radiation
reaction term $\frac{2q^{2}}{3c^{3}}v_{R}^{k\,\prime}$ among others. The
imaginary part was the angular momentum conservation law with a very natural
looking flux expression of the form:
%
\begin{equation*}
J^{\prime}=\mathrm{Flux}_{\mathrm{E\& M\;dipole}}+\mathrm{Flux}_{\mathrm{Grav}}+\mathrm{Flux}_{\mathrm{E\& M\;quad}}.
\end{equation*}
%
The first flux term is \textit{identical} to that calculated from classical
electromagnetic theory

\item Using the kinematic expression for the Bondi momentum in a second
Bianchi identity (\ref{AsyBI2}), we obtained a second-order ODE
for the center of mass world line that could be identified with Newton's
second law with radiation reaction forces and recoil forces, $M_{\mathrm{B}}\xi_{R}^{\prime\prime}=F$.

\item From Bondi's mass/energy loss theorem we obtained the correct energy
flux from the electromagnetic dipole and quadrupole radiation as well as the
gravitational quadrupole radiation.

\item From the specialized case where the two world lines coincide and the
definitions of spin and magnetic moment, we obtained the Dirac gyromagnetic
ratio, $g=2$. In addition, we find the classical electrodynamic radiation
reaction term with the correct numerical factors. In this case we have the
identifications of the meaning of the complex position vector:
$\xi^{i}=\xi_{R}^{i}+i\xi_{I}^{i}$.
\begin{eqnarray*}
\xi_{R}^{i} &=& \text{center-of-mass position} \\
S^{i} &=&Mc\xi_{I}^{i}= \text{spin angular momentum} \\
D_{E}^{i} &=&q\xi_{R}^{i}= \text{electric dipole moment} \\
D_{M}^{i} &=&q\xi_{I}^{i}= \text{magnetic dipole moment}
\end{eqnarray*}

\end{enumerate}

\begin{remark}
In the past, most discussions of angular momentum make use of group theoretical ideas with Noether theorem type arguments, via the BMS group and the Lorentz subgroup, to define angular momentum.  Unfortunately this has been beset with certain difficulties; different authors get slightly different numerical factors in their definitions, with further ambiguities arising from the supertranslation freedom of the BMS group. (See the discussion after Eq.~(\ref{SC27})) Our approach is very different from the group theoretical approach in that we come to angular momentum directly from the dynamics of the Einstein equations (the asymptotic Bianchi Identities). We use the angular momentum definition from linear theory, Eq.~(\ref{SC27}), (agreed to by virtually all) and then supplement it via conservation equations (the flux law) obtained directly from the Bianchi Identities.  We have a unique one-parameter family of cuts coming from the complex world line defining the complex center of mass.  This is a geometric structure with no ambiguities.  However, another ambiguity does arise by asking which Bondi frame should be used in the description of angular momentum; this is the ambiguity of what coordinate system to use.
\end{remark}


\subsection{Issues and open questions}

\begin{enumerate}

\item A particularly interesting issue raised by our equations is that of the
run-away (unstable) behavior of the equations of motion for a charged
particle (with or without an external field). We saw in Eq.~(\ref{newton2nd})
that there was a driving term in the equation of motion depending on the
electric dipole moment (or the real center of charge). This driving term
was totally independent of the real center of mass and thus does not lead to
the classical instability. However, if we restrict the complex center of
charge to be the same as the complex center of mass (a severe, but very
attractive restriction leading to $g=2$), then the innocuous driving dipole
term becomes the classical radiation reaction term -- suggesting instability.
(Note that in this coinciding case there was no model building -- aside
from the coinciding lines -- and no mass renormalization.)

A natural question then is: does this unstable behavior really remain? In
other words, is it possible that the large number of extra terms in the
gravitational radiation reaction or the momentum recoil force might
stabilize the situation? Answering this question is extremely difficult. If
the gravitational effects do not stabilize, then -- at least in this special
case -- the Einstein--Maxwell equations are unstable, since the run-away
behavior would lead to an infinite amount of electromagnetic dipole energy
loss.

An alternative possible resolution to the classical run-away problem is
simply to say that the classical electrodynamic model is wrong; and that one
must treat the center of charge as different from the center of mass with
its own dynamics.


\item In our approximations, it was assumed that the complex world line
yielded cuts of $\mathfrak{I}^{+}$ that were close to Bondi cuts. At the
present we do not have any straightforward means of finding the world lines
and their associated cuts of $\mathfrak{I}^{+}$ that are far from the Bondi
cuts.

\item As mentioned earlier, when the gravitational and electromagnetic
world lines coincide we find the rather surprising result of the Dirac value
for the gyromagnetic ratio. Though this appears to be a
significant result, we unfortunately do not have any deeper understanding of it.

\item Is it possible that the complex structures that we have been seeing and
using are more than just a technical device to organize ideas, and
that they have a deeper significance? One direction to explore this is
via Penrose's twistor and asymptotic twistor theory. It is known that much of
the material described here is closely related to twistor theory; an example
is the fact that asymptotic shear-free NGCs are really a special case of the
Kerr theorem, an important application of twistor theory (see
Appendix~\ref{appendixA}). This connection is being further explored.

\item With much of the kinematics and dynamics of \textit{ordinary classical
mechanics} sitting in our results, i.e., in classical GR, is it possible that
ordinary particle quantization could play a role in understanding quantum
gravity? Attempts along this line have been made~\cite{Tait, Bergmann} but,
so far, without much success.


\item An interesting issue, not yet fully explored but potentially
important, is what more can be said about the $\mathcal{H}$-space structures
associated with the special regions (the $\mathcal{H}$-space ribbon of~(\ref{real slicing}) that are related to the real cuts of null infinity.  We touch on this briefly below.

\item Another issue to be explored comes from the \textit{duality} between
the complex $\mathcal{H}$-space light-cones and the \textit{real shear-free
but twisting NGCs} in the real physical spacetime. From either one the
other can be determined. It appears as if one might be able to reinterpret
(almost) all the $\mathcal{H}$-space structures in terms of real structures
associated with the optical parameters of the \textit{twisting NGCs} and
the real slicings associated with the ribbons. This reinterpretation would
likely result in lost geometric simplicity and elegance -- but perhaps would
avoid the mysterious use of the complex $\mathcal{H}$-space for physical
identifications.

\item As a final remark, we want to point out that there is an issue that we
have ignored: do the asymptotic solutions of the Einstein equations
that we have discussed and used throughout this work really exist? By
`really exist' we mean the following: are there, in sufficiently general
circumstances, Cauchy surfaces with physically-given data such that their
evolution yields these asymptotic solutions? We have tacitly
assumed throughout, with physical justification but no rigorous mathematical
justification, that the full interior vacuum Einstein equations do lead to
these asymptotic situations. However, there has been a great deal of deep and
difficult analysis~\cite{Helmut, HelmutII, Corvino} showing, in fact, that
large classes of solutions to the Cauchy problem with physically-relevant
data do lead to the asymptotic behavior we have discussed. Recently there
has been progress made on the same problem for the Einstein--Maxwell
equations. 
\end{enumerate}



\subsection{New interpretations and future directions}

Throughout this review, we have focused on classical general
relativity as expressed in the Newman--Penrose (or spin-coefficient)
formalism; yet despite this highly classical setting, the geometric
structures discussed here bear a striking resemblance to ideas from
more ambitious areas of theoretical physics.  In particular, the
relationship between real, twisting asymptotically shear-free NGCs in
asymptotically flat spacetime and complex, twist-free NGCs in
(complex) $\mathcal{H}$-space appears to form a dual pair.  The
complexified asymptotic boundary $\mathfrak{I}^{+}_{\mathbb{C}}$ acts
as the translator between these two descriptions: from
$\mathfrak{I}^{+}_{\mathbb{C}}$ we determine the complexified
congruence in $\mathcal{H}$-space via the angle fields $L$ and
$\tilde{L}$ or the real congruence (in the real spacetime) via $L$ and
$\overline{L}$.  Imposing reality conditions in the former case gives
an open world sheet (the `$\mathcal{H}$-space ribbon' discussed in the
text) for the NGC's complex source.  In the real case, the twisting
congruence's caustic set in Minkowski space (which is interpreted as
its real source) forms a closed loop propagating in real
time~\cite{AdamoNewman4}, or a closed world sheet.  In the
asymptotically flat case, we cannot follow twisting congruences back
to their real caustic set, and they must be represented by the dual
`ribbon'.  

In both cases, we see that the congruence's source is a structure which has a natural interpretation as a \emph{classical string}, with the boundary $\mathfrak{I}^{+}_{\mathbb{C}}$ interpolating between the two descriptions.  This is suggestive of the so-called `holographic principle', which aims to equate a theory in a $d$-dimensional compact space with another theory defined on its $d-1$-dimensional boundary~\cite{'tHooft:1993gx,Bousso:2002ju}.  In practice, this can allow one to interpolate between a `physical' theory in one space and a `dual' theory living on its boundary (or \textit{vice versa}). In our case, the `physical' information is the real, twisting NGC in the asymptotically flat spacetime; $\mathfrak{I}^{+}_{\mathbb{C}}$ acts as a lens into $\mathcal{H}$-space, which serves as the virtual image space where physical data (such as the mass, linear momentum, angular momentum, etc.) is computed by the methods reviewed here.  Hence, it is tempting to refer to $\mathfrak{I}^{+}_{\mathbb{C}}$ as the `holographic screen' for some application of the holographic principle to general relativity.  The presence of classical string-like structures on both sides of the duality makes such a possibility all the more intriguing.

This should be contrasted with the most well-known instance of the holographic principle: the AdS/CFT correspondence~\cite{Maldacena:1997re,Witten:1998qj,Aharony:1999ti}. Here, the AdS-boundary acts as the holographic screen between a type IIB string theory in $AdS_{5}\times S^{5}$ (the virtual image space) and maximally supersymmetric Yang--Mills theory in real four-dimensional Minkowski spacetime (other versions exist, but all involve some supersymmetry). It is interesting that we appear to be describing an instance of the holographic principle that requires no supersymmetry, although 't~Hooft's original work relating the planar limit of gauge theories to string-type theories did not use supersymmetry either~\cite{'tHooft:1973jz}.  In 't~Hooft's work, an extra dimension for string propagation enters the picture due to anomaly cancellation in the same way that the extra dimensions of $AdS_{5}\times S^5$ allow for anomaly cancellation in a full supersymmetric string theory.  In our investigations, one can think of the analytic continuation of $\mathfrak{I}^{+}$ to $\mathfrak{I}^{+}_{\mathbb{C}}$ in the same manner: instead of cancelling a (quantum) anomaly, the `extra dimensions' arising from analytic continuation allow us to reinterpret the real twisting NGC in terms of a simpler geometric structure, namely the complex light-cones in $\mathcal{H}$-space.

It is worth noting that this is not the first time that there has been a suggested connection between structures in asymptotically flat spacetimes and the holographic principle.  Most prior studies have attempted to understand such a duality in terms of the BMS group, which serves as the symmetry group of the asymptotic boundary~\cite{Arcioni:2003xx}.  Loosely speaking, these studies take their cue more directly from the AdS/CFT correspondence: by studying fields living on $\mathfrak{I}^{+}$ which carry representations of the BMS group, one hopes to reconstruct the full interior of the spacetime `holographically'.  It would be interesting to see how, if at all, our methodology relates to this program of research.

Additionally, as we have mentioned throughout this review (and further elaborated in Appendix~\ref{appendixA}), the nature of many of the objects studied here is highly twistorial.  This is essentially because $\mathcal{H}$-space is a complex vacuum spacetime equipped with an anti-self-dual metric, and hence possesses a curved twistor space by Penrose's non-linear graviton construction~\cite{PropHspace}.  It would be interesting to know if our procedure for identifying the complex center of mass (and/or charge) in an asymptotically flat spacetime could be phrased purely in terms of twistor theory.  Furthermore, the past several years have seen dramatic progress in using twistor theory to study gauge theories and their scattering amplitudes \cite{Adamo:2011pv}.  These techniques may provide the most promising route for connecting our work with any quantized version of gravity, as illustrated by the recent twistorial derivation of the tree-level MHV graviton scattering amplitude~\cite{Mason:2008jy}.  

While the interpretations we have suggested here are far from precise, they suggest a myriad of further directions which research in this area could take.  It would be truly fascinating for a topic as old as asymptotically flat spacetimes to make meaningful contact with ambitious new areas of mathematics and physics such as holography or twistorial scattering theory.


\newpage

\section{Acknowledgements}
\label{acknowledgments}

We are happy to note and acknowledge the great deal of detailed help and
understanding that we have received over the years from Gilberto
Silva-Ortigoza, our co-author on many of the earlier papers on the present
subject. Many others have contributed to our understanding of
asymptotically-flat spacetimes. Prime among them are Roger Penrose,
Andrzej Trautman, Jerzy Lewandowski, Pawel Nurowski, Paul Tod, Lionel
Mason, J\"{o}rg Frauendiener and Helmut Friedrich. We thank them
all. 

We also want to point out the very early work of Brian
Bramson~\cite{Bramson:1975}, William Hallidy and Malcolm
Ludvigsen~\cite{Hallidy:1979}. Their ideas, though not fully
developed, are close to the ideas developed here, and should be viewed
as precusors or preliminary formulations of our theory. We apologize
for not having noticed them earlier and giving them more credit.


\newpage

\appendix

\section{Twistor Theory}
\label{appendixA}

Throughout this review, the study of the asymptotic
gravitational field has been at the heart of all our investigations, and there is a natural connection between asymptotic gravitational fields and twistor theory. We give here a brief overview
of Penrose's asymptotic twistor theory (see, e.g.,~\cite{PenroseRindler2, PenroseTwist, HuggettTod}) and its connection to the good-cut equation and
the study of asymptotically shear-free NGCs at $\mathfrak{I}^{+}$; for a
more in depth exposition of this connection, see~\cite{Twistors1, Adamo:2010ey}. 

Let $\mathcal{M}$ be any asymptotically-flat spacetime manifold, with conformal
future null infinity $\mathfrak{I}^{+}$, coordinatized by $(u_{\mathrm{B}},\zeta,\bar{\zeta})$. We can consider the complexification of $\mathfrak{I}^{+}$, referred to as $\mathfrak{I}_{\mathbb{C}}^{+}$, which is in turn
coordinatized by $(u_{\mathrm{B}},\zeta,\tilde{\zeta})$, where now $u_{\mathrm{B}}\in \mathbb{C}$ and $\tilde{\zeta}$ is different, but close to $\bar{\zeta}$. Assuming analytic asymptotic Bondi shear $\sigma^{0}(u_{\mathrm{B}},\zeta,\bar{\zeta})$, it can then be analytically continued to $\mathfrak{I}_{\mathbb{C}}^{+}$, i.e., we can consider $\sigma^{0}(u_{\mathrm{B}},\zeta,\tilde{\zeta})$. We have seen in Section~\ref{good-cut-eq}
that solutions to the good-cut equation
\begin{equation}
\eth^{2}G=\sigma^{0}(G,\zeta, \tilde{\zeta})
\label{1}
\end{equation}
yield a four complex parameter family of solutions, given by
\begin{equation}
u_{\mathrm{B}}=Z(z^{a};\zeta, \tilde{\zeta}).
\label{2}
\end{equation}
In our prior discussions, we interpreted these solutions as defining a four
(complex) parameter family of surfaces on $\mathfrak{I}_{\mathbb{C}}^{+}$
corresponding to each choice of the parameters $z^{a}$.
In order to force agreement with the conventional description of Penrose's
asymptotic twistor theory we must use the \textit{complex conjugate} good-cut equation
\begin{equation}
\overline{\eth}^{2}\overline{G}=\overline{\sigma}^{0}(\overline{G},\zeta, \tilde{\zeta}),
\label{ccGoodCut}
\end{equation}
whose properties are identical to that of the good-cut equation. Its
solutions, written as
\begin{equation}
u_{\mathrm{B}}=\overline{Z}(\overline{z}^{a};\zeta, \tilde{\zeta}),
\label{two***}
\end{equation}
define complex two-surfaces in $\mathfrak{I}_{\mathbb{C}}^{+}$ for
fixed $\overline{z}^{a}$. If we fix $\zeta =\zeta_{0}\in \mathbb{C}$, then
Eq.~(\ref{ccGoodCut}) becomes an ordinary second-order
differential equation with solutions describing curves in
$(u_{\mathrm{B}},\tilde{\zeta})$ space. Hence, each solution to this
ODE is given by specifying initial conditions for $\tilde{G}$ and
$\partial_{\tilde{\zeta}}\tilde{G}$ at some arbitrary initial point,
$\tilde{\zeta}=\tilde{\zeta}_{0}$.

Note that it is \textit{not
  necessary} that $\tilde{\zeta}_{0}= \bar{\zeta}_{0}$ on
$\mathfrak{I}_{\mathbb{C}}^{+}$. However, we chose this initial point
to be the complex conjugate of the constant $\zeta_{0}$, i.e., we take
$\overline{G}$ and its first $\widetilde{\zeta}$ derivative at
$\widetilde{\zeta}=\overline{\zeta}_{0}$ as the \textit{initial}
conditions. Then the initial conditions for Eq.~(\ref{ccGoodCut}) can be written as~\cite{Twistors1}
\begin{eqnarray}
u_{\mathrm{B}0} &=&\overline{G}(\zeta_{0},\bar{\zeta}_{0}),  \label{ICs1} \\
\overline{L}_{0} &=&\overline{\eth}\overline{G}(\zeta_{0},\bar{\zeta}_{0})=P_{0}\frac{\partial \overline{G}}{\partial \bar{\zeta}_{0}}(\zeta_{0},\bar{\zeta}_{0}),  \notag
\end{eqnarray}
with $P_{0}=$ $1+\zeta_{0}\tilde{\zeta}_{0}$. Asymptotic projective twistor
space, denoted $\mathbb{P}\mathfrak{T}$, is the space of all curves in $\mathfrak{I}_{\mathbb{C}}^{+}$ generated by initial condition triplets $(u_{\mathrm{B}0},\zeta_{0},\overline{L}_{0})$~\cite{PenroseRindler2}: an asymptotic
projective twistor is the curve corresponding to $(u_{\mathrm{B}0},\zeta_{0},\overline{L}_{0})$. A particular subspace of $\mathbb{P}\mathfrak{T}$,
called null asymptotic projective twistor space ($\mathbb{P}\mathfrak{N}$),
is the family of curves generated by initial conditions, which lie on (real) 
$\mathfrak{I}^{+}$; that is, at the initial point, $\tilde{\zeta}_{0}=$ $\bar{\zeta}_{0}$, the curve should cross the real $\mathfrak{I}^{+}$, i.e.,
should be real, $u_{\mathrm{B}0}=\bar{u}_{\mathrm{B}0}$. Equivalently, an element of $\mathbb{P}\mathfrak{N}$ can be said to intersect its dual curve (the solution
generated by the complex conjugate initial conditions) at $\tilde{\zeta}_{0}=\bar{\zeta}_{0}$. The effect of this is to reduce the
three-dimensional complex twistor space to five real dimensions. In standard
notation, asymptotic projective twistors are defined in terms of their three
complex twistor coordinates, $(\mu^{0},\mu^{1},\zeta)$~\cite{PenroseRindler2}. These twistor coordinates may be re-expressed in terms of
the asymptotic twistor curves by
\begin{eqnarray}
\mu^{0} &=&u_{\mathrm{B}0}-\bar{L}_{0}\bar{\zeta}_{0},
\label{Twistorcoord} \\
\mu^{1} &=&\bar{L}_{0}+\zeta_{0}u_{\mathrm{B}0},  \notag \\
\zeta &=&\zeta_{0}.  \notag
\end{eqnarray}

By only considering the twistor initial conditions $\tilde{\zeta}_{0}=$ $\bar{\zeta}_{0}$, we can drop the initial value notation, and just let $u_{\mathrm{B}0}=u_{\mathrm{B}}$ and $\tilde{\zeta}=$ $\bar{\zeta}$. The connection of
twistor theory with \textit{shear-free} NGCs takes the form of the \textit{flat-space} Kerr theorem~\cite{PenroseRindler2,Twistors1}:

\begin{thm}[Kerr Theorem]
Any analytic function on $\mathbb{PT}$ (projective
  twistor space) generates a shear-free NGC in
  Minkowski space.
\end{thm}

Any analytic function $F(\mu^{0},\mu^{1},\zeta)$ on projective twistor space generates a shear-free NGC
in Minkowski space by obtaining the $\overline{L}=\overline{L}(u_{\mathrm{B}},\zeta, \overline{\zeta})$, which defines the
congruence via solving the algebraic equation
\begin{equation*}
F(\mu^{0},\mu^{1},\zeta)=F(u_{\mathrm{B}}-\overline{L}\overline{\zeta},\overline{L}+\zeta u_{\mathrm{B}},\zeta )=0.
\end{equation*}
It automatically satisfies the complex conjugate shear-free condition 
\begin{equation*}
\overline{\eth}\overline{L}+\overline{L}\dot{\overline{L}}=0.
\end{equation*}

We are interested in a version of the Kerr theorem that yields the regular
\emph{asymptotically} shear-free NGCs. Starting with the general four-parameter
solution to Eq.~(\ref{ccGoodCut}), i.e., $u_{\mathrm{B}}=\overline{Z}(\overline{z}^{a};\zeta, \tilde{\zeta})$, we chose an arbitrary world line $\overline{z}^{a}=\xi^{a}(\tau)$, so that we have
\begin{eqnarray}
u_{\mathrm{B}} &=&\overline{Z}(\xi^{a}(\tau),\zeta, \bar{\zeta})=\overline{G}(\tau, \zeta, \bar{\zeta}),
\label{ccCuts} \\
\overline{L}(\tau,\zeta, \bar{\zeta}) &=&\overline{\eth}_{(\tau)}\overline{G}(\tau,\zeta, \bar{\zeta}).  \notag
\end{eqnarray}
By inserting these into the twistor coordinates, Eq.~(\ref{Twistorcoord}), we
find
\begin{eqnarray}
\mu^{0}(\tau,\zeta, \bar{\zeta}) &=&u_{\mathrm{B}}-\bar{L}\bar{\zeta}=\overline{G}-\bar{\zeta}\bar{\eth}_{(\tau)}\overline{G},
\label{GCcoords} \\
\mu^{1}(\tau,\zeta, \bar{\zeta}) &=&\bar{L}+\zeta u_{\mathrm{B}}=\bar{\eth}_{(\tau)}\overline{G}+\zeta \overline{G}.
\label{GCcoords2}
\end{eqnarray}
The $\mu^{0}$ and $\mu^{1}$ are now functions of $\tau$ and $\zeta$, while the $\bar{\zeta}$ is now to be treated as a fixed quantity, the complex conjugate
of $\zeta$, and not as an independent variable. By eliminating $\tau$ in
Eqs.~(\ref{GCcoords}) and~(\ref{GCcoords2}), we obtain a single
function of $\mu^{0}$, $\mu^{1}$, and $\zeta$: namely, $F(\mu^{0},\mu^{1},\zeta)=0$. Thus, the \textit{regular} asymptotically shear-free NGCs are
described by a special class of twistor functions. This is a special case of
a generalized version of the Kerr theorem~\cite{PenroseRindler2, Twistors1}. 


\newpage

\section{CR Structures}
\label{appendixB}

A CR manifold $\mathcal{N}$ is a differentiable manifold endowed with an additional structure called its `CR structure'; formally this is a complex distribution (i.e., a sub-bundle $L\subset T\mathcal{N}\otimes \mathbb{C}$) which is formally integrable and almost Lagrangian~\cite{Dragomir:2006}.  More concretely, the CR structure can be described by a set of vectors or 1-forms on $\mathcal{N}$ defined up to a particular gauge freedom.  In the context of this review, we are interested in the case where $\mathcal{N}$ is a real three-manifold.

A CR structure on a real three manifold $\mathcal{N}$, with local
coordinates $x^{a}$, can be given intrinsically by equivalence classes of
one-forms, one real, one complex and its complex conjugate~\cite{CR1}. If we
denote the real one-form by $l$ and the complex one-form by $m$, then these
are defined up to the transformations:
\begin{eqnarray}
l &\rightarrow &a(x^{a})l,
\label{Trans1} \\
m &\rightarrow &f(x^{a})m+g(x^{a})l.  \notag
\end{eqnarray}

The $(a,f,g)$ are functions on $\mathcal{N}$: $a$ is nonvanishing and real, 
$f$ and $g$ are complex function with $f$ nonvanishing. We further
require that there be a three-fold linear-independence relation between
these one-forms~\cite{CR1}:
\begin{equation}
l\wedge m\wedge \bar{m}\neq 0.
\label{Lin}
\end{equation}

Any three-manifold with a CR structure is referred to as a three-dimensional
CR manifold. There are special classes (referred to as embeddable) of
three-dimensional CR manifolds that can be directly embedded into $\mathbb{C}^{2}$. We show how the choice of any specific asymptotically shear-free NGC
induces an embeddable CR structure on $\mathfrak{I}^{+}$. Though there are several 
ways of
arriving at this CR structure, the simplest way is to look at the asymptotic
null tetrad system associated with the asymptotically shear-free NGC, i.e.,
look at the ($l^{\ast a}$, $m^{\ast a}$, $\overline{m}^{\ast a}$, $n^{\ast a}$) of
Eq.~(\ref{null-rot*}). The associated dual one-forms, restricted
to $\mathfrak{I}^{+}$ (after a conformal rescaling of $m$), become
(with a slight notational dishonesty),

\begin{eqnarray}
l^{\ast} &=&du_{\mathrm{B}}-\frac{L}{1+\zeta \bar{\zeta}}d\zeta -\frac{\bar{L}}{1+\zeta \bar{\zeta}}d\bar{\zeta},
\label{one-forms} \\
m^{\ast} &=&\frac{d\overline{{\zeta}}}{1+\zeta \bar{\zeta}},\qquad \overline{m}^{\ast}=\frac{d{\zeta}}{1+\zeta \bar{\zeta}},  \notag
\end{eqnarray}
with $L=L(u_{\mathrm{B}},\zeta, \bar{\zeta})$, satisfying the shear-free condition.
(This same result could have been obtained by manipulating the exterior
derivatives of the twistor coordinates, Eq.~(\ref{Twistorcoord}).)
The dual vectors -- also describing the CR structure -- are
\begin{eqnarray}
\overline{\mathfrak{M}} &=&P\frac{\partial}{\partial \zeta}+L\frac{\partial}{\partial u_{\mathrm{B}}}=\eth_{(u_{_{B}})}+L\frac{\partial}{\partial u_{\mathrm{B}}},
\label{duals} \\
\mathfrak{M} &=&P\frac{\partial}{\partial \overline{\zeta}}+\overline{L}\frac{\partial}{\partial u_{\mathrm{B}}}=\eth_{(u_{_{B}})}+\overline{L}\frac{\partial}{\partial u_{\mathrm{B}}},  \notag \\
\mathfrak{L} &=&\frac{\partial}{\partial u_{\mathrm{B}}}.  \notag
\end{eqnarray}

Therefore, for the situation discussed here, where we have singled out a unique
asymptotically shear-free NGC and associated complex world line, we have
a uniquely chosen CR structure induced on $\mathfrak{I}^{+}$.
To see how our three manifold, $\mathfrak{I}^{+}$, can be embedded
into $\mathbb{C}^{2}$ we introduce the CR equation~\cite{CR2}
\begin{equation*}
\overline{\mathfrak{M}}K\equiv \eth_{(u_{\mathrm{B}})}K+L\frac{\partial}{\partial
u_{\mathrm{B}}}K=0
\end{equation*}
and seek two independent (complex) solutions,
$K_{1}=K_{1}(u_{\mathrm{B}},\zeta, \overline{\zeta}),
K_{2}=K_{2}(u_{\mathrm{B}}, \zeta, \overline{\zeta})$ that define the
embedding of $\mathfrak{I}^{+}$ into $\mathbb{C}^{2}$ with coordinates
$(K_{1}, K_{2})$.
We have immediately that $K_{1}=\overline{\zeta} = x-iy$ is a
solution. The second solution is also easily found; we see directly
from Eq.~(\ref{CREq})~\cite{ScriCR},
\begin{equation}
\eth_{(u_{\mathrm{B}})}T+L\dot{T}=0,
\end{equation}
that 
\begin{equation*}
\tau =T(u_{\mathrm{B}},\zeta, \overline{\zeta}),
\end{equation*}
the inverse to $u_{\mathrm{B}}=X(\tau ,\zeta, \overline{\zeta})$, is a
CR function and that we can consider $\mathfrak{I}^{+}$ to be embedded
in the $\mathbb{C}^{2}$ of $(\tau ,\overline{\zeta})$.

An important structure associated to any embeddable CR manifold of codimension one is its Levi form; this determines a metric on the CR structure as a bundle on the manifold~\cite{Dragomir:2006}.  As we have just discussed, $\mathfrak{I}^{+}$ is just such a CR manifold, and one can show that its Levi form (in the CR structure induced by an asymptotically shear-free NGC) is proportional to the twist of the congruence.  Hence, any CR structure on $\mathfrak{I}^+$ which is generated by a congruence with its source on a real world line $\xi^{a}(s)\in\mathbb{M}$ is Levi-flat~\cite{AdamoNewman4}.

In the context of this review, the important observation is that the physical content of asymptotically shearfree NGCs is encoded in the corresponding CR structure.  This gives a physical interpretation for CR structures in the setting of asymptotically flat spacetimes.  It would be interesting for future research to study the relationship between our findings and those of~\cite{Hill:2008}, which demonstrates how the Einstein equations for algebraically special spacetimes can be realized in terms of the embeddable CR structures associated with their PNDs.


\newpage

\section{Tensorial Spin-\textit{s} Spherical Harmonics}
\label{appendixC}


Some time ago, the generalization of ordinary spherical harmonics
$Y_{lm}(\zeta, \bar{\zeta})$ to spin-weighted functions
$_{(s)}Y_{lm}(\zeta, \bar{\zeta})$ (e.g.,~\cite{Harmonics1,Edth,BMS})
was developed to allow for harmonic expansions of spin-weighted
functions on the sphere. In this paper we have instead used the
tensorial form of these spin-weighted harmonics, the \textit{tensorial
  spin-s spherical harmonics}, which are formed by taking appropriate
linear combinations of the $_{(s)}Y_{lm}(\zeta, \bar{\zeta})$~\cite{Spins}:
\begin{equation*}
Y_{l\,i...k}^{s}=\mathop{\displaystyle \sum } K_{l\, i...k\, (s)}^{sm}Y_{lm},
\end{equation*}
where the indices obey $|s|\leq l$, and the number of spatial indices (i.e., 
$i...k$) is equal to $l$. Explicitly, these tensorial spin-weighted
harmonics can be constructed directly from the parametrized Lorentzian null
tetrad, Eq.~(\ref{l.hat})-(\ref{m.hat}):
\begin{eqnarray}
\hat{l}^{a} &=&\frac{\sqrt{2}}{2(1+\zeta \bar{\zeta})}\left( 1+\zeta
\bar{\zeta},\zeta +\bar{\zeta}, i\bar{\zeta}-i\zeta, -1+\zeta \bar{\zeta}\right),
\label{Tets} \\
\hat{n}^{a} &=&\frac{\sqrt{2}}{2(1+\zeta \bar{\zeta})}\left( 1+\zeta \bar{\zeta},-(\zeta +\bar{\zeta}),i\zeta -i\bar{\zeta},1-\zeta \bar{\zeta}\right),  \notag \\
\hat{m}^{a} &=&\frac{\sqrt{2}}{2(1+\zeta \bar{\zeta})}\left( 0,1-\bar{\zeta}^{2},-i(1+\bar{\zeta}^{2}),2\bar{\zeta}\right) ,  \notag \\
P &\equiv &1+\zeta \bar{\zeta}.  \notag
\end{eqnarray}
Taking the spatial parts of their duals, we obtain the one-forms
\begin{eqnarray}
l_{i} &=&\frac{-1}{\sqrt{2}P}\left( \zeta +\bar{\zeta},-i(\zeta -\bar{\zeta}),-1+\zeta \bar{\zeta}\right) ,
\label{Spacialtets} \\
n_{i} &=&\frac{1}{\sqrt{2}P}\left( \zeta +\bar{\zeta},-i(\zeta +\bar{\zeta}),-1+\zeta \bar{\zeta}\right) ,  \notag \\
m_{i} &=&\frac{-1}{\sqrt{2}P}\left( 1-\bar{\zeta}^{2},-i(1+\bar{\zeta}^{2}),2\bar{\zeta}\right) ,  \notag \\
c_{i} &=&l_{i}-n_{i}=-\sqrt{2}i\epsilon_{ijk}m_{j}\bar{m}_{k}.  \notag
\end{eqnarray}
From this we define $Y_{l\, i...k}^{l}$ as~\cite{Spins}
\begin{eqnarray}
Y_{l\, i...k}^{l} &=&m_{i}m_{j}...m_{k},
\label{Harmonics 1} \\
Y_{l\, i...k}^{-l} &=&\bar{m}_{i}\bar{m}_{j}...\bar{m}_{k}.  \notag
\end{eqnarray}
The other harmonics are determined by the action of the $\eth$-operator on
the forms, Eq.~(\ref{Spacialtets}), (with complex conjugates) via
\begin{eqnarray}
\eth l &=&m,
\label{eth.on.tet} \\
\eth m &=&0,  \notag \\
\eth n &=&-m,  \notag \\
\eth c &=&2m,  \notag \\
\eth \overline{m} &=&n-l=-c.  \notag
\end{eqnarray}
Specifically, the spin-$s$ harmonics are defined by
\begin{eqnarray}
Y_{l\, i...k}^{s} &=&\bar{\eth}^{l-s}\left( Y_{l\, i...k}^{l}\right) ,
\label{Harmonics 2} \\
Y_{l\, i...k}^{-|s|} &=&\eth^{l-|s|}\left( Y_{l\, i...k}^{-l}\right).  \notag
\end{eqnarray}

We now present a table of the tensorial spherical harmonics up to $l=2$, in
terms of the tetrad. Higher harmonics can be found in~\cite{Spins}.

\begin{center}
\begin{tabular}{l}
\midrule
$l=0$ \\ \addlinespace[6pt]
$Y_{0}^{0}=1$ \\ \addlinespace[6pt]
\midrule
$l=1$ \\ \addlinespace[6pt]
$Y_{1i}^{1} \phantom{a} =m_{i}$, \\ 
$Y_{1i}^{0} \phantom{a} =-c_{i}$, \\
$Y_{1i}^{-1} =\bar{m}_{i}$ \\ \addlinespace[6pt]
\midrule
$l=2$ \\ \addlinespace[6pt]
$Y_{2ij}^{2}=m_{i}m_{j}$, \quad
$Y_{2ij}^{1}=-\left(c_{i}m_{j}+m_{i}c_{j}\right)$, \quad 
$Y_{2ij}^{0}=3c_{i}c_{j}-2\delta_{ij}$ \\
$Y_{2ij}^{-2}=\bar{m}_{i}\bar{m}_{j}$, \quad 
$Y_{2ij}^{-1}=-\left(c_{i}\bar{m}_{j}+\bar{m}_{i}c_{j}\right)$ \\
\end{tabular}
\end{center}

In addition, it is useful to give the explicit relations between these
different harmonics in terms of the $\eth$-operator and its
conjugate. Indeed, we can see generally that applying $\eth$ once
raises the spin index by one, and applying $\bar{\eth}$ lowers the
index by one. This in turn means that 
\begin{eqnarray*}
\eth Y_{l\, i...k}^{l} &=&0, \\
\bar{\eth}Y_{l\, i...k}^{-l} &=&0.
\end{eqnarray*}
Other relations for $l\leq 2$ are given by
\begin{eqnarray*}
\bar{\eth}Y_{1i}^{1} &=&Y_{1i}^{0}=\eth Y_{1i}^{-1}, \\
\eth Y_{1i}^{0} &=&-2Y_{1i}^{1}, \\
\bar{\eth}Y_{1i}^{0} &=&-2Y_{1i}^{-1},
\end{eqnarray*}

\begin{eqnarray*}
\bar{\eth}Y_{2ij}^{2} &=&Y_{2ij}^{1}, \\
\bar{\eth}^{2}Y_{2ij}^{2} &=&Y_{2ij}^{0}=\eth^{2}Y_{2ij}^{-2}, \\
\eth Y_{2ij}^{0} &=&-6Y_{2ij}^{1}, \\
\eth Y_{2ij}^{1} &=&-4Y_{2ij}^{2}.
\end{eqnarray*}

Finally, due to the nonlinearity of the theory, we
have been forced throughout this review to consider products of the tensorial spin-$s$ spherical
harmonics while expanding nonlinear expressions. These products can be
expanded as a linear combination of individual harmonics using
Clebsch--Gordon expansions. The explicit expansions for products of
harmonics with $l=1$ or $l=2$ are given below (we omit higher products due
to the complexity of the expansion expressions). Further products can be
found in~\cite{Spins,PhysicalContent}.


\subsection{Clebsch--Gordon expansions}

\begin{eqnarray*}
Y_{1i}^{1}Y_{1j}^{0} &=&\frac{i}{\sqrt{2}}\epsilon_{ijk}Y_{1k}^{1}+\frac{1}{2}Y_{2ij}^{1}, \\
Y_{1i}^{1}Y_{1j}^{-1} &=&\frac{1}{3}\delta_{ij}-\frac{i\sqrt{2}}{4}\epsilon
_{ijk}Y_{1k}^{0}-\frac{1}{12}Y_{2ij}^{0}, \\
Y_{1i}^{0}Y_{1j}^{0} &=&\frac{2}{3}\delta_{ij}+\frac{1}{3}Y_{2ij}^{0}
\end{eqnarray*}

\begin{eqnarray*}
Y_{1i}^{1}Y_{2ij}^{2} &=&Y_{3ijk}^{3}, \\
Y_{1i}^{0}Y_{2jk}^{0} &=&-\frac{4}{5}\delta_{kj}Y_{1i}^{0}+\frac{6}{5}\left( \delta_{ij}Y_{1k}^{0}+\delta_{ik}Y_{1j}^{0}\right) +\frac{1}{5}Y_{3ijk}^{0}, \\
Y_{1i}^{1}Y_{2jk}^{0} &=&\frac{2}{5}Y_{1i}^{1}\delta_{jk}-\frac{3}{5}Y_{1j}^{1}\delta_{ik}-\frac{3}{5}Y_{1k}^{1}\delta_{ij}+\frac{i}{\sqrt{2}}\left( \epsilon_{ikl}Y_{2jl}^{1}+\epsilon_{ijl}Y_{2kl}^{1}\right) +\frac{2}{5}Y_{3ijk}^{1}, \\
Y_{1i}^{1}Y_{2jk}^{1} &=&-\frac{1}{6}\eth \left(Y_{1i}^{1}Y_{2jk}^{0}\right) , \\
Y_{2ij}^{-1}Y_{1k}^{1} &=&\frac{3}{10}Y_{1i}^{0}\delta_{jk}+\frac{3}{10}Y_{1j}^{0}\delta_{ik}-\frac{1}{5}Y_{1k}^{0}\delta_{ij}+\frac{i\sqrt{2}}{12}\left( \epsilon_{jkl}Y_{2il}^{0}+\epsilon_{ikl}Y_{2lj}^{0}\right) -\frac{1}{30}Y_{3ijk}^{0}, \\
Y_{1i}^{0}Y_{2jk}^{1} &=&-\frac{2}{5}Y_{1i}^{1}\delta_{jk}+\frac{3}{5}Y_{1j}^{1}\delta_{ik}+\frac{3}{5}Y_{1k}^{1}\delta_{ij}-\frac{i}{3\sqrt{2}}\left( \epsilon_{ikl}Y_{2jl}^{1}+\epsilon_{ijl}Y_{2kl}^{1}\right) +\frac{4}{15}Y_{3ijk}^{1}, \\
Y_{2ij}^{2}Y_{1k}^{-1} &=&\frac{3}{10}Y_{1i}^{0}\delta_{jk}+\frac{3}{10}Y_{1j}^{0}\delta_{ik}-\frac{1}{5}Y_{1k}^{0}\delta_{ij}-\frac{i\sqrt{2}}{12}\left( \epsilon_{jkl}Y_{2il}^{0}+\epsilon_{ikl}Y_{2lj}^{0}\right) -\frac{1}{30}Y_{3ijk}^{0}, \\
Y_{2ij}^{2}Y_{1k}^{0} &=&\eth \left( Y_{2ij}^{2}Y_{1k}^{-1}\right)
\end{eqnarray*}
\newline

The Clebsch--Gordon expansions involving two $l=2$ harmonics have been
used in the text. They are fairly long and are not given here but can be
found in~\cite{Spins}.


\newpage

\section[$\mathcal{H}$-Space Metric]{\boldmath$\mathcal{H}$-Space Metric}
\label{appendixD}


In the following, the derivation of the $\mathcal{H}$-space metric of~(\ref{H metric}) is given.
We begin with the cut function, $u_{\mathrm{B}}=Z(\xi^{a}(\tau),\zeta, \overline{\zeta})=G(\tau, \zeta,\overline{\zeta})$ that satisfies the
good-cut equation $\eth^{2}Z=\sigma^{0} (Z,\zeta, \overline{\zeta})$. The ($\zeta, \overline{\zeta})$ are (for the time being) completely independent
of each other, though $\overline{\zeta}$ is to be treated as being
`close' the complex conjugate of $\zeta$. Taking the gradient of $Z(z^{a},\zeta, \overline{\zeta})$, multiplied by an arbitrary four vector $v^{a}$ (i.e., $V=v^{a}Z,_{a}$), we see that it satisfies the linear good
cut equation,
\begin{eqnarray}
\eth^{2}Z,_{a} &=&\sigma^{0} ,_{Z}Z,_{a}  \label{Linear GCE} \\
\eth^{2}V &=&\sigma^{0} ,_{Z}V.  \nonumber
\end{eqnarray}
Let $V_{0}$ be a particular solution, and assume for the moment that
the general solution can be written as
\begin{equation}
Z,_{a}=V_{0}l_{a}^{\ast}
 \label{assume}
\end{equation}
with the four components of $l_{a}^{\ast}$ to be determined. Substituting
Eq.~(\ref{assume}) into the linearized good-cut equation, we have 
\begin{eqnarray*}
\eth^{2}(V_{0}l_{a}^{\ast}) &=&\sigma^{0} ,_{Z}V_{0}l_{a}^{\ast}, \\
\eth(l_{a}^{\ast}\eth (V_{0})+V_{0}\eth l_{a}^{\ast})
&=&\sigma^{0},_{Z}V_{0}l_{a}^{\ast}, \\
l_{a}^{\ast}\eth^{2}(V_{0})+2\eth V_{0}\eth
l_{a}^{\ast}+V_{0}\eth^{2}l_{a}^{\ast} &=&\sigma^{0} ,_{Z}V_{0}l_{a}^{\ast}, \\
2\eth V_{0}\eth l_{a}^{\ast}+V_{0}\eth^{2}l_{a}^{\ast} &=&0, \\
2V_{0}\eth V_{0}\eth l_{a}^{\ast}+V_{0}^{2}\eth^{2}l_{a}^{\ast} &=&0, \\
\eth V_{0}^{2}\eth l_{a}^{\ast}+V_{0}^{2}\eth^{2}l_{a}^{\ast} &=&0, \\
\eth (V_{0}^{2}\eth l_{a}^{\ast}) &=&0,
\end{eqnarray*}
which integrates immediately to
\begin{equation}
V_{0}^{2}\eth l_{a}^{\ast}=m_{a}^{\ast}  \label{first integral}
\end{equation}
where the $m_{a}^{\ast}$ are three independent $l=1$, $s=1$ functions. 

By taking linear combinations they can be written as
\begin{equation*}
m_{a}^{\ast}=T_{a}^{b}\hat{m}_{b}=T_{a}^{b}\eth \hat{l}_{b}
\end{equation*}
where $\hat{l}_{a}$ is our usual $\hat{l}_{a}=\frac{\sqrt{2}}{2}\left(1,-\frac{\zeta +\overline{\zeta}}{1+\zeta \overline{\zeta}},-\frac{i(\overline{\zeta -}\zeta )}{1+\zeta \overline{\zeta}},\frac{1-\zeta 
\overline{\zeta}}{1+\zeta \overline{\zeta}}\right)$, and the coefficients $T_{a}^{b}$ are functions only of the coordinates $z^{a}$. Assuming that the
monopole term in $V^{2}$ is sufficiently large so that it has no zeros and
then by rescaling $V$ we can write $V^{-2}$ as a monopole plus higher
harmonics in the form
\begin{equation*}
V_{0}^{\ -2}=1+\eth W,
\end{equation*}
where $W$ is a spin-wt $s=-1$ quantity. From Eq.~(\ref{first integral}), we
obtain
\begin{equation*}
\eth l_{a}^{*} = V_{0}^{-2}m^{*}_{a} = (1+\eth W)m_{a}^{*} = m_{a}^{*}+\eth(Wm^{*}_{a}) = T^{b}_{a}\eth\hat{l}_{b}+T^{b}_{a}\eth(W \hat{m}_{b}) =T^{b}_{a}\eth(\hat{l}_{b}+W\hat{m}_{b}),
\end{equation*}
which integrates to
\begin{equation}
l_{a}^{\ast}=T_{a}^{b}(\hat{l}_{a}+W\hat{m}_{a}).
\label{l star}
\end{equation}

The general solution to the linearized good-cut equation is thus
\begin{eqnarray}
Z,_{a} &=&V_{0}l_{a}^{\ast}=V_{0}T_{a}^{b}(\hat{l}_{b}+W\hat{m}_{b}),
\label{linear solution} \\
V &=&v^{a}Z,_{a}=V_{0}v^{a}T_{a}^{b}(\hat{l}_{b}+W\hat{m}_{b}).  \nonumber
\end{eqnarray}

We now demonstrate that
\begin{eqnarray}
(g_{ab}v^{a}v^{b})^{-1} &=&(8\pi )^{-1}\int V^{-2}d\Omega ,   \label{Hmetric}
\\
d\Omega  &=&4i\frac{d\zeta \wedge d\overline{\zeta}}{(1+\zeta 
\overline{\zeta})^{2}}.
\end{eqnarray}

In the integral of~(\ref{Hmetric}), we replace the independent variables $(\zeta,\overline{\zeta})$ by
\begin{equation}
\zeta^{\ast}=\frac{\zeta +W}{1-W\overline{\zeta}},\ \qquad \overline{\zeta}^{\ast}=\overline{\zeta}.
\label{zeta*}
\end{equation}
After some algebraic manipulation we obtain
\begin{equation}
d\Omega^{\ast}=V_{0}^{-2}d\Omega ,
\label{area element}
\end{equation}
and (surprisingly)
\begin{equation}
(\hat{l}_{a}+W\hat{m}_{a})=L_{a}^{\ast} \equiv
  \frac{\sqrt{2}}{2}\left(1,-\frac{\zeta^{\ast}+\overline{\zeta}}{1+\zeta^{\ast}\overline{\zeta}},
  -\frac{i(\overline{\zeta}-\zeta^{\ast})}{1+\zeta^{\ast}\overline{\zeta}},
  \frac{1-\zeta^{\ast}\overline{\zeta}}{1+\zeta^{\ast}\overline{\zeta}}\right),
\label{L*}
\end{equation}
so that
\begin{equation}
V=V_{0}v^{a}T_{a}^{b}L_{b}^{\ast}.
\label{V canonical}
\end{equation}

Inserting Eqs.~(\ref{zeta*}), (\ref{area element}) and (\ref{V canonical})
into (\ref{Hmetric}) we obtain 
\begin{eqnarray}
(g_{ab}v^{a}v^{b})^{-1} &=&(8\pi )^{-1}\int (V_{0}v^{a}T_{a}^{b}L_{b}^{\ast})^{-2}V_{0}^{2}d\Omega^{\ast},  \label{metric*} \\
&=&(8\pi)^{-1}\int (v^{a}T_{a}^{b}L_{b}^{\ast})^{-2}d\Omega^{\ast}, \nonumber \\
&=&(8\pi)^{-1}\int (v^{\ast b}L_{b}^{\ast})^{-2}d\Omega^{\ast}. 
\nonumber
\end{eqnarray}

Using the form Eq.~(\ref{L*}) the last integral can be easily evaluated
(most easily done using  $\theta$ and $\varphi$) leading to
\begin{eqnarray}
(g_{ab}v^{a}v^{b})^{-1} &=&(\eta_{ab}v^{\ast a}v^{\ast b})^{-1}=(T_{a}^{c}T_{b}^{d}\eta_{cd}v^{a}v^{b})^{-1},  \label{metric**} \\
g_{ab} &=&T_{a}^{c}T_{b}^{d}\eta_{cd},  \nonumber
\end{eqnarray}
In particular, note that when $\sigma^{0}=0$ (i.e., the case of an \emph{everywhere} shear-free NGC) $T^{a}_{b}=\delta^{a}_{b}$ and the metric on $\mathcal{H}$-space reduces to the complex Minkowski metric, as claimed throughout the text. We can go a step further by taking the
derivative of Eq.~(\ref{metric*}) with respect to $v^{a}$ to
find the covariant form of $v$, namely 
\begin{equation*}
\frac{v_{a}}{(g_{ab}v^{a}v^{b})^{2}}=\frac{g_{ab}v^{b}}{(g_{ab}v^{a}v^{b})^{2}}=(8\pi )^{-1}\int (v^{a}T_{a}^{b}L_{b}^{\ast})^{-3}T_{a}^{b}L_{b}^{\ast}d\Omega^{\ast}.
\end{equation*}


\newpage

\section{Shear-Free Congruences from Complex World Lines}
\label{appendixE}

In this appendix, we show that the family of complex light-cones with apex on a complex
world line in complex Minkowski space $\mathbb{M}_{\mathbb{C}}$ have null
generators that form a real shear-free null geodesic congruence in real
Minkowski space~\cite{AdamoNewman4}.
\begin{thm}
There exists a mapping from the arbitrary complex-analytic world line $\xi^{a}(\tau)\in\mathbb{M}_{\mathbb{C}}$ to the real shear-free NGC in $\mathbb{M}$ given by complex null displacements.
\end{thm}

\textbf{Proof:} We first recall from Section~\ref{shear-free-NGC} that \textit{regular} real
shear-free NGCs in $\mathbb{M}$ are parametrically given by
\begin{equation}
x^{a}=u_{\mathrm{B}}(\hat{l}^{a}+\hat{n}^{a})-L\overline{\hat{m}}^{a}-\bar{L}\hat{m}^{a}+(r^{\ast}-r_{0})\hat{l}^{a}
\label{shearfreeNGC*}
\end{equation}
with
\begin{eqnarray}
u_{\mathrm{B}} &=&\xi^{b}(\tau)\widehat{l}_{b},\ \ \ \tau =T(u_{\mathrm{B}},\zeta,\bar{\zeta})  \label{standard} \\
L(u_{\mathrm{B}},\zeta,\bar{\zeta}) &=&\xi^{a}(\tau)\hat{m}_{a}(\zeta,\bar{\zeta}),  \notag \\
\overline{L}(u_{\mathrm{B}},\zeta,\bar{\zeta}) &=&\overline{\xi}^{a}(\tau)\overline{\hat{m}}_{a}(\zeta,\bar{\zeta}).  \notag
\end{eqnarray}
The $\tau$ is taken so that $u_{\mathrm{B}}$ is real via Eq.~(\ref{real-tau}): $\tau =s+i\Lambda (s,\zeta,\bar{\zeta})$, and $r_{0}$ is the arbitrary
origin for the affine parameter along each geodesic of the congruence.

Beginning with the world line, $\xi^{a}(\tau)$, we add to it a specific
complex null displacement (to be constructed) 
\begin{eqnarray*}
L^{a} &=&L_{0}^{a}(u_{\mathrm{B}},\zeta,\bar{\zeta})+rL_{1}^{a}(u_{\mathrm{B}},\zeta,\bar{\zeta}),  \\
L^{a}L_{a} &=&L_{1}^{a}L_{1a}=L_{0}^{a}L_{0a}=0
\end{eqnarray*}
parametrized by the real variable $r$. We will show that the curve (complex null
geodesic) given by
\begin{equation*}
x^{a}=\xi^{a}(T(u_{\mathrm{B}},\zeta,\bar{\zeta}))+L_{0}^{a}(u_{\mathrm{B}},\zeta,\bar{\zeta})+rL_{1}^{a}(u_{\mathrm{B}},\zeta,\bar{\zeta})
\end{equation*}
with fixed $(u_{\mathrm{B}},\zeta,\bar{\zeta})$ but varying $r,$ is
identical to that given by Eq.~(\ref{shearfreeNGC*}).

This is demonstrated by taking the world line, $\xi^{a}(\tau)$, written in
terms of its components ($\xi^{b}l_{b},\xi^{b}n_{b},\xi^{b}m_{b},\xi^{b}\overline{m}_{b}$) as
\begin{equation}
\xi^{a}(\tau)=\xi^{b}(\tau)\hat{l}_{b}\hat{n}^{a}+\xi^{b}(\tau)\widehat{n}_{b}\hat{l}^{a}-\xi^{b}(\tau)\hat{m}_{b}\overline{\hat{m}}^{a}-\xi^{b}(\tau)\overline{\hat{m}}_{b}\hat{m}^{a}
\label{2***}
\end{equation}
and replacing the $n^{a}$ by the identity
\begin{eqnarray*}
\hat{n}^{a} &=&\sqrt{2}t^{a}-\hat{l}^{a}. \\
t^{a} &=&\delta_{0}^{a}.
\end{eqnarray*}
This yields
\begin{equation}
\xi^{a}(\tau)=\sqrt{2}\xi^{b}(\tau)\hat{l}_{b}t^{a}+\sqrt{2}\xi^{b}(\tau)t_{b}\hat{l}^{a}-2\xi^{b}(\tau)\hat{l}_{b}\widehat{l}^{a}-\xi^{b}(\tau)\hat{m}_{b}\overline{\hat{m}}^{a}-\xi^{b}(\tau)\overline{\hat{m}}_{b}\hat{m}^{a}.
\label{2}
\end{equation}
Using the relations, Eq.~(\ref{Ltwiddle}), etc.,
\begin{eqnarray*}
u_{\mathrm{B}} &=&\xi^{b}(\tau)\hat{l}_{b},\ \ \ \tau =T(u_{\mathrm{B}},\zeta,\bar{\zeta}) \\
L(u_{\mathrm{B}},\zeta,\bar{\zeta}) &=&\xi^{b}(\tau)\hat{m}_{b} \\
\widetilde{L}(u_{\mathrm{B}},\zeta,\bar{\zeta}) &=&\xi^{a}(\tau)\overline{\hat{m}}_{a}(\zeta,\bar{\zeta}),
\end{eqnarray*}
Eq.~(\ref{2}) becomes
\begin{equation}
\xi^{a}(\tau)=\sqrt{2}u_{\mathrm{B}}t^{a}+\sqrt{2}\xi^{0}(\tau)\hat{l}^{a}-2u_{\mathrm{B}}\hat{l}^{a}-L\overline{\hat{m}}^{a}-\tilde{L}\hat{m}^{a}.
\label{2&}
\end{equation}

By adding the complex null vector (displacement),  
\begin{equation*}
L^{a}=(\tilde{L}-\overline{L})\hat{m}^{a}+(r-r_{0}+2u_{\mathrm{B}}-\sqrt{2}\xi^{0}(\tau))\hat{l}^{a}
\end{equation*}
to both sides of Eq.~(\ref{2&}), we obtain
\begin{eqnarray*}
x^{a} &\equiv &\xi^{a}(\tau)+L^{a} \\
x^{a} &=&\xi^{a}(\tau)+(\tilde{L}-\overline{L})\hat{m}^{a}+\left(r-r_{0}+2u_{\mathrm{B}}-\sqrt{2}\xi^{0}(\tau)\right)\hat{l}^{a} \\
x^{a} &=&u_{\mathrm{B}}t^{a}-L\overline{\hat{m}}^{a}-\overline{L}\hat{m}^{a}+(r-r_{0})\hat{l}^{a}
\end{eqnarray*}
To complete our task we now restrict the values of $\tau$ to those that
produce a real $u$, namely 
\begin{equation*}
\tau \rightarrow \tau^{(\mathrm{R})}=s+i\Lambda (s,\zeta,\bar{\zeta})
\end{equation*}
and restrict $r$ to the real.

We see that by adding a null ray, combinations of $\hat{m}^{a}$ and $\hat{l}^{a}$, directly to the complex world line $\xi^{a}(\tau)$, we obtain a
mapping of the complex world line directly to the real shear-free NGC,
Eq.~(\ref{shearfreeNGC*}). Note that when the affine parameter, $r$,
is chosen (complex) as $r=r_{0}-2u+\sqrt{2}\xi^{0}(\tau)$ the $\hat{l}^{a}$ term drops out and
we have the `point' $\xi^{a}(\tau)$ surrounded by the embedded complex
sphere, $z^{a}=\xi^{a}(\tau)+L_{0}^{a}(u,\zeta,\bar{\zeta})=\xi^{a}(\tau
)+(\tilde{L}-\overline{L})\hat{m}^{a}$. The ray can be thought of as having
its origin on this surface.

We thus have the explicit relationship between the complex world line and the
shear-free NGC, completing the proof.

\newpage


\section{The Generalized Good-Cut Equation}
\label{appendixF}

Throughout this work, the Good-Cut Equation (GCEq) has played a major role in allowing us to study shear-free and asymptotically shear-free NGCs in asymptotically flat spacetimes.  In this context, the GCEq is a partial differential equation on a topologically $S^2$ cut of $\mathfrak{I}^{+}$; due to the freedom in the choice of conformal factor on the two-sphere in the conformal compactification of asymptotically flat spacetimes, we can always take the space of null generators of $\mathfrak{I}^{+}$ to be a \emph{metric} two-spheres.  However, one can imagine solving the GCEq on a surface which is only conformal to a metric two-sphere, we refer to such a PDE as the `Generalized' GCEq, or G$^2$CEq for short.  In this appendix, we briefly motivate why one could be interested in the G$^2$CEq, and then prove that it can be reduced to the GCEq on the metric two-sphere by a coordinate transformation (this is essentially a proof of the conformal invariance of the GCEq)~\cite{Adamo:2010ey}.

The study of horizons in the interior of spacetime is an important topic in a variety of areas, particularly quantum gravity.  One interesting class of null horizons are the so-called `vacuum non-expanding horizons', which are null 3-surfaces in a spacetime that have vanishing divergence and shear, and are topologically $\mathbb{R}\times S^{2}$~\cite{Ashtekar:2001jb,Ashtekar:2004cn}.  In analogy with the setting on $\mathcal{I}^{+}$ discussed in the body of this review, one can look for null geodesic congruences in the interior of a spacetime which have vanishing shear at their intersection with a vacuum non-expanding horizon.  It has been shown that such `horizon-shear-free' NGCs are described, where they `cut' the horizon, by a good-cut equation on the topologically $S^2$ cut.  Since we cannot freely rescale objects in the interior of the spacetime, this means that horizon-shear-free NGCs are described by the G$^2$CEq~\cite{Adamo:2009cd}.

Consider an arbitrary vacuum non-expanding horizon $\mathfrak{H}$ with
associated G$^2$CEq.  As in the asymptotic case, we consider the
complexification $\mathfrak{H}_{\mathbb{C}}$ of the horizon when
looking for solutions to the G$^2$CEq, and make use of local
Bondi-like coordinates $(u,\zeta,\overline{\zeta})$.  The
($\zeta,\overline{\zeta}$), which label the null generators of
$\mathfrak{H,}$ are the stereographic coordinates on the $S^{2}$
portion of $\mathfrak{H}$ ($S^{2}$ need not be a metric sphere); while
the coordinate $u$ parametrizes the cross-sections of $\mathfrak{H.}$
For $\mathfrak{H}_{\mathbb{C}}$, the $u$ is allowed to take complex values close to the real,
while $\overline{\zeta}$ goes over to an independent variable \textit{close} to the complex conjugate of $\zeta$. The context should make it clear when $\overline{\zeta}$ is actually the complex conjugate of $\zeta$. The
distinction between the GCEq and the G$^2$CEq is that the former lives
on a 3-surface $\mathfrak{H}$ whose $u=constant$ cross-sections are metric
spheres, while for the latter equation the 2-surface metric is arbitrary.

As mentioned earlier, the 3-surface $\mathfrak{H}$ is described by an $S^{2}$ worth of null geodesics with the cross sections given by $u$~=~constant. The
metric of the two-surface cross-sections are expressed in stereographic
coordinates ($\zeta,\overline{\zeta}$) so that the metric takes the
conformally flat form:
\begin{equation}
ds^{2}=\frac{4d\zeta d\overline{\zeta}}{P^{2}(\zeta,\overline{\zeta})},
\label{conformally flat}
\end{equation}
with $P(u,\zeta,\overline{\zeta})$ an arbitrary smooth non-vanishing
function on the $(\zeta,\overline{\zeta})$-sphere, the extended complex
plane (Riemann sphere). In the special case of a \textit{metric sphere} we
take 
\begin{equation*}
P=P_{0}\equiv 1+\zeta \overline{\zeta},
\end{equation*}
while in general we write 
\begin{equation}
P=V(u,\zeta,\overline{\zeta})P_{0}.
\label{VP}
\end{equation}
The G$^2$CEq contains the general $P$, while the special case using 
$P_{0}$ yields the GCEq.

For the most general situation, the G$^2$CEq can be written as a
differential equation for the function $u=G(\zeta,\overline{\zeta})$:
\begin{equation}
\overline{\eth}^{2}G\equiv \partial_{_{\overline{\zeta}}}(V^{2}P_{0}^{2}\partial_{\overline{\zeta}}G)=\overline{\sigma} (G,\zeta,\overline{\zeta}),
\label{G2CEq.I}
\end{equation}
or
\begin{equation}
P_{0}^{2}\partial_{\overline{\zeta}}^{2}G+2[P_{0}^{2}V^{-1}\partial_{_{\overline{\zeta}}}V+P_{0}\zeta]\partial_{\overline{\zeta}}G=V^{-2}\overline{\sigma} (G,\zeta,\overline{\zeta}).
\label{G2CEq}
\end{equation}
When $V=1$ we have the GCEq:
\begin{equation*}
\overline{\eth}_{0}^{2}G\equiv \partial_{_{\overline{\zeta}}}(P_{0}^{2}\partial_{\overline{\zeta}}G)=\overline{\sigma} (G,\zeta,\overline{\zeta}). 
\end{equation*}
When the arbitrary spin-weight-2 function, $\overline{\sigma} (G,\zeta,\bar{\zeta})$ vanishes, we have the homogeneous G$^{2}$CEq:
\begin{equation}
\partial_{_{\overline{\zeta}}}(P^{2}\partial_{\overline{\zeta}}G)=0.
\label{hGCEq}
\end{equation}

It is now shown how, by a coordinate transformation of the (independent)
complex stereographic coordinates ($\zeta,\overline{\zeta}$), G$^2$CEq
can be transformed into the GCEq. It must be remembered from our notation
that $\overline{\zeta}^{\ast}$ (or $\overline{\zeta})$ is close to, but
is not necessarily, the complex conjugate of $\zeta^{\ast}$ (or $\zeta$).

First rewrite the GCEq with stereographic coordinates
($\zeta^{\ast},\overline{\zeta}^{\ast}$) as
\begin{eqnarray}
\overline{\eth}_{0\ast}^{2}G &=&\partial_{_{\overline{\zeta}^{\ast}}}(P_{0}^{\ast 2}\partial_{_{\overline{\zeta}^{\ast}}}G)=\overline{\sigma}^{\ast}(G,\zeta^{\ast},\overline{\zeta}^{\ast}),  \label{C} \\
P_{0}^{\ast} &=&1+\zeta^{\ast}\overline{\zeta}^{\ast},  \label{P*}
\end{eqnarray}
and the G$^2$CEq as
\begin{equation}
\overline{\eth}^{2}G=\partial_{_{\overline{\zeta}}}(V^{2}P_{0}^{2}\partial_{\overline{\zeta}}G)=\overline{\sigma} (G,\zeta,\overline{\zeta}).
\label{D**}
\end{equation}

We now apply the coordinate transformation
\begin{eqnarray}
\overline{\zeta}^{\ast} &=&\frac{\overline{\zeta}+W}{1-W\zeta}\equiv
N(\zeta,\overline{\zeta}),   \label{CT} \\
\zeta^{\ast} &=&\zeta,
\end{eqnarray}
with $W$ (a spin-weight 1 function) defined from
\begin{eqnarray*}
V^{-2} &=&1+\overline{\eth}_{0}W=1+P_{0}\partial_{_{\overline{\zeta}}}W-W\zeta, \\
P_{0} &=&1+\zeta \overline{\zeta},
\end{eqnarray*}
to Eq.~(\ref{D**}). Substituting the derived relations,
\begin{eqnarray*}
P_{0}^{\ast} &=&1+\zeta \overline{\zeta}^{\ast}=\frac{1+\zeta \overline{\zeta}}{1-W\zeta}=\frac{P_{0}}{1-W\zeta}, \\
\partial_{_{\overline{\zeta}}}G &=&\partial_{_{\overline{\zeta}^{\ast}}}G\cdot \partial_{_{\overline{\zeta}}}N, \\
\partial_{_{\overline{\zeta}}}^{2}G &=&\partial_{_{\overline{\zeta}^y{\ast}}}^{2}G\cdot (\partial_{_{\overline{\zeta}}}N)^{2}+\partial_{_{\overline{\zeta}^{\ast}}}G\cdot \partial_{_{\overline{\zeta}}}^{2}N, \\
\partial_{_{\overline{\zeta}}}N &=&\frac{V^{-2}-W\zeta}{(1-W\zeta)^{2}},
\\
\partial_{_{\overline{\zeta}}}^{2}N &=&\frac{2\zeta\lbrack
V^{-2}-1]\partial_{_{\overline{\zeta}}}W}{(1-W\zeta)^{3}}+\frac{\zeta
\partial_{_{\overline{\zeta}}}W}{(1-W\zeta)^{2}}-\frac{2V^{-3}\partial_{_{\overline{\zeta}}}V}{(1-W\zeta)^{2}},
\end{eqnarray*}
into Eq.~(\ref{D**}), we have, after a bit of algebra,
\begin{eqnarray*}
\overline{\eth}_{0\ast}^{2}G &=&\partial_{_{\overline{\zeta}^{\ast}}}(P_{0}^{\ast 2}\partial_{_{\overline{\zeta}^{\ast}}}G) \\
&=&F(\zeta^{\ast},\overline{\zeta}^{\ast})\sigma (G,\zeta (\zeta^{\ast},\overline{\zeta}^{\ast}),\overline{\zeta}^{\ast}(\zeta^{\ast},\overline{\zeta}^{\ast})) \\
&\equiv &\overline{\sigma}^{\ast}(G,\zeta^{\ast},\overline{\zeta}^{\ast}),
\end{eqnarray*}
namely Eq.~(\ref{C}), the GCEq.

Hence, we see that the G$^2$CEq is equivalent to the GCEq via the
coordinate transformation (\ref{CT}). This means that the study of the G$^2$CEq on a general 3-surface $\mathfrak{H}$ can be reduced to the study
of the properties of the GCEq on a 3-surface whose cross-sections are metric
spheres.

\begin{remark}
As in the main text, solutions to the GCEq or G$^2$CEq, $u=G(\zeta,\bar{\zeta})$, known as `good-cut functions',
describe cross-sections of $\mathfrak{H}$ that are referred to as `good
cuts.' From the tangents to these good cuts, $L=\overline{\eth}G$, one can
construct null directions (pointing out of $\mathfrak{H}$) into the
spacetime itself that determine a NGC whose shear vanishes at $\mathfrak{H}$.
\end{remark}


\newpage

\bibliography{refs}

\begin{thebibliography}{10}

\bibitem{Adamo:2011pv}
Adamo, T., Bullimore, M., Mason, L., and Skinner, D., ``Scattering amplitues
  and {W}ilson loops in twistor space'', {\em J. Phys.A}, {\bf A44}, 454008,
  (2011).
  {\small[\href{http://dx.doi.org/10.1088/1751-8113/44/45/454008}{DOI}]},
  {\small[\href{http://arxiv.org/abs/arXiv:1104.2890}{{arXiv:1104.2890}}]}.

\bibitem{AdamoNewman1}
Adamo, T.M., and Newman, E.T., ``The gravitational field of a radiating
  electromagnetic dipole'', {\em Class. Quantum Grav.}, {\bf 25}, 245005,
  (2008).
  {\small[\href{http://dx.doi.org/10.1088/0264-9381/25/24/245005}{DOI}]},
  {\small[\href{http://arxiv.org/abs/arXiv:0807.3537}{{arXiv:0807.3537}}]}.

\bibitem{AdamoNewman3}
Adamo, T.M., and Newman, E.T., ``Asymptotically stationary and static
  spacetimes and shear free null geodesic congruences'', {\em Class. Quantum
  Grav.}, {\bf 26}, 155003, (2009).
  {\small[\href{http://dx.doi.org/10.1088/0264-9381/26/15/155003}{DOI}]},
  {\small[\href{http://arxiv.org/abs/arXiv:0906.2409}{{arXiv:0906.2409}}]}.

\bibitem{AdamoNewman2}
Adamo, T.M., and Newman, E.T., ``Electromagnetically induced gravitational
  perturbations'', {\em Class. Quantum Grav.}, {\bf 26}, 015004, (2009).
  {\small[\href{http://dx.doi.org/10.1088/0264-9381/26/1/015004}{DOI}]},
  {\small[\href{http://arxiv.org/abs/arXiv:0807.3671}{{arXiv:0807.3671}}]}.

\bibitem{Adamo:2009cd}
Adamo, T.M., and Newman, E.T., ``Vacuum non-expanding horizons and shear-free
  null geodesic congruences'', {\em Class. Quantum Grav.}, {\bf 26}, 235012,
  (2009).
  {\small[\href{http://dx.doi.org/10.1088/0264-9381/26/23/235012}{DOI}]},
  {\small[\href{http://arxiv.org/abs/arXiv:0908.0751}{{arXiv:0908.0751}}]}.

\bibitem{Adamo:2010ey}
Adamo, T.M., and Newman, E.T., ``The Generalized Good Cut Equation'', {\em
  Class. Quantum Grav.}, {\bf 27}, 245004, (2010).
  {\small[\href{http://dx.doi.org/10.1088/0264-9381/27/24/245004}{DOI}]},
  {\small[\href{http://arxiv.org/abs/arXiv:1007.4215}{{arXiv:1007.4215}}]}.

\bibitem{AdamoNewman4}
Adamo, T.M., and Newman, E.T., ``The real meaning of complex Minkowski-space
  world-lines'', {\em Class. Quantum Grav.}, {\bf 27}, 075009, (2010).
  {\small[\href{http://dx.doi.org/10.1088/0264-9381/27/7/075009}{DOI}]},
  {\small[\href{http://arxiv.org/abs/arXiv:0911.4205}{{arXiv:0911.4205}}]}.

\bibitem{Adamo:2011cx}
Adamo, T.M., and Newman, E.T., ``Light cones in relativity: Real, complex and
  virtual, with applications'', {\em Phys.Rev.}, {\bf D83}, 044023, (2011).
  {\small[\href{http://dx.doi.org/10.1103/PhysRevD.83.044023}{DOI}]},
  {\small[\href{http://arxiv.org/abs/arXiv:1101.1052}{{arXiv:1101.1052}}]}.

\bibitem{Aharony:1999ti}
Aharony, O., Gubser, S.~S., Maldacena, J.~M., Ooguri, H., and Oz, Y., ``Large
  {N} field theories, string theory and gravity'', {\em Phys.Rept.}, {\bf 323},
  183--386, (2000).
  {\small[\href{http://dx.doi.org/10.1016/S0370-1573(99)00083-6}{DOI}]},
  {\small[\href{http://arxiv.org/abs/hep-th/9905111}{{hep-th/9905111}}]}.

\bibitem{Arcioni:2003xx}
Arcioni, G., and Dappiaggi, C., ``Exploring the holographic principle in
  asymptotically flat spacetimes via the {B}{M}{S} group'', {\em Nucl. Phys.},
  {\bf B674}, 553--592, (2003).
  {\small[\href{http://dx.doi.org/10.1016/j.nuclphysb.2003.09.051}{DOI}]},
  {\small[\href{http://arxiv.org/abs/hep-th/0306142}{{hep-th/0306142}}]}.

\bibitem{Spacelike1}
Arnowitt, R., Deser, S., and Misner, C.W., ``Energy and the Criteria for
  Radiation in General Relativity'', {\em Phys. Rev.}, {\bf 118}, 1100--1104,
  (1960). {\small[\href{http://dx.doi.org/10.1103/PhysRev.118.1100}{DOI}]},
  {\small[\href{http://adsabs.harvard.edu/abs/1960PhRv..118.1100A}{ADS}]}.

\bibitem{Aronson}
Aronson, B., and Newman, E.T., ``Coordinate systems associated with
  asymptotically shear-free null congruences'', {\em J. Math. Phys.}, {\bf 13},
  1847--1851, (1972).
  {\small[\href{http://dx.doi.org/10.1063/1.1665919}{DOI}]}.

\bibitem{Ashtekar:2001jb}
Ashtekar, A., Beetle, C., and Lewandowski, J., ``Geometry of generic isolated
  horizons'', {\em Class. Quantum Grav.}, {\bf 19}, 1195--1225, (2002).
  {\small[\href{http://dx.doi.org/10.1088/0264-9381/19/6/311}{DOI}]},
  {\small[\href{http://arxiv.org/abs/gr-qc/0111067}{{gr-qc/0111067}}]}.

\bibitem{Ashtekar:2004cn}
Ashtekar, A., and Krishnan, B., ``Isolated and dynamical horizons and their
  applications'', {\em Living Rev. Relativity}, {\bf 7}, lrr-2004-10, (2004).
  URL (cited on 28 April 2011):
  \newline\url{http://www.livingreviews.org/lrr-2004-10}.

\bibitem{Bergmann}
Bergmann, P.G., ``Non-Linear Field Theories'', {\em Phys. Rev.}, {\bf 75},
  680--685, (1949).
  {\small[\href{http://dx.doi.org/10.1103/PhysRev.75.680}{DOI}]},
  {\small[\href{http://adsabs.harvard.edu/abs/1949PhRv...75..680B}{ADS}]}.

\bibitem{Bondi}
Bondi, H., van~der Burg, M.G.J., and Metzner, A.W.K., ``Gravitational Waves in
  General Relativity. VII. Waves from Axi-Symmetric Isolated Systems'', {\em
  Proc. R. Soc. London, Ser. A}, {\bf 269}, 21--52, (1962).
  {\small[\href{http://dx.doi.org/10.1098/rspa.1962.0161}{DOI}]},
  {\small[\href{http://adsabs.harvard.edu/abs/1962RSPSA.269...21B}{ADS}]}.

\bibitem{Bousso:2002ju}
Bousso, R., ``The holographic principle'', {\em Rev. Mod. Phys.}, {\bf 74},
  825--874, (2002).
  {\small[\href{http://dx.doi.org/10.1103/RevModPhys.74.825}{DOI}]},
  {\small[\href{http://arxiv.org/abs/hep-th/0203101}{{hep-th/0203101}}]}.

\bibitem{Bramson}
Bramson, B., ``Do electromagnetic waves harbour gravitational waves?'', {\em
  Proc. R. Soc. London, Ser. A}, {\bf 462}, 1987--2000, (2006).
  {\small[\href{http://dx.doi.org/10.1098/rspa.2006.1658}{DOI}]}.

\bibitem{Bramson:1975}
Bramson, B.~D., ``Relativistic Angular Momentum for Asymptotically Flat
  Einstein-Maxwell Manifolds'', {\em Proc. R. Soc. London, Ser. A}, {\bf 341},
  463--490, (1975).
  {\small[\href{http://dx.doi.org/10.1098/rspa.1975.0004}{DOI}]}.

\bibitem{HelmutII}
Chru{\'{s}}ciel, P.T., and Friedrich, H., eds., {\em The Einstein Equations and
  the Large Scale Behavior of Gravitational Fields: 50 Years of the Cauchy
  Problem in General Relativity}, (Birkh\"auser, Basel; Boston, 2004).
  {\small[\href{http://books.google.com/books?id=t0D_YmZpTucC}{Google Books}]}.

\bibitem{Corvino}
Corvino, J., and Schoen, R.M., ``On the asymptotics for the vacuum Einstein
  constraint equations'', {\em J. Differ. Geom.}, {\bf 73}, 185--217, (2006).
  {\small[\href{http://arxiv.org/abs/gr-qc/0301071}{{gr-qc/0301071}}]}.

\bibitem{Dragomir:2006}
Dragomir, S., and Tomassini, G., {\em Differential Geometry and Analysis on CR
  Manifolds}, (Birkh\"{a}user, Boston; Basel; Berlin, 2006), 1st edition.

\bibitem{FrauendienerLR}
Frauendiener, J., ``Conformal Infinity'', {\em Living Rev. Relativity}, {\bf
  7}, lrr-2004-1, (2004). URL (cited on 31 July 2009):
  \newline\url{http://www.livingreviews.org/lrr-2004-1}.

\bibitem{Helmut}
Friedrich, H., ``On the Existence of $n$-Geodesically Complete or Future
  Complete Solutions of Einstein's Field Equations with Smooth Asymptotic
  Structure'', {\em Commun. Math. Phys.}, {\bf 107}, 587--609, (1986).
  {\small[\href{http://dx.doi.org/10.1007/BF01205488}{DOI}]}.

\bibitem{Tait}
Frittelli, S., Kozameh, C.N., Newman, E.T., Rovelli, C., and Tate, R.S.,
  ``Fuzzy spacetime from a null-surface version of general relativity'', {\em
  Class. Quantum Grav.}, {\bf 14}, A143--A154, (1997).
  {\small[\href{http://dx.doi.org/10.1088/0264-9381/14/1A/012}{DOI}]},
  {\small[\href{http://arxiv.org/abs/gr-qc/9603061}{{gr-qc/9603061}}]}.

\bibitem{Reps}
Frittelli, S., and Newman, E.T., ``Pseudo-Minkowskian coordinates in
  asymptotically flat space-times'', {\em Phys. Rev. D}, {\bf 55}, 1971--1976,
  (1997). {\small[\href{http://dx.doi.org/10.1103/PhysRevD.55.1971}{DOI}]},
  {\small[\href{http://adsabs.harvard.edu/abs/1997PhRvD..55.1971F}{ADS}]}.

\bibitem{GGV}
Gel'fand, I.M., Graev, M.I., and Vilenkin, N.Y., {\em Generalized Functions,
  Vol. 5: Integral geometry and representation theory}, (Academic Press, New
  York; London, 1966).

\bibitem{Edth}
Goldberg, J.N., Macfarlane, A.J., Newman, E.T., Rohrlich, F., and Sudarshan,
  E.C.G., ``Spin-$s$ Spherical Harmonics and $\eth$'', {\em J. Math. Phys.},
  {\bf 8}, 2155--2161, (1967).
  {\small[\href{http://dx.doi.org/10.1063/1.1705135}{DOI}]}.

\bibitem{GoldbergSachs}
Goldberg, J.N., and Sachs, R.K., ``A Theorem on Petrov Types'', {\em Acta Phys.
  Pol.}, {\bf 22}, 13--23, (1962). Republished as 10.1007/s10714-008-0722-5.

\bibitem{Hallidy:1979}
Hallidy, W., and Ludvigsen, M., ``Momentum and Angular Momentum in the H-space
  of Asymptotically flat, Einstein-Maxwell Space-times'', {\em Gen. Relativ.
  Gravit.}, {\bf 10}, 7--30, (1979).
  {\small[\href{http://dx.doi.org/10.1007/BF00757019}{DOI}]}.

\bibitem{HansenNewman}
Hansen, R.O., and Newman, E.T., ``A complex Minkowski space approach to
  twistors'', {\em Gen. Relativ. Gravit.}, {\bf 6}, 361--385, (1975).
  {\small[\href{http://dx.doi.org/10.1007/BF00761970}{DOI}]}.

\bibitem{PropHspace}
Hansen, R.O., Newman, E.T., Penrose, R., and Tod, K.P., ``The Metric and
  Curvature Properties of $\mathcal{H}$-Space'', {\em Proc. R. Soc. London,
  Ser. A}, {\bf 363}, 445--468, (1978).
  {\small[\href{http://dx.doi.org/10.1098/rspa.1978.0177}{DOI}]},
  {\small[\href{http://adsabs.harvard.edu/abs/1978RSPSA.363..445H}{ADS}]}.

\bibitem{Harmonics1}
Held, A., Newman, E.T., and Posadas, R., ``The Lorentz Group and the Sphere'',
  {\em J. Math. Phys.}, {\bf 11}, 3145--3154, (1970).
  {\small[\href{http://dx.doi.org/10.1063/1.1665105}{DOI}]}.

\bibitem{Hill:2008}
Hill, C.~D., Lewandowski, J., and Nurowski, P, ``Einstein's equations and the
  embedding of 3-dimensional {C}{R} manifolds'', {\em Indiana Univ. Math. J.},
  {\bf 57}, 3131--3176, (2008).
  {\small[\href{http://dx.doi.org/10.1512/iumj.2008.57.3473}{DOI}]},
  {\small[\href{http://arxiv.org/abs/arXiv:0709.3660}{{arXiv:0709.3660}}]}.

\bibitem{HuggettTod}
Hugget, S.A., and Tod, K.P., {\em An Introduction to Twistor Theory}, London
  Mathematical Society Student Texts, vol.~4, (Cambridge University Press,
  Cambridge; New York, 1994), 2nd edition.
  {\small[\href{http://books.google.com/books?id=bA1hlTPH0Y0C}{Google Books}]}.

\bibitem{Greens}
Ivancovich, J., Kozameh, C.N., and Newman, E.T., ``Green's functions of the edh
  operators'', {\em J. Math. Phys.}, {\bf 30}, 45--52, (1989).
  {\small[\href{http://dx.doi.org/10.1063/1.528587}{DOI}]}.

\bibitem{HspaceR}
Ko, M., Newman, E.T., and Tod, K.P., ``$\mathcal{H}$-Space and Null Infinity'',
  in Esposito, F.P., and Witten, L., eds., {\em Asymptotic Structure of
  Space-Time}, Proceedings of a Symposium on Asymptotic Structure of Space-Time
  (SOASST), held at the University of Cincinnati, Ohio, June 14\,--\,18, 1976,
  pp. 227--271, (Plenum Press, New York, 1977).

\bibitem{EDRad}
Kozameh, C.N., and Newman, E.T., ``Electromagnetic dipole radiation fields,
  shear-free congruences and complex centre of charge world lines'', {\em
  Class. Quantum Grav.}, {\bf 22}, 4667--4678, (2005).
  {\small[\href{http://dx.doi.org/10.1088/0264-9381/22/22/002}{DOI}]},
  {\small[\href{http://arxiv.org/abs/gr-qc/0504093}{{gr-qc/0504093}}]}.

\bibitem{Footprints}
Kozameh, C.N., and Newman, E.T., ``The large footprints of H-space on
  asymptotically flat spacetimes'', {\em Class. Quantum Grav.}, {\bf 22},
  4659--4665, (2005).
  {\small[\href{http://dx.doi.org/10.1088/0264-9381/22/22/001}{DOI}]},
  {\small[\href{http://arxiv.org/abs/gr-qc/0504022}{{gr-qc/0504022}}]}.

\bibitem{UCF}
Kozameh, C.N., Newman, E.T., Santiago-Santiago, J.G., and Silva-Ortigoza, G.,
  ``The universal cut function and type II metrics'', {\em Class. Quantum
  Grav.}, {\bf 24}, 1955--1979, (2007).
  {\small[\href{http://dx.doi.org/10.1088/0264-9381/24/8/004}{DOI}]},
  {\small[\href{http://arxiv.org/abs/gr-qc/0612004}{{gr-qc/0612004}}]}.

\bibitem{RTmetrics}
Kozameh, C.N., Newman, E.T., and Silva-Ortigoza, G., ``On the physical meaning
  of the Robinson--Trautman--Maxwell fields'', {\em Class. Quantum Grav.}, {\bf
  23}, 6599--6620, (2006).
  {\small[\href{http://dx.doi.org/10.1088/0264-9381/23/23/002}{DOI}]},
  {\small[\href{http://arxiv.org/abs/gr-qc/0607074}{{gr-qc/0607074}}]}.

\bibitem{PhysicalContent}
Kozameh, C.N., Newman, E.T., and Silva-Ortigoza, G., ``On extracting physical
  content from asymptotically flat spacetime metrics'', {\em Class. Quantum
  Grav.}, {\bf 25}, 145001, (2008).
  {\small[\href{http://dx.doi.org/10.1088/0264-9381/25/14/145001}{DOI}]},
  {\small[\href{http://arxiv.org/abs/arXiv:0802.3314}{{arXiv:0802.3314}}]}.

\bibitem{LL}
Landau, L.D., and Lifshitz, E.M., {\em The classical theory of fields},
  (Pergamon Press; Addison-Wesley, Oxford; Reading, MA, 1962), 2nd edition.

\bibitem{CR1}
Lewandowski, J., and Nurowski, P., ``Algebraically special twisting
  gravitational fields and CR structures'', {\em Class. Quantum Grav.}, {\bf
  7}, 309--328, (1990).
  {\small[\href{http://dx.doi.org/10.1088/0264-9381/7/3/007}{DOI}]}.

\bibitem{CR2}
Lewandowski, J., Nurowski, P., and Tafel, J., ``Einstein's equations and
  realizability of CR manifolds'', {\em Class. Quantum Grav.}, {\bf 7},
  L241--L246, (1990).
  {\small[\href{http://dx.doi.org/10.1088/0264-9381/7/11/003}{DOI}]}.

\bibitem{Lind}
Lind, R.W., ``Shear-free, twisting Einstein-Maxwell metrics in the
  Newman-Penrose formalism'', {\em Gen. Relativ. Gravit.}, {\bf 5}, 25--47,
  (1974). {\small[\href{http://dx.doi.org/10.1007/BF00758073}{DOI}]}.

\bibitem{Maldacena:1997re}
Maldacena, J.~M., ``The large {N} limit of superconformal field theories and
  supergravity'', {\em Adv.Theor.Math.Phys.}, {\bf 2}, 231--252, (1998).
  {\small[\href{http://dx.doi.org/10.1023/A:1026654312961}{DOI}]},
  {\small[\href{http://arxiv.org/abs/hep-th/9711200}{{hep-th/9711200}}]}.

\bibitem{Mason:2008jy}
Mason, L.~J., and Skinner, D., ``Gravity, Twistors and the {M}{H}{V}
  Formalism'', {\em Commun. Math. Phys.}, {\bf 294}, 827--862, (2010).
  {\small[\href{http://dx.doi.org/10.1007/s00220-009-0972-4}{DOI}]},
  {\small[\href{http://arxiv.org/abs/arXiv:0808.3907}{{arXiv:0808.3907}}]}.

\bibitem{Hspace}
Newman, E.T., ``Heaven and Its Properties'', {\em Gen. Relativ. Gravit.}, {\bf
  7}, 107--111, (1976).
  {\small[\href{http://dx.doi.org/10.1007/BF00762018}{DOI}]}.

\bibitem{Maxwell}
Newman, E.T., ``Maxwell fields and shear-free null geodesic congruences'', {\em
  Class. Quantum Grav.}, {\bf 21}, 3197--3221, (2004).
  {\small[\href{http://dx.doi.org/10.1088/0264-9381/21/13/007}{DOI}]}.

\bibitem{Twistors1}
Newman, E.T., ``Asymptotic twistor theory and the Kerr theorem'', {\em Class.
  Quantum Grav.}, {\bf 23}, 3385--3392, (2006).
  {\small[\href{http://dx.doi.org/10.1088/0264-9381/23/10/009}{DOI}]},
  {\small[\href{http://arxiv.org/abs/gr-qc/0512079}{{gr-qc/0512079}}]}.

\bibitem{Newman:2011im}
Newman, E.T., ``Newton's second law, radiation reaction and type {I}{I}
  {E}instein-{M}axwell fields'', {\em Class. Quantum Grav.}, {\bf 28}, 245003,
  (2011).
  {\small[\href{http://dx.doi.org/10.1088/0264-9381/28/24/245003}{DOI}]},
  {\small[\href{http://arxiv.org/abs/arXiv:1109.4106}{{arXiv:1109.4106}}]}.

\bibitem{KerrNewman}
Newman, E.T., Couch, E., Chinnapared, K., Exton, A., Prakash, A., and Torrence,
  R., ``Metric of a Rotating, Charged Mass'', {\em J. Math. Phys.}, {\bf 6},
  918--919, (1965). {\small[\href{http://dx.doi.org/10.1063/1.1704351}{DOI}]}.

\bibitem{ScriCR}
Newman, E.T., and Nurowski, P., ``CR structures and asymptotically flat
  spacetimes'', {\em Class. Quantum Grav.}, {\bf 23}, 3123--3127, (2006).
  {\small[\href{http://dx.doi.org/10.1088/0264-9381/23/9/022}{DOI}]},
  {\small[\href{http://arxiv.org/abs/gr-qc/0511119}{{gr-qc/0511119}}]}.

\bibitem{NPF}
Newman, E.T., and Penrose, R., ``An Approach to Gravitational Radiation by a
  Method of Spin Coefficients'', {\em J. Math. Phys.}, {\bf 3}, 566--578,
  (1962). {\small[\href{http://dx.doi.org/10.1063/1.1724257}{DOI}]},
  {\small[\href{http://adsabs.harvard.edu/abs/1962JMP.....3..566N}{ADS}]}.

\bibitem{BMS}
Newman, E.T., and Penrose, R., ``Note on the Bondi--Metzner--Sachs Group'',
  {\em J. Math. Phys.}, {\bf 7}, 863--870, (1966).
  {\small[\href{http://dx.doi.org/10.1063/1.1931221}{DOI}]},
  {\small[\href{http://adsabs.harvard.edu/abs/1966JMP.....7..863N}{ADS}]}.

\bibitem{NewmanPenrose2009}
Newman, E.T., and Penrose, R., ``Spin-coefficient formalism'', {\em
  Scholarpedia}, {\bf 4}(6), 7445, (2009). URL (cited on 30 July 2009):
  \newline\url{http://www.scholarpedia.org/article/Spin-coefficient_formalism}.

\bibitem{RTmetrics2}
Newman, E.T., and Posadas, R., ``Motion and Structure of Singularities in
  General Relativity'', {\em Phys. Rev.}, {\bf 187}, 1784--1791, (1969).
  {\small[\href{http://dx.doi.org/10.1103/PhysRev.187.1784}{DOI}]},
  {\small[\href{http://adsabs.harvard.edu/abs/1969PhRv..187.1784N}{ADS}]}.

\bibitem{Spins}
Newman, E.T., and Silva-Ortigoza, G., ``Tensorial spin-s harmonics'', {\em
  Class. Quantum Grav.}, {\bf 23}, 497--509, (2006).
  {\small[\href{http://dx.doi.org/10.1088/0264-9381/23/2/014}{DOI}]},
  {\small[\href{http://arxiv.org/abs/gr-qc/0508028}{{gr-qc/0508028}}]}.

\bibitem{NewmanTod}
Newman, E.T., and Tod, K.P., ``Asymptotically flat space-times'', in Held, A.,
  ed., {\em General Relativity and Gravitation: One Hundred Years After the
  Birth of Albert Einstein}, vol.~2, pp. 1--36, (Plenum Press, New York, 1980).

\bibitem{NUT}
Newman, E.T., and Unti, T.W.J., ``Behavior of Asymptotically Flat Empty
  Spaces'', {\em J. Math. Phys.}, {\bf 3}, 891--901, (1962).
  {\small[\href{http://dx.doi.org/10.1063/1.1724303}{DOI}]},
  {\small[\href{http://adsabs.harvard.edu/abs/1962JMP.....3..891N}{ADS}]}.

\bibitem{Scri1}
Penrose, R., ``Asymptotic Properties of Fields and Space-Times'', {\em Phys.
  Rev. Lett.}, {\bf 10}, 66--68, (1963).
  {\small[\href{http://dx.doi.org/10.1103/PhysRevLett.10.66}{DOI}]},
  {\small[\href{http://adsabs.harvard.edu/abs/1963PhRvL..10...66P}{ADS}]}.

\bibitem{Scri2}
Penrose, R., ``Zero Rest-Mass Fields Including Gravitation: Asymptotic
  Behaviour'', {\em Proc. R. Soc. London, Ser. A}, {\bf 284}, 159--203, (1965).
  {\small[\href{http://dx.doi.org/10.1098/rspa.1965.0058}{DOI}]},
  {\small[\href{http://adsabs.harvard.edu/abs/1965RSPSA.284..159P}{ADS}]}.

\bibitem{PenroseTwist}
Penrose, R., ``Twistor Algebra'', {\em J. Math. Phys.}, {\bf 8}, 345--366,
  (1967). {\small[\href{http://dx.doi.org/10.1063/1.1705200}{DOI}]}.

\bibitem{BMS2}
Penrose, R., ``Relativistic symmetry groups'', in Barut, A.O., ed., {\em Group
  Theory in Non-Linear Problems}, Proceedings of the NATO Advanced Study
  Institute, held in Istanbul, Turkey, August 7\,--\,18, 1972, NATO ASI Series
  C, vol.~7, pp. 1--58, (Reidel, Dordrecht; Boston, 1974).

\bibitem{Spinors}
Penrose, R., and Rindler, W., {\em Spinors and space-time, Vol. 1: Two-spinor
  calculus and relativistic fields}, Cambridge Monographs on Mathematical
  Physics, (Cambridge University Press, Cambridge; New York, 1984).
  {\small[\href{http://books.google.com/books?id=CzhhKkf1xJUC}{Google Books}]}.

\bibitem{PenroseRindler2}
Penrose, R., and Rindler, W., {\em Spinors and space-time, Vol. 2: Spinor and
  twistor methods in space-time geometry}, Cambridge Monographs on Mathematical
  Physics, (Cambridge University Press, Cambridge; New York, 1986).
  {\small[\href{http://books.google.com/books?id=fzgIuOozIb8C}{Google Books}]}.

\bibitem{Petrov}
Petrov, A.Z., ``The Classification of Spaces Defining Gravitational Fields'',
  {\em Gen. Relativ. Gravit.}, {\bf 32}, 1665--1685, (2000).
  {\small[\href{http://dx.doi.org/10.1023/A:1001910908054}{DOI}]}.

\bibitem{Pirani}
Pirani, F.A.E., ``Invariant Formulation of Gravitational Radiation Theory'',
  {\em Phys. Rev.}, {\bf 105}(3), 1089--1099, (1957).
  {\small[\href{http://dx.doi.org/10.1103/PhysRev.105.1089}{DOI}]}.

\bibitem{Robinson}
Robinson, I., ``Null Electromagnetic Fields'', {\em J. Math. Phys.}, {\bf 2},
  290--291, (1961). {\small[\href{http://dx.doi.org/10.1063/1.1703712}{DOI}]}.

\bibitem{RobinsonTrautman}
Robinson, I., and Trautman, A., ``Some spherical gravitational waves in general
  relativity'', {\em Proc. R. Soc. London, Ser. A}, {\bf 265}, 463--473,
  (1962). {\small[\href{http://dx.doi.org/10.1098/rspa.1962.0036}{DOI}]}.

\bibitem{Sachs}
Sachs, R.K., ``Gravitational Waves in General Relativity. VIII. Waves in
  Asymptotically Flat Space-Time'', {\em Proc. R. Soc. London, Ser. A}, {\bf
  270}, 103--126, (1962).
  {\small[\href{http://dx.doi.org/10.1098/rspa.1962.0206}{DOI}]},
  {\small[\href{http://adsabs.harvard.edu/abs/1962RSPSA.270..103S}{ADS}]}.

\bibitem{Peeling}
Sachs, R.K., ``Gravitational radiation'', in DeWitt, C.M., and DeWitt, B.,
  eds., {\em Relativity, Groups and Topology}, Lectures delivered at Les
  Houches during the 1963 session of the Summer School of Theoretical Physics,
  University of Grenoble, pp. 523--562, (Gordon and Breach, New York, 1964).

\bibitem{Spacelike2}
Sommers, P., ``The geometry of the gravitational field at spacelike infinity'',
  {\em J. Math. Phys.}, {\bf 19}, 549--554, (1978).
  {\small[\href{http://dx.doi.org/10.1063/1.523698}{DOI}]},
  {\small[\href{http://adsabs.harvard.edu/abs/1978JMP....19..549S}{ADS}]}.

\bibitem{Szabados}
Szabados, L.B., ``Quasi-Local Energy-Momentum and Angular Momentum in General
  Relativity'', {\em Living Rev. Relativity}, {\bf 12}, lrr-2009-4, (2009). URL
  (cited on 31 July 2009):
  \newline\url{http://www.livingreviews.org/lrr-2009-4}.

\bibitem{'tHooft:1973jz}
't~Hooft, G., ``A Planar Diagram Theory for Strong Interactions'', {\em Nucl.
  Phys.}, {\bf B72}, 461, (1974).
  {\small[\href{http://dx.doi.org/10.1016/0550-3213(74)90154-0}{DOI}]}.

\bibitem{'tHooft:1993gx}
't~Hooft, G., ``Dimensional reduction in quantum gravity'', (1993).
  {\small[\href{http://arxiv.org/abs/gr-qc/9310026}{{gr-qc/9310026}}]}.

\bibitem{Witten:1998qj}
Witten, E., ``Anti-de {S}itter space and holography'', {\em Adv. Theor. Math.
  Phys.}, {\bf 2}, 253--291, (1998).
  {\small[\href{http://arxiv.org/abs/hep-th/9802150}{{hep-th/9802150}}]}.

\end{thebibliography}

\end{document}